\documentclass[11pt]{article}
\pdfoutput=1
\usepackage{graphicx}
\usepackage{amsmath,amssymb,enumerate}
\usepackage{slashed}
\usepackage{amsfonts}
\usepackage{geometry}
\usepackage{color}
\usepackage{appendix}
\usepackage{resizegather}
\usepackage{bm}
\usepackage{subfig}
\usepackage{cite}
\definecolor{nicered}{rgb}{0.7,0.1,0.1}
\definecolor{nicegreen}{rgb}{0.1,0.5,0.1}

\usepackage{array}
\newcolumntype{L}[1]{>{\raggedright\let\newline\\\arraybackslash\hspace{0pt}}m{#1}}
\newcolumntype{C}[1]{>{\centering\let\newline\\\arraybackslash\hspace{0pt}}m{#1}}
\newcolumntype{R}[1]{>{\raggedleft\let\newline\\\arraybackslash\hspace{0pt}}m{#1}}

\usepackage[normalem]{ulem}
\usepackage{setspace}

\def\wdt{\widetilde}
\def\({\left(}
\def\){\right)}
\def\[{\left[}
\def\]{\right]}

\newcommand{\ba}{\begin{array}}
\newcommand{\ea}{\end{array}}
\newcommand{\bd}{\begin{displaymath}}
\newcommand{\ed}{\end{displaymath}}
\newcommand{\be}{\begin{equation}}
\makeatletter
\newcommand*{\rom}[1]{\expandafter\@slowromancap\romannumeral #1@}
\makeatother
\newcommand{\ee}{\end{equation}}
\def\bt{\begin{table}}
\def\et{\end{table}}
\def\bc{\begin{center}}
\def\ec{\end{center}}
\def\bi{\begin{itemize}}
\def\ei{\end{itemize}}
\def\bw{\begin{widetext}}
\def\ew{\end{widetext}}

\def\bea{\begin{eqnarray}}
\def\eea{\end{eqnarray}}
\def\beas{\begin{eqnarray*}}
\def\eeas{\end{eqnarray*}}

\def\e{\ell}

\def\N0{\widetilde{\chi}^0}

\def\g{\gamma}
\def\d{\delta}
\def\e{\epsilon}

\def\m{\mu}

\def\D{\Delta}

\def\s{\sigma}

\def\ov {\overline}

\def\l{\lambda}

\allowdisplaybreaks

\begin{document}


\onehalfspacing

\renewcommand{\thefootnote}{\fnsymbol{footnote}}

\thispagestyle{empty}

\begin{flushright}
OSU-HEP-14-12
\end{flushright}

\vspace*{2.0cm}
\begin{center}
{\LARGE Higgs Boson Spectra in Supersymmetric Left-Right
Models \\[0.1in]}
\end{center}

\vspace*{0.2in}

\begin{center}
{\Large  K.S. Babu}$^{\footnotesize {a,b}}$\footnote{Email:
babu@okstate.edu}
{\it \large and} {\Large  Ayon Patra}$^{\footnotesize b,c}$\footnote{Email: ayon@okstate.edu}
\end{center}

\begin{center}
\it{ $^a$Kavli Institute for Theoretical Physics, University of California,\\ Santa Barbara, CA 93106, USA}
\end{center}

\begin{center}
{$^{\footnotesize b}$\large \it Department of Physics, Oklahoma State University, \\Stillwater, Oklahoma 74078, USA}
\end{center}

\vspace{-0.2in}
\begin{center}
{$^{\footnotesize c}$\large \it Institute of Convergence Fundamental Studies, \\Seoul National University of Science and Technology, Seoul 139-743, Korea}

\end{center}

\renewcommand{\thefootnote}{\arabic{footnote}}
\setcounter{footnote}{0}

\begin{abstract}

We present a comprehensive analysis of the Higgs boson spectra in several versions of the supersymmetric left--right model based on the gauge symmetry $SU(3)_c \times SU(2)_L \times SU(2)_R \times U(1)_{B-L}$.  A variety of symmetry breaking sectors
are studied, with a focus on the constraints placed on model parameters by the lightest neutral CP even Higgs boson mass $M_h$.
The breaking of $SU(2)_R$ symmetry is achieved by Higgs fields transforming either as triplets or doublets, and the
electroweak symmetry breaking is triggered by either bi--doublets or doublets.  The Higgs potential is analyzed with or without a gauge singlet Higgs field present. Seesaw models of Type I and Type II, inverse seesaw models, universal seesaw models and an $E_6$ inspired alternate left--right model are included in our analysis.  Several of these models lead to the tree--level relation $M_h \leq \sqrt{2}\,m_W$ (rather than $M_h \leq m_Z$ that arises in the MSSM), realized when the $SU(2)_R$ symmetry breaking scale is of order TeV. With such an enhanced upper limit, it becomes possible to accommodate a Higgs boson of mass 126 GeV with relatively light stops that mix negligibly. In models with Higgs triplets, a doubly charged scalar remains light below a TeV with its mass arising entirely from radiative corrections.  We carry out the complete one--loop calculation for its mass induced by the Majorana Yukawa couplings and show the consistency of the framework. We argue that these models prefer a low $SU(2)_R$ breaking scale.  Other theoretical and phenomenological implications of these models are briefly discussed.

\end{abstract}

\newpage

\section{Introduction}

Models based on the left--right symmetric gauge group $G_{3221} \equiv SU(3)_c \times SU(2)_L \times SU(2)_R \times U(1)_{B-L}$ \cite{lr}
are attractive extensions of the Standard Model (SM) with several interesting features.  At the fundamental level Parity
is a good symmetry in these models. The observed Parity violation in weak interactions is explained by the spontaneous breaking of $SU(2)_R \times U(1)_{B-L}$ down to $U(1)_Y$ of the SM at a scale $v_R$ well above the masses of the $W$ and $Z$ bosons.  The gauge structure requires the
existence of the right--handed neutrino, and thus leads naturally to small neutrino masses via the seesaw mechanism.  In fact,
with the right--handed neutrino included, $G_{3221}$ is the maximal flavor--blind gauge symmetry that can be realized at a scale of order TeV, relevant to the ongoing LHC experiments.\footnote{There is a natural embedding of $G_{3221}$ into the Pati--Salam symmetry $G_{422} \equiv SU(4)_c \times SU(2)_L \times SU(2)_R$ \cite{ps}, however, the scale of $G_{422}$ symmetry breaking must be of order $10^5$ GeV or above, from
$K_L \rightarrow \mu e$ decay constraints.  Embedding $G_{3221}$ (or $G_{422}$) into the unified symmetry group of $SO(10)$ is very natural, but that
symmetry breaking scale must be of order $10^{15}$ GeV, from constraints on nucleon decay and gauge coupling unification.}  Because
of Parity invariance these models can potentially solve the strong CP problem \cite{strongcp} without introducing a global Peccei--Quinn symmetry and
the resulting axion.

Supersymmetric versions of left--right gauge models, denoted here as SUSYLR models, preserve the merits of $G_{3221}$ noted above, and in addition, can solve the gauge hierarchy problem.  These models can have a natural dark matter candidate in the
lightest supersymmetric particle, with an unbroken $R$-parity emerging from the $U(1)_{B-L}$ gauge symmetry \cite{mohapR}.
It has been noted that the puzzle of small phases in the SUSY breaking sector
(arising from electric dipole moment constraints) has a natural explanation
in  SUSY left--right models, by virtue of Parity symmetry \cite{susycp}.  Several versions of the SUSYLR models have been proposed and
studied in the literature, with differing Higgs boson sectors used for symmetry breaking \cite{bhm,ma,km,susylr,bm0,amz}.
Here we undertake a systematic study of the Higgs potential in various realizations of these models, focussing on the
lightest neutral Higgs boson mass $M_h$.  In many cases
we find that the tree--level constraint $M_h \leq m_Z$  of the MSSM is modified to the less stringent constraint $M_h \leq \sqrt{2}\, m_W$ \cite{bhm}. In some models this limit is relaxed even further.  
This difference in the upper limit arises from the non-decoupling $D$--terms of $SU(2)_R \times U(1)_{B-L}$, which occurs
when the symmetry breaking scale $v_R$ and the SUSY breaking scale are of the same order. Thus, these models
would predict additional $W^\pm_R$ and $Z_R$ gauge bosons within reach of LHC experiments, in addition to SUSY particles in the parameter regime where the upper limit on $M_h$ is relaxed.
In the MSSM heavy stops ($m_{\tilde{t}} > 2$ TeV) with large mixing are typically needed in order to accommodate the Higgs boson of mass 126 GeV
discovered recently at LHC.  Such a large mass of the stop puts the gauge hierarchy problem in a different perspective, since
some amount of tuning would be required. With the increased tree--level mass of $M_h$, SUSYLR models would allow for the stops
to be much lighter and less mixed, and thus would alleviate the tuning problem.

Our analysis focuses on two basic classes of SUSYLR models which have been developed in the literature.
In one class Higgs triplets are introduced for $SU(2)_R$  symmetry breaking along with $SU(2)_L \times
SU(2)_R$ bi-doublets which break the electroweak symmetry \cite{km,susylr,bm0,amz}.  Fermion mass generation is via direct Yukawa couplings in this class of models,
including the Majorana mass of the right-handed neutrino.  In a second class, Higgs doublets are used to break $SU(2)_R$ symmetry, with $SU(2)_L$
doublets and/or $SU(2)_L \times SU(2)_R$ bi-doublets breaking the electroweak symmetry.  Additional fermions are necessary in this class for fermion mass
generation, at least in the neutrino sector.  A specific example studied incorporates the
inverse seesaw mechanism for neutrino masses with the inclusion of gauge singlet fermions. Another
example, termed alternate left--right model \cite{ma,bhm}, has an $E_6$ inspired particle spectrum.  A third example uses
a universal seesaw mechanism for quarks and leptons by introducing vector--like $SU(2)_L \times SU(2)_R$
gauge singlet quarks and leptons \cite{bm2,universal}.  In each of the cases listed above,
we also allow for the presence of a Higgs singlet scalar, which would admit the possibility of $SU(2)_R \times U(1)_{B-L}$ symmetry breaking down to $U(1)_Y$ in the SUSY limit.
In the absence of such a gauge singlet Higgs field this symmetry breaking scale would be of the same order as the supersymmetry breaking scale, which is shown to be consistent with experimental limits.
In certain cases with Higgs singlets we present approximate analytic expressions for the limit on $M_h$ that interpolate between the limit $M_h \leq  m_Z$ of the MSSM if the $SU(2)_R$ breaking scale is much higher than the
SUSY breaking scale, and the limit $M_h \leq \sqrt{2}\, m_W$ which arises if this scale is comparable to the SUSY breaking scale.

Non--decoupling $D$--term effects on the lightest Higgs boson mass in extensions of the MSSM have been studied by various
authors.  In Ref. \cite{bhm} symmetry breaking in SUSYLR models with an $E_6$ inspired particle spectrum was studied and
a relation $M_h \leq \sqrt{2} \,m_W$ was derived.   In Ref. \cite{amz} symmetry breaking of SUSYLR models with
Higgs triplets was studied and an enhancement of $M_h$ compared to the MSSM result was observed.  Ref. \cite{bdkt} has studied extended gauge
sectors, including an extra $SU(2)$ added to the SM gauge symmetry.  In some cases there is an unknown gauge coupling, which was chosen so that it remains perturbative all the way to a GUT scale, and significant increase in $M_h$ was observed.  In Ref. \cite{malinsky}
non--decoupling effects of an additional $U(1)$ gauge symmetry was studied, which also showed a modest increase in $M_h$.
Our aim in this paper is to systematically study the Higgs boson sectors of various realizations of SUSYLR models, which
has some overlap with some of the earlier studies.  In one case we reproduce and generalize the results of Ref. \cite{bhm}. In another case studied, where we provide an analytic formula for the upper limit on $M_h$ that interpolates between the decoupling and non--decoupling limits of left--right symmetry, our results agree roughly with the numerical results of Ref. \cite{amz}.
We provide complete listings of the Higgs spectra for each case studied.  Many of the examples, such as the
inverse seesaw model and the universal seesaw model, are analyzed in the SUSYLR framework for the first time here.

When gauge singlets that couple to the MSSM Higgs fields are present in the theory, additional $F$--term contributions to
$M_h$ arise.  In several cases this contribution is non--decoupling, a well-known case being the
next to minimal supersymmetric standard model (NMSSM) \cite{ellwanger}.  Modest increase in
$M_h$ can arise from this contribution, although we find the non--decoupling $D$--term to be somewhat more significant in the SUSYLR models.

There are direct and indirect limits on the mass of the $W_R^\pm$ gauge boson of left--right symmetric models with or without supersymmetry.  CMS collaboration has obtained a lower limit on $M_{W_R}$ that ranges from 1.8 TeV to 3 TeV, depending on the mass of the right--handed neutrino, if the gauge coupling of $W_R$ is the same as that of the Standard Model
$W$ boson \cite{cms}.  This limit is obtained with 19.6 $fb^{-1}$ data collected at 8 TeV by looking for an excess in the $eejj$ and $\mu \mu jj$ channels in $pp$ collision \cite{keung}.  For a discussion of limits on the $W_R$ mass in the minimal left-right symmetric model see Ref. \cite{miha}.  Indirect limits on $W_R$ mass arise from box diagram contributions
to $K^0-\overline{K^0}$ mixing which is found in the minimal model to be $m_{W_R} > 1.8$ TeV \cite{soni}.
These limits are all compatible with the scale of left--right symmetry breaking comparable to that of SUSY breaking.

In minimal SUSYLR models with Higgs triplets used for $SU(2)_R$ symmetry breaking, a doubly charged scalar remains
light with its mass below a TeV.  This state acquires its mass entirely through radiative corrections.  The consistency
of such a framework was shown in Ref. \cite{bm0}.  In this paper we carry out a complete one--loop calculation of the
doubly charged Higgs boson mass in these models arising from the Majorana Yukawa couplings.  We demonstrate the finiteness
of the mass, and show that if the right--handed symmetry breaking scale is taken to be much above the SUSY breaking
scale, the squared mass of this field would be negative.  Minimal models with Higgs triplets would be suggestive of low energy $SU(2)_R$ breaking.

The remainder of the paper is organized as follows.  In Sec. II we introduce the various versions of SUSY left-right
model.  Here we list the Higgs content and explain how realistic fermion masses, including neutrino masses, are generated.  In Sec. III we analyze the Higgs potentials of SUSYLR models with triplet
scalars breaking the $SU(2)_R \times U(1)_{B-L}$ symmetry.  Various scenarios are discussed here.  For electroweak symmetry breaking we allow for one or two bi-doublets.  We also allow for a gauge singlet that facilitates LR symmetry breaking in the SUSY limit. We focus on the lightest neutral Higgs boson mass and derive the tree--level constraint $M_h \leq \sqrt{2}\,m_W$
in one case.   In Sec. IV we analyze inverse seesaw models which utilize Higgs doublets and bidoublets.  Sec. V has our results on the universal seesaw models which contain
only Higgs doublet fields.  In Sec. VI, an $E_6$ inspired SUSYLR model is studied in detail.  Sec. VII is devoted to the calculation of one-loop radiative corrections to the doubly charged Higgs boson mass in SUSYLR models with triplet Higgs.
This particle is predicted to be light, below a TeV, regardless of the scale of SUSY breaking.  Sec. VIII have our
conclusions and some discussions. We collect several results relevant for the symmetry breaking analyses in the Appendix.

\section {Variety of Supersymmetric Left--Right Models} \label{sec2}

In this section we introduce various realizations of the SUSY left--right model.  All these realizations have an extended
gauge symmetry which is $SU(3)_c \times SU(2)_L \times SU(2)_R \times U(1)_{B-L}$. We develop and analyze different Higgs boson sectors that break this symmetry down to the SM symmetry and then to $U(1)_{\rm em}$.
The right-handed $SU(2)_R$ symmetry breaking can be achieved either by Higgs triplets or by Higgs doublets, while the electroweak symmetry may be broken by Higgs bidoublets or by $SU(2)_L$ doublets.
In each case we allow for the possibility of a gauge singlet scalar field as well, which would enable the
breaking of $SU(2)_R \times U(1)_{B-L}$ down to $U(1)_Y$  in the supersymmetric limit.  If such a gauge singlet Higgs scalar is not present, which is also studied, then the scale of $SU(2)_R$ symmetry breaking should be comparable to the scale
of SUSY breaking. We also investigate the Higgs boson spectrum of an alternate left--right symmetric model motivated by
$E_6$ unification.

Each of these models  has a common chiral fermion sector consisting of three families of quark and lepton superfields given as\footnote{The identification of normal quarks and leptons will be slightly different in the $E_6$ motivated SUSYLR model.}
\begin{eqnarray}
\!\!Q\!\!&=&\!\!\begin{pmatrix}
u\\ d \end{pmatrix}  \sim \left (3,2, 1, \frac13 \right ),~~~~~~~~
Q^c\!=\!\begin{pmatrix}
d^c\\-u^c
\end{pmatrix} \sim \left ( 3^{\ast},1, 2, -\frac13
\right ),\nonumber \\
L&=&\begin{pmatrix}
\nu\\ e \end{pmatrix} \sim\left ( 1,2, 1, -1 \right ),~~~~~~~~
L^c=\begin{pmatrix}
e^c \\ -\nu^c \end{pmatrix} \sim \left ( 1,1, 2, 1 \right ),
\label{eq:matter}
\end{eqnarray}
where the quantum numbers under
$SU(3)_c \times SU(2)_L \times SU(2)_R \times U(1)_{B-L}$ gauge group are listed.

We require that each of the models studied must meet four basic criteria. First, a mechanism for $SU(2)_R$ breaking consistent with the experimental limits on the $W_R^\pm$ and $Z_R$ gauge boson masses must be present. Second, the model must be able to generate realistic quark and lepton masses. Third, there must be a mechanism to generate small neutrino masses.
This could be Type I or Type II seesaw, inverse seesaw or a universal seesaw mechanism.
Fourth, there should be an unbroken $R$-parity that provides a dark matter candidate.  In models with Higgs triplets, this
last requirement turns out to be automatic, but with Higgs doublets fields, an additional $Z_2$ symmetry will be assumed
that distinguishes lepton doublets from the Higgs doublets, as is usually done within the MSSM framework.

 We now describe briefly each of these models, with a more extended discussion on the Higgs boson spectrum delegated to subsequent sections.

\subsection{Models with Higgs triplets and bidoublets}\label{2.1}

This scenario satisfies our requirements for a consistent model in the most straightforward way. An $SU(2)_R$ triplet Higgs field $\D^c(1,1,3,-2)$ is introduced, which breaks the gauge symmetry once its neutral component acquires a nonzero vacuum expectation value (VEV). It also couples to the right-handed neutrinos and generates Majorana masses for them.
Two bidoublet fields $\Phi_a(1,2,2,0)$ are introduced which can have Yukawa couplings with the quarks and leptons, generating their (Dirac) masses and CKM mixing angles. (With only one bidoublet, the CKM mixing angles would
all vanish.  We shall also study the simpler case of having only one bidoublet in the theory, which would ascribe the
CKM mixings to soft SUSY breaking \cite{bdm1}.)
Being a supersymmetric theory the right-handed $\D^c(1,1,3,-2)$ field must be accompanied by another $SU(2)_R$ triplet field $\ov \D^c(1,1,3,+2)$ for anomaly cancellation and for achieving symmetry breaking consistently.
In a left-right symmetric model, the right-handed triplets must also be accompanied by left-handed triplet partners $\D(1,3,1,2)$ and $\ov \D(1,3,1,-2)$ for parity conservation.
We allow for cases with and without an extra gauge singlet scalar field $S(1,1,1,0)$. In the absence of the field $S$, it is not possible to break the $SU(2)_R$ symmetry in the supersymmetric limit.  This is a viable possibility, as we shall see.
The presence of the singlet field $S$ would enable decoupling the two symmetry breaking scales, allowing the $SU(2)_R$ symmetry to be broken at a much higher scale than SUSY breaking. Thus, the Higgs boson fields in this model are given as
\begin{eqnarray}
\Delta(1,3,1,2)&=&\begin{pmatrix}
\frac{\delta^{+}}{\sqrt{2}} & \delta^{++}\\ \delta^{0} & -\frac{\delta^{+}}{\sqrt{2}}  \end{pmatrix},~~~~
\overline{\Delta}(1,3,1,-2)=\begin{pmatrix}
\frac{\overline{\delta}^{-}}{\sqrt{2}} & \overline{\delta}^{0}\\ \overline{\delta}^{--} & -\frac{\overline{\delta}^{-}}{\sqrt{2}}  \end{pmatrix},\nonumber \\
\Delta^{c}(1,1,3,-2)&=&\begin{pmatrix}
\frac{\delta^{c^{-}}}{\sqrt{2}} & \delta^{c^{0}}\\ \delta^{c^{--}} & -\frac{\delta^{c^{-}}}{\sqrt{2}}  \end{pmatrix},~~~~
\overline{\Delta}^{c}(1,1,3,2)=\begin{pmatrix}
\frac{\overline{\delta}^{c^{+}}}{\sqrt{2}} & \overline{\delta}^{c^{++}}\\ \overline{\delta}^{c^{0}} & -\frac{\overline{\delta}^{c^{+}}}{\sqrt{2}}  \end{pmatrix},\nonumber \\
\Phi_i(1,2,2,0)&=&{\begin{pmatrix}
\phi^{+}_1 & \phi^{0}_{2} \\ \phi^{0}_{1} & \phi^{-}_{2} \end{pmatrix}_i}~~(i=1,2),~~~~ S(1,1,1,0).
\label{eq:triphig}
\end{eqnarray}
The nonzero VEVs of various component fields are denoted as
\begin{equation}
\left< \d^{c^0} \right> = v_R,~~~~ \left< \ov \d^{c^0} \right> = \ov v_R,~~~~ \left< \phi_{1_i}^0 \right> = v_{u_i},~~~~ \left< \phi^0_{2_i} \right> = v_{d_i}~.
\label{vev}
\end{equation}
None of the other fields acquire vacuum expectation value. In particular, $\langle \delta^0 \rangle = 0$ in this model,
so that there is no type II seesaw contribution to the neutrino masses.  We shall take the limit where $v_R, \ov v_R >> v_u, v_d$.

The Yukawa couplings in the model are given by the superpotential
\begin{eqnarray}
W_Y &=& \sum_{j=1}^2 \left( Y_q^{(j)} Q^T \tau_2 \Phi_j \tau_2 Q^c + Y_l^{(j)} L^T \tau_2 \Phi_j \tau_2 L^c \right) + i \frac{f}{2} L^T \tau_2 \D L + i \frac{f^c}{2} {L^c}^T \tau_2 \D^c L^c ,
\end{eqnarray}
where $Y_q^j$ and $Y_l^j$ are the quark and lepton Yukawa coupling matrices and $f$ is the Majorana  Yukawa coupling matrix which generates large Majorana masses for right-handed neutrinos. This superpotential is invariant under parity transformation under which $\Phi \rightarrow \Phi^{\dagger}, \D \rightarrow \D^{c^*}, \ov \D \rightarrow \ov \D^{c^*}, S \rightarrow S^*,Q \rightarrow Q^{c^*}, L \rightarrow L^{c^*}, \theta \rightarrow \ov{\theta}$, along with
 $W_L^\pm \rightarrow W_R^{\pm *}$. Parity invariance requires the Yukawa coupling matrices $Y_q^j$ and $Y_l^j$ to be hermitian and $f^c=f$. Once the various fileds acquire VEVs as shown in Eq. (\ref{vev}), the following mass matrices for fermions will
 be induced:
 \begin{eqnarray}
 M_u &=& Y_q^{(1)} v_{u_1} + Y_q^{(2)} v_{u_2}, ~~~ M_d = Y_q^{(1)} v_{d_1} + Y_q^{(2)} v_{d_2},\nonumber \\
 M_\nu^D &=& Y_l^{(1)} v_{u_1} + Y_l^{(2)} v_{u_2},~~~  M_\ell = Y_l^{(1)} v_{d_1} + Y_l^{(2)} v_{d_2},\nonumber \\
  M_R &=& f v_R
 \end{eqnarray}
 for the up quarks, down quarks, neutrino Dirac, charged lepton and right-handed Majorana neutrino sectors.
 Note that with only a single bidoublet scalar, the up and down quark mass matrices would become proportional, resulting
 in vanishing CKM angles.

Left--right symmetric models predict the existence of new gauge bosons, one charged $W_R^\pm$ and one neutral $Z_R$.
In the limit where the right-handed symmetry breaking VEVs are much bigger than the electroweak symmetry breaking VEVs, we can neglect the mixing between the left-handed and the right-handed gauge bosons and obtain relations for the masses of the heavier $W_R^\pm$ and $Z_R$ bosons.  They are given by
\begin{equation}
M^2_{W_R^\pm} \simeq \frac{1}{2} g_R^2 (2 v_R^2+2\ov v_R^2+v_{u_i}^2+v_{d_i}^2),
\end{equation}
and
\begin{equation}
M^2_{Z_R} \simeq \frac{g_R^2}{2 \cos^2 \theta_W \cos 2\theta_W} \left[ 4 (v_R^2+\ov v_R^2)\cos^4 \theta_W+(v_{u_i}^2+v_{d_i}^2)\cos^2 2\theta_W\right],
\end{equation}
where $g_R$ is the $SU(2)_R$ gauge coupling, $\theta_W$ is the weak mixing angle, and
the index $i$ is summed over the number of bidoublets in the model.
 Since the gauge boson masses must be consistent with the experimental limits, these expressions will be relevant in setting lower limits on the right-handed symmetry breaking scale $v_R,\,\ov v_R$.

\subsection{Inverse seesaw model}

$SU(2)_R$ gauge symmetry can be broken by Higgs doublet fields, instead of Higgs triplets of the previous subsection.
This would simplify the analysis of the Higgs boson sector considerably. Two bidoublet fields are assumed to be present in the Higgs spectrum as before, which can generate quark and charged lepton masses and CKM mixings. Unlike the Higgs triplet field of the previous subsection, the Higgs doublet fields do not directly couple to fermions, and so
right-handed neutrinos would not receive heavy Majorana masses needed for the seesaw mechanism.
This situation is remedied by slightly complicating the fermion sector with the introduction of gauge singlet neutral fermions  $N_i$ ($i=1-3$), one for each generation, along with the chiral matter fields that are given in Eq.~(\ref{eq:matter}).
The $N$'s can have
gauge invariant Majorana masses which in turn will generate small masses for the ordinary neutrinos via an inverse seesaw mechanism \cite{inverse}.  The Higgs sector of this model consists of the following fields:
\begin{eqnarray}
H_L(1,2,1,-1) &=& \begin{pmatrix}
H_L^0 \\ H_L^{-}  \end{pmatrix},
\ov{H}_L(1,2,1,1) = \begin{pmatrix}
\ov{H}_L^+ \\ \ov{H}_L^0  \end{pmatrix},
H_R(1,1,2,1) = \begin{pmatrix}
H_R^+ \\ H_R^{0}  \end{pmatrix},\nonumber \\
\ov{H}_R(1,1,2,-1) &=&\begin{pmatrix}
\ov H_R^0 \\ \ov H_R^{-}  \end{pmatrix},
\Phi_a(1,2,2,0)= \begin{pmatrix}
\phi^{+}_1 & \phi^{0}_{2} \\ \phi^{0}_{1} & \phi^{-}_{2} \end{pmatrix}_a (a=1,2).
\label{eq:2dh}
\end{eqnarray}
Here the $SU(2)_R$ doublet $H_R(1,1,2,1)$ is accompanied by $\ov{H}_R(1,1,2,-1)$ for anomaly cancelation,
while $H_L(1,2,1,1)+\ov{H}_L(1,2,1,-1)$ are their parity partners.
The VEVs of the neutral components of these fields are parametrized as
\begin{equation}
\left< H_L^0 \right> = v_L,~~~ \left< \ov H_L^0 \right> = \ov v_L,~~~\left< H_R^0 \right> = v_R,~~~ \left< \ov H_R^0 \right> = \ov v_R,~~~ \left< \phi_{1_i}^0 \right> = v_{u_i},~~~ \left< \phi^0_{2_i} \right> = v_{d_i}.
\label{vev2}
\end{equation}
As noted earlier, in this model, a $Z_2$ symmetry is assumed that distinguishes the $H_L$ field and the lepton doublet
$L$ (and similarly $H_R$ and $L^c$ fields).  Under this $Z_2$, the lepton fields are odd, while the Higgs doublet fields
are all even.

The superpotential relevant for quark and lepton mass generation in this case is given as
\begin{eqnarray}
W_Y = \sum_{j=1}^2 \left( Y_q^{(j)} Q^T \tau_2 \Phi_j \tau_2 Q^c + Y_l^{(j)} L^T \tau_2 \Phi_j \tau_2 L^c \right) + i f L^T \tau_2 \ov H_L N + i f^c {L^c}^T \tau_2 \ov H_R N +\frac 12 \mu_N N N
\end{eqnarray}
where $Y_q^{(j)}$ and $Y_l^{(j)}$ are the quark and lepton Yukawa coupling matrices, $f$ and $f^c$ are the
analogs of Dirac Yukawa couplings matrices with the singlet neutrino and $\mu_N$ is the lepton number violating Majorana mass term for $N$.
Under parity symmetry $N$ transforms as $N \rightarrow N^*$.
From parity invariance we see that $Y_q^{(j)}$ and $Y_l^{(j)}$  are hermitian, $f^c=f$ and $\mu_N$ is real.  The $3 N_g \times 3 N_g$ neutrino mass matrix is given as (with $N_g$ being the number of generations)
\begin{equation}
M_\nu = \begin{pmatrix} 0&  Y_l^{(i)} v_{1_i}&f \ov v_L \\ Y_l^{(i) T} v_{1_i} & 0 & f \ov v_R \\ f^T \ov v_L & f^T \ov v_R & \mu_N \end{pmatrix}.
\end{equation}
This is the left--right symmetric realization of the inverse seesaw mechanism \cite{inverse,barr}.
In the limit of small $\ov v_L$ and $\mu_N$, the light neutrino masses would vanish.  Thus, the smallness of
these two parameters would provide an understanding of small neutrino masses.  Note that the condition $\ov v_L$ being
small for explaining small neutrino masses is something specific to the left--right symmetric realization of the
inverse seesaw mechanism.

The heavy gauge boson masses in this model are given as:
\begin{equation}
M^2_{W_R^\pm} \simeq \frac{1}{2} g_R^2 (v_R^2+\ov v_R^2+v_{u_i}^2+v_{d_i}^2),
\label{eq:WR}
\end{equation}
and
\begin{equation}
M^2_{Z_R} \simeq \frac{g_R^2}{2 \cos^2 \theta_W \cos 2\theta_W} \left[(v_R^2+\ov v_R^2)\cos^4 \theta_W+(v_{u_i}^2+v_{d_i}^2)\cos^2 2\theta_W + (v_L^2+\ov v_L^2)\sin^4 \theta_W \right],
\label{eq:ZR}
\end{equation}
where $g_R$ is the $SU(2)_R$ gauge coupling, $\theta_W$ is the weak mixing angle, the index $i$ is summed over the number of bidoublets in the model and the VEVs are given
 in Eq.~(\ref{vev2}). Comparing these expressions for the masses with the experimental limit on the heavy gauge bosons we will be able to set lower limits on $v_R$ and $\ov v_R$.

\subsection{Universal seesaw model}

One can choose an even simpler Higgs boson sector  compared to the Higgs sector of the inverse seesaw model
in order to achieve the desired symmetry breaking. A doublet field $H_R(1,1,2,1)$ is sufficient to break the $SU(2)_R$ symmetry and analogously a doublet field $H_L(1,2,1,-1)$ suffices for $SU(2)_L$ symmetry breaking. Anomaly cancelation, needed in
Higgs sectors of supersymmetric models, would require the addition of a $\ov H_R(1,1,2,-1)$ and $\ov H_L(1,2,1,1)$ fields. This Higgs boson sector, without any bidoublet fields, will not be able to generate quark and lepton masses.
Additional vectorlike quarks and leptons which are $SU(2)_L \times SU(2)_R$ singlets are introduced for this purpose \cite{universal,bm2}. This scenario is termed ``universal seesaw", since all quarks and leptons
would acquire masses via a generalized seesaw involving mixing with the heavy fermions.
The chiral matter sector in this case would consist of the quarks and leptons listed in Eq.~(\ref{eq:matter}) along with a set of heavy singlet quarks and lepton fields, one per generation, denoted as  $P(3,1,1,\frac{4}{3})$, $N(3,1,1,-\frac{2}{3})$ and $E(1,1,1,-2)$ and their conjugates $P^c(3,1,1,-\frac{4}{3}), N^c(3,1,1,\frac{2}{3})$ and $E^c(1,1,1,2)$.
There can also be a neutral singlet lepton denoted as $R(1,1,1,0)$ for generating small neutrino masses via the seesaw mechanism, however, this is not essential as tiny Dirac masses for the neutrinos can be generated at the two-loop level arising through the mixed exchange of from $W_L$ and $W_R$ gauge bosons without breaking lepton number \cite{bh}. The Higgs sector of the universal seesaw model is given by
\begin{eqnarray}
H_L(1,2,1,-1) &=& \begin{pmatrix}
{H_L}^0 \\ {H_L}^{-}  \end{pmatrix},
\ov{H}_L(1,2,1,1) = \begin{pmatrix}
{\ov{H}_L}^+ \\ {\ov{H}_L}^{0}  \end{pmatrix},\nonumber \\
H_R(1,1,2,1) &=& \begin{pmatrix}
{H_R}^+ \\ {H_R}^{0}  \end{pmatrix},
\ov{H}_R(1,1,2,-1) =\begin{pmatrix}
{\ov H_R}^0 \\ {\ov H_R}^{-}  \end{pmatrix}.
\label{eq:2ch}
\end{eqnarray}
The absence of bidoublet fields prevents any direct Yukawa coupling between the left-handed and the right-handed fermion fields
of Eq. (\ref{eq:matter}). The Higgs doublet fields can couple quarks with the vectorlike quarks and leptons with the vectorlike
leptons.  The Yukawa superpotential of the model is given by
\begin{eqnarray}
W_Y &=& y_u Q \ov H_L P^c - y_d Q H_L N^c - y_l L H_L E^c + y_\nu L \ov H_L R \notag \\
&+& y_u^c Q^c \ov H_R P - y_d^c Q^c H_R N - y_l^c L^c H_R E + y_\nu^c L^c \ov H_R R\notag \\
&+& m_u P P^c + m_d N N^c + m_l E E^c + \frac12 m_R R R
\end{eqnarray}
where $y_i$ and $y^c_i$ stand for $3\times3$ Yukawa coupling matrices and $m_i$ are the heavy singlet fermions Majorana mass matrices. Under parity symmetry, $P \rightarrow P^{c*},\, N \rightarrow N^{c*}$ and
$E \rightarrow E^{c*}$.  Parity invariance then requires $y_i^c = y_i^*$ and $m_i$ to be real.  The fermion mass matrix
for the up--qurk, down--quark and charged lepton sectors are given by
\begin{eqnarray}
M_u = \left(\begin{matrix} 0 & y_u \ov v_L \cr y_u^\dagger \ov v_R & m_u \end{matrix}\right);~~
M_d = \left(\begin{matrix} 0 & y_d  v_L \cr y_d^\dagger  v_R & m_d \end{matrix}\right);~~
M_l = \left(\begin{matrix} 0 & y_l  v_L \cr y_l^\dagger  v_R & m_l \end{matrix}\right)~.
\end{eqnarray}
Here $m_u$ is multiplied by $(u,\,P)$ from the left and $(u^c,\,P^c)$ from the right, and so on.
The vacuum expectation values are defined as $\langle H_L^0 \rangle = v_L,\,\langle \ov H_L^0 \rangle =
\ov v_L,\,  \langle H_R^0 \rangle = v_R,\,\langle \ov H_R^0 \rangle =
\ov v_R$. The light fermion mass can be obtained as $m_{\rm down} \sim |y_d|^2  v_L  v_R/m_d$, etc.
Note that the determinant of the quark mass matrices are real, by virtue of
parity symmetry, provided that the VEVs are real.  CP violation will occur via CKM mixings, since the Yukawa couplings themselves are not real.
This feature of the model has been used to explain the strong CP problem without the use of axions \cite{bm0}.

The heavy gauge boson masses can be obtained from Eq.~(\ref{eq:WR}) and Eq.~(\ref{eq:ZR}) by setting $v_{u_i}= v_{d_i} =0$.

We investigate two variations of this model with and without a singlet Higgs boson. The upper limit on the mass of the lightest CP-even Higgs boson will be quite different in the two cases as we show later.

\subsection{$E_6$ motivated left-right supersymmetric model}

This model is motivated by the low energy manifestation of heterotic superstring theory where the matter supermultiplets belong to the {\bf{27}} representation of $E_6$ group. The particle content of this representation under the subgroup  $SU(3)_c\times SU(2)_L\times SU(2)_R\times U(1)$ is given as:
\begin{align}
&Q = (u,d):(3,2,1,\frac{1}{3}),~~ d^c:(\ov{3},1,1,\frac{2}{3}),~~Q^c = (h^c,u^c):(\ov{3},1,2,-\frac{1}{3}), \nonumber \\
&L^c = (e^c,n):(1,1,2,1),~~F = \begin{pmatrix}\nu_e&E^c\\e&N^c_E \end{pmatrix} :(1,2,2,0),~~ h:(3,1,1-\frac{2}{3}),\notag \\
&\psi = (\nu_E,E):(1,2,1,-1),~~N:(1,1,1,0).
\end{align}
We can define an $R$-parity  in this case under which the $\{u,d,\nu_e,e\}$ fields are even while the $\{h,E,\nu_E,N_E^c,n\}$ fields are odd. The $W_R^\pm$ gauge boson is also odd under this $R$-parity. The superpartners of these fields have opposite R-parity. The fermions and the gauge bosons have odd and even $R$-parity respectively, except for  the $W_R^\pm$ gauge boson which is odd as it links particles of opposite $R$-parity.

The Higgs fields of this model can be identified as:
\begin{eqnarray}
H_L(1,2,1,-1) &=&\begin{pmatrix}
{H_L}^0 \\ {H_L}^{-}  \end{pmatrix}=\begin{pmatrix}\tilde{\nu}_E\\ \tilde{E}\end{pmatrix},~~
H_R(1,1,2,1) = \begin{pmatrix}
{H_R}^+ \\ {H_R}^{0}  \end{pmatrix}=\begin{pmatrix}\tilde{e}^c\\ \tilde{n} \end{pmatrix}, \notag \\
\Phi(1,2,2,0)&=&\begin{pmatrix}
\phi^{+}_1 & \phi^{0}_{2} \\ \phi^{0}_{1} & \phi^{-}_{2} \end{pmatrix}=\begin{pmatrix}\tilde{E}^c&\tilde{N}^c_E\\\tilde{\nu}_e&\tilde{e} \end{pmatrix}.~~~~
\label{eq:2ah}
\end{eqnarray}

The Yukawa interaction terms in the superpotential are given as
\begin{equation}
W_{Y}= \l_1 Q d^c \psi+\l_2 Q Q^c F+\l_3h Q^c L^c+\l_4 F L^c \psi+\l_5 F N F+\l_6h d^c N.
\end{equation}
This generates masses for quarks and leptons as well as CKM mixings. A small neutrino mass can be generated by the mixing of the $n, \nu_E$ and the $N_E^c$ fields with $\nu_e$ and $N$ fields. The neutrino mass matrix in the basis $(\nu_e, N,\nu_E,N^c_E,n)$ takes the form
\begin{equation}
m_{\nu} = \begin{pmatrix}
0 & \lambda_5 \left< \widetilde{N}^c_E \right> & 0 & 0 & \lambda_5 \left< \widetilde{N}\right>\\
\lambda_5 \left< \widetilde{N}^c_E\right> & 0 & 0 & 0 & \lambda_5 \left< \widetilde{\nu}_e \right> \\
0 & 0 & 0&\l_4 \left< \widetilde{n} \right>&\l_4 \left< \widetilde{N}^c_E \right>\\
0 & 0 & \l_4 \left< \widetilde{n} \right>&0&\l_4 \left< \widetilde{\nu}_E \right>\\
\lambda_5 \left< \widetilde{N}\right> & \lambda_5 \left< \widetilde{\nu}_e \right>  & \l_4 \left< \widetilde{N}^c_E \right>&\l_4 \left< \widetilde{\nu}_E \right>&0
\end{pmatrix}~.
\end{equation}
In the limit of vanishing $\lambda_5$, $\nu_e$ form an almost Dirac neutrino with $N$.  The VEV $\langle N \rangle$ can be small, since only the $\lambda_5$ coupling can induce its VEV, which should be small for neutrino masses.  In our symmetry
breaking analysis we shall keep $\langle N \rangle=0$.

The heavy gauge boson masses can be obtained from Eq.~(\ref{eq:WR}) and Eq.~(\ref{eq:ZR}) by setting $\ov v_R$  and $\ov v_L$ to be zero.

\section {Symmetry Breaking with Higgs Triplet fields} \label{case1}

In this section we analyze the Higgs sectors of a class of models which use Higgs triplets for $Su(2)_R$ symmetry breaking.
Here we construct the relevant superpotential for symmetry breaking, compute the Higgs potential and from it the Higgs boson
spectrum.  The mixing of Higgsinos with the gauginos is also analyzed.  We concentrate on the lightest neutral CP even Higgs boson mass $M_h$ and study how it gets modified and its effect on the parameter space of each model.

Four models are studied under this class. The first case has a pair of Higgs triplets, one bidoublet and a gauge singlet
in the spectrum.  In the second case, we integrate out the singlet field of case one, but keep its effective non-renormalizable interactions.  The third case has no singlet field at all. Case four has two triplets and two bidoublets, which is fully realistic for fermion mass generation, including CKM mixing angles.  The cases with only one bidoublet field should be thought
of as being special cases of this case, with or without an additional singlet.

\subsection{Case with a pair of triplets, a bidoublet and a gauge singlet} \label{case1d}

We first analyze the case with the triplet Higgs fields $\D,\ov \D,\D^c, \ov \D^c$, one bidoublet Higgs field $\Phi$ and a singlet Higgs field $S$. The quantum numbers and compositions of these fields are shown in Eq. (\ref{eq:triphig}). 
For a fully realistic model we need two bidoublet fields to generate the quark mixings, but for simplicity we will only use a single bidoublet in our calculations in this section. This does not significantly affect the Higgs boson masses as will be shown in a later section.  The most general superpotential terms involving only the Higgs boson fields in this case is given as:
\begin{eqnarray}
W&=& S\left[{\text{Tr}}(\lambda \Delta \overline{\Delta})+{\text{Tr}}(\lambda ^c \Delta^{c}\overline{\Delta}^{c}) +\dfrac{\lambda^{\prime}}{2} {\text{Tr}}(\Phi^{T}\tau_{2}\Phi\tau_{2})-M^2 \right] \notag \\
&+& \text{Tr} \left[ \mu_1 \D \ov \D+\mu_2 \D^c \ov \D^c+\frac{\mu}{2} \left( \Phi^T \tau_2 \Phi \tau_2 \right) \right] + \frac{\mu_S}{2} S^2+\frac{\kappa}{3} S^3,
\end{eqnarray}
where $\l^c = \l^*$, $\mu_1 = \mu_2^*$ and $\l^{\prime}, M^2,\mu$ and $\mu_S$ are real from parity invariance.

The Higgs potential consists of the $F$-terms, $D$-terms and soft supersymmetry-breaking terms, 
\begin{equation}
V_{Higgs} = V_F+V_D+V_{Soft}.
\end{equation}

In this case, the relevant terms in the Higgs potential are given by:
\begin{eqnarray}
\label{eq:1df}
V_F&=&  {\text{Tr}} \left| (\lambda \Delta \overline{\Delta})+(\lambda^* \Delta^{c}\overline{\Delta}^{c}) +\frac{\lambda^\prime}{2}(\Phi^{T}\tau_{2}\Phi\tau_{2})-M^2+\mu_S S+\kappa S^2 \right|^2+\text{Tr} \left| \mu \Phi + \lambda ^ \prime S \Phi \right|^2 \nonumber \\
&+&	\text{Tr} \left[ \left| \mu_1 \D + \lambda S \D \right| ^2 + \left| \mu_1 \ov{\D} + \lambda S \ov{\D} \right| ^2 + \left| \mu_1^* \D^c + \lambda^* S \D^c \right| ^2 \right. \notag \\
&+& \left. \left| \mu_1^* \ov \D^c + \lambda^* S \ov \D^c \right| ^2  \right] , \\
\label{eq:two}
V_D&=&\frac{g_L^2}{8}\sum \limits_{a=1}^3 \left|{\text{Tr}}(2\D^\dagger \tau_a \D+2\ov\D^\dagger \tau_a \ov\D+\Phi^\dagger \tau_a \Phi)\right|^2\nonumber \\
&+&\frac{g_R^2}{8}\sum \limits_{a=1}^3  \left|{\text{Tr}}(2{\D^c}^\dagger \tau_a \D^c+2\ov{\D^c}^\dagger \tau_a \ov\D^c+\Phi^* \tau_a \Phi^T)\right|^2 \nonumber \\
&+& \frac{g_V^2}{2}\left|{\text{Tr}}(\D^\dagger \D- \ov\D^\dagger \ov\D-{\D^c}^\dagger \D^c+ \ov{\D^c}^\dagger \ov\D^c)\right|^2, \\
V_{Soft}&=&m_1^2{\text{Tr}}({\D^c}^\dagger\D^c)+m_2^2{\text{Tr}}({\ov{\D}^c}^\dagger\ov\D^c)+m_3^2{\text{Tr}}(\D^\dagger\D)+m_4^2{\text{Tr}}(\ov\D^\dagger\ov\D)\nonumber \\
&+&m_S^2 |S|^2+ m_5^2 {\text{Tr}}(\Phi^\dagger \Phi)
+\left[\lambda A_{\lambda} S {\text{Tr}}(\D \ov\D +\D^c \ov{\D}^c)+h.c.\right]\nonumber \\&+&[\lambda^\prime A_{\lambda^\prime} S {\text{Tr}}(\Phi^{T}\tau_{2}\Phi\tau_{2})+h.c.]+(\lambda C_{\lambda} M^2 S +h.c.) + \left( \mu_S B_S S^2 + h.c. \right) \notag \\
&+& \left[ \mu_1 B_1 \text{Tr} \left( \D \ov \D \right)+\mu_2 B_2 \text{Tr} \left( \D^c \ov \D^c \right)+\mu B \text{Tr} \left( \Phi^T \tau_2 \Phi \tau_2 \right) +\kappa A_\kappa S^3+ h.c. \right].
\label{eq:pot1d}
\end{eqnarray}

The soft mass terms $m_1^2$ and $m_3^2$  (and similarly $m_2^2$ and $m_4^2$) should be equal with exact parity symmetry, but
we shall allow for soft breaking of parity in these dimension two terms.  With exact parity, consistent symmetry breaking cannot be achieved in this model, as there would be unwanted massless modes if parity is forced to be broken spontaneously. We use this potential to calculate the Higgs boson mass-squared matrices for the charged, neutral CP-even and neutral CP-odd Higgs bosons. For simplicity we will assume all the parameters in the potential to be real. The vacuum structure that we choose is given by:
\begin{equation} \label{eq:vev1a}
\left<\D^c\right> = \begin{pmatrix}
0 & v_R \\ 0 & 0  \end{pmatrix},~~
\left<\ov\D^c\right> = \begin{pmatrix}
0&0\\\ov{v}_R e^{i \phi_R} & 0  \end{pmatrix},~~
\left<\Phi\right> ={\begin{pmatrix}
0 & v_2 \\ v_1 e^{i \phi_1}  & 0 \end{pmatrix}}, \left< S \right> = v_S e^{i \phi_S}.
\end{equation}
while the neutral components of $\D$ and $\ov{\D}$ fields do not get any vacuum expectation value. In the absence of the singlet field $S$, parity conservation dictated that all the soft supersymmetry breaking parameters must be real.  The presence of the singlet does result in a nonzero phase but for simplicity we assume this to be zero. If all parameters in the Higgs potential are real, there should be a minimum that preserves CP invariance.  We focus on this minimum.
So for the present case $\phi_R=0$, $\phi_1=0$ and $\phi_S = 0$. This choice of phases negates the mixing between the scalar and the pseudo-scalar Higgs bosons but does not significantly affect the mass of the lightest CP-even Higgs boson. The values of $v_R$ and $\ov v_R$ are of the order of the right-handed $W_R^\pm$ mass, while $v_1$ and $v_2$ are of electroweak scale and hence $v_R,\ov v_R >> v_1, v_2$. 

We first look at the CP-even Higgs boson which is the main focus of this section. To easily identify the field corresponding to the lightest eigenvalue, we take a linear combination of the Higgs fields so that only two of the newly defined fields get non-zero vacuum expectation values -- one at the high right-handed symmetry breaking scale and the other at the lower electroweak symmetry breaking scale.  The field redefinition that we use is given as:
\begin{align}  \label{eq:basis1d}
\rho_1&=\frac{v_1 \phi_1^0 +v_2 \phi_2^0}{\sqrt{v_1^2+v_2^2}},~ \rho_2=\frac{v_2\phi_1^0-v_1 \phi_2^0}{\sqrt{v_1^2+v_2^2}},~ \rho_3=\frac{v_R {\d^c}^0 +\ov v_R {\ov \d^c}^0}{\sqrt{v_R^2+\ov v_R^2}},~ \rho_4=\frac{\ov v_R{\d^c}^0-v_R {\ov \d^c}^0}{\sqrt{v_R^2+\ov v_R^2}}.
\end{align}
In this rotated basis we calculate the mass matrix subject to the following minimization conditions:
\begin{align}
0= &  v_1 [4 m_5^2 + g_L^2 (v_1^2 - v_2^2) +
     g_R^2 (v_1^2- v_2^2 + 2 v_R^2 - 2 \ov v_R^2)]
      - 8 \lambda^\prime A_{\lambda^\prime}  v_2 v_S  - 8 \mu B v_2 \notag \\
      &+4 {\lambda^\prime} v_2 (M^2-\lambda v_R \ov v_R+ \lambda^\prime v_1 v_2 - \mu_S v_S-\kappa v_S^2) + 4 v_1(\mu+ \lambda^\prime v_S)^2, \notag \\
0= &  v_2 [4 m_5^2 + g_L^2 (v_2^2 - v_1^2) +
     g_R^2 (v_2^2 - v_1^2 - 2 v_R^2 + 2 \ov v_R^2)]
      - 8 \lambda^\prime A_{\lambda^\prime}  v_1 v_S  - 8 \mu B v_1 \notag \\
      &+4 {\lambda^\prime} v_1 (M^2-\lambda v_R \ov v_R+ \lambda^\prime v_1 v_2 - \mu_S v_S-\kappa v_S^2) + 4 v_2(\mu+ \lambda^\prime v_S)^2, \notag \\
0=&  2 m_1^2 v_R + g_R^2 v_R ( v_1^2 - v_2^2 + 2 v_R^2 - 2 \ov v_R^2) +
 2 [g_V^2 v_R (v_R^2 -\ov v_R^2) + \lambda A_\lambda \ov v_R v_S +\mu_1 B_2 \ov v_R \notag \\
 & + v_R \left(\lambda v_S + \mu_1 \right) ^2 +\lambda \ov v_R  (- M^2 +\lambda v_R \ov v_R-\lambda^ \prime v_1 v_2 + \mu_S v_S+\kappa v_S^2 ), \nonumber \\
0=&  2 m_2^2 \ov v_R + g_R^2 \ov v_R ( -v_1^2 + v_2^2 - 2 v_R^2 + 2 \ov v_R^2) +
 2 [g_V^2 \ov v_R (-v_R^2 +\ov v_R^2) + \lambda A_\lambda v_R v_S +\mu_1 B_2 v_R \notag \\
 & + \ov v_R \left(\lambda v_S + \mu_1 \right) ^2 +\lambda v_R  (- M^2 +\lambda v_R \ov v_R-\lambda^ \prime v_1 v_2 + \mu_S v_S+\kappa v_S^2 ) , \nonumber \\
0=&      2 \left[ m_S^2 v_S + C_\lambda M^2 \lambda -
 2 \lambda^\prime A_{\lambda^\prime} v_1 v_2 + \lambda
    A_\lambda v_R \ov v_R + {\lambda^\prime}^2 (v_1^2 + v_2^2) + \lambda^2 (v_R^2 +
      \ov v_R^2)] v_S + \mu_S B_S v_S \right. \notag \\
      &  \mu \lambda^\prime (v_1^2+v_2^2)+  \lambda \mu_1 (v_R^2+\ov v_R^2) + 3 \kappa A_\kappa v_S^2  (\mu_S+2 \kappa v_S) \left( -M^2 + \lambda v_R \ov v_R -\lambda^\prime v_1 v_2 + \mu_S v_S + \kappa v_S^2 \right)  {\Big]}.
\label{eq:min1d}
\end{align}

We first look at the CP even scalar Higgs boson masses. 
We get a $5\times 5$ mass-squared matrix in the basis $({\text{Re}}\rho_1,{\text{Re}}\rho_2,{\text{Re}}\rho_3,{\text{Re}}\rho_4,{\text{Re}}{S})$ where one of the eigenvalues would remain light. The relevant terms in this $5\times 5$ mass-squared matrix are given as:
\begin{eqnarray} \label{eq:masmat1d}
M_{11}& =&  \frac{g_L^2 (v_1^2 - v_2^2)^2 + g_R^2 (v_1^2 - v_2^2)^2 +
   8 v_1^2 v_2^2 {\lambda^\prime}^2}{2 (v_1^2 + v_2^2)},\notag \\
   M_{12}& =& \frac{ v_1 v_2 (v_1^2 - v_2^2) (g_L^2 + g_R^2 - 2 {\lambda^\prime}^2)}{ (v_1^2 + v_2^2)},\notag \\
M_{13}& =& \frac{g_R^2 (v_1^2 - v_2^2) (v_R^2 -\ov v_R^2) -
 4 \lambda \lambda^\prime v_1 v_2 v_R \ov v_R}{\sqrt{(v_1^2 + v_2^2)(v_R^2 +\ov v_R^2)}},  \notag \\
 M_{14}& =& \frac{2[ g_R^2 (v_1^2 - v_2^2)v_R \ov v_R -
  \lambda \lambda^\prime v_1 v_2  (v_R^2 -\ov v_R^2)]}{\sqrt{(v_1^2 + v_2^2)(v_R^2 +\ov v_R^2)}}, \notag \\
  M_{15} &=& \frac{2 \lambda^\prime [-2 A_{\lambda^\prime} v_1 v_2  +
   (v_1^2 + v_2^2) (v_S \lambda^\prime+\mu)-(\mu_S+2 \kappa v_S) v_1 v_2]}{\sqrt{v_1^2 + v_2^2}}, \notag \\
M_{22}&=&{\Big{[}}(2 g_L^2 + 2 g_R^2) v_1^2 v_2^2 + 2 m_5^2 (v_1^2 + v_2^2) +
  {\lambda^{\prime}}^2(v_1^2 - v_2^2)^2 +2 {\lambda^\prime}^2 v_S^2(v_1^2 +v_2^2) \notag \\
 &+&  4 \lambda^\prime \mu v_S (v_1^2 +  v_2^2)+
 2 \mu^2(v_1^2 + v_2^2){\Big{]}}/(v_1^2 + v_2^2), \notag \\
M_{23}&=& \frac{2 \left[ g_R^2 v_1 v_2 (v_R^2 -\ov v_R^2) + \l \l^\prime (v_1^2 -
      v_2^2) v_R \ov v_R\right]}{\sqrt{(v_1^2 + v_2^2) ( v_R^2 +\ov v_R^2)}}, \notag \\
 M_{24}&=& \frac{4 g_R^2 v_1 v_2 v_R \ov v_R - \l \l^\prime (v_1^2 -
      v_2^2) (v_R^2- \ov v_R^2)}{\sqrt{(v_1^2 + v_2^2)(
 v_R^2 +\ov v_R^2)}}, \notag \\
 M_{25}&=&\frac{\lambda^\prime (v_1^2 - v_2^2) (2 A_{\lambda^\prime} + \mu_S+2 \kappa v_S)}{\sqrt{v_1^2 + v_2^2}},\notag \\
   M_{33} &=&\frac{2\left[ \left( g_R^2+g_V^2\right)(v_R^2 - \ov v_R^2)^2+2 \lambda^2 v_R^2 \ov v_R^2 \right]}{v_R^2+\ov v_R^2}, \notag \\
   M_{34}&=& \frac{2 v_R \ov v_R (v_R^2 - \ov v_R^2)^2 \left(2 g_R^2+2g_V^2+\lambda^2 \right)}{v_R^2+\ov v_R^2}, \notag \\
   M_{35} &=& \frac{2 \lambda  \left[ A_\lambda v_R \ov v_R+ (\lambda v_S+\mu_1)(v_R^2+\ov v_R^2)
+v_R \ov v_R(\mu_S+2 \kappa v_S) \right]}{\sqrt{v_R^2+\ov v_R^2}}, \notag \\
M_{44}&=&\left[8 (g_R^2+g_V^2) v_R^2 \ov v_R^2 + (m_1^2+m_2^2 ) (v_R^2 + \ov v_R^2)  + \lambda^2(v_R^2-\ov v_R^2)^2 \right. \notag \\
&+& \left. 2( \lambda v_S+\mu_1 )^2 (v_R^2 + \ov v_R^2) \right]/(v_R^2 + \ov v_R^2),\notag \\
M_{45} &=& -\frac{\lambda(v_R^2 - \ov v_R^2) (A_\lambda + \mu_S+2 \kappa v_S)}{\sqrt{v_R^2 + \ov v_R^2}},\notag \\
   M_{55} &=& m_S^2 + {\lambda^\prime}^2(v_1^2 + v_2^2) + \lambda^2 (v_R^2 +\ov v_R^2)+ \mu_S^2 +2 \mu_S B_S \notag \\
   &+& 2 \kappa[-M^2+\l v_R \ov v_R-\l^{\prime} v_1 v_2+3(A_\kappa+\mu_S+\kappa v_s) v_S.
\end{eqnarray}
From our choice of basis, we can guess that the $M_{11}$ element of the mass-matrix, along with the corrections from the off-diagonal elements, would approximately be the lightest eigenvalue for this matrix. We calculate the corrections to lightest eigenvalue coming from the off-diagonal $M_{12},M_{13},M_{14}$ and $M_{15}$ elements. It can be seen that the $M_{12}$ element is proportional to the square of the lighter VEVs that break electroweak symmetry while the diagonal $M_{22}$ element comes out to be proportional to the square of the heavy VEV $v_R$. Hence the $M_{12}$ term gives a negligible correction to the lightest eigenvalue, of order $v^4/v_R^2$, which is negligible. Noting that the off-diagonal corrections to the lightest eigenvalue are negative definite, in order to derive an upper limit on the lightest eigenvalue we choose parameters $\lambda$, $A_{\lambda^\prime}$ and $A_{\lambda}$ such that they make $M_{13}$, $M_{15}$ and $M_{35}$ zero respectively. 
There is no longer any freedom left to make $M_{14}$ also vanish. However, parametrically we can choose the soft mass parameter
$m_1$ to be much bigger than $v_R, \ov v_R$, in which case the off-diagonal corrections from $M_{14}$ would be suppressed.  
Thus we see that he upper limit on the lightest eigenvalue is simply $M_{11}$.\footnote{If we choose $\mu_S$ to be much greater than all the other mass scales in the model, we get back the familiar result where the upper limit of the neutral Higgs mass is bound by $M_Z$.} This gives us the largest allowed value of $M_h^2$ to be
\begin{equation}
M_{h_{tree}}^2 = 2 M_W^2 \cos^2 2\beta+ \lambda^2 v^2 \sin^2 2 \beta,
\label{limit1}
\end{equation}
where $\tan\beta = \frac{v_1}{v_2}$, $v^2 = v_1^2+v_2^2$ and we have assumed that the $SU(2)_R$ gauge coupling ($g_R)$ is equal to the $SU(2)_L$ gauge coupling $g_L$. Choosing a larger or smaller value of $g_R$ will lead to a larger or smaller value of the tree level Higgs boson mass respectively. For example, for $g_R^2=2g_L^2$ we get $M_{h_{tree}}^2 = 3 M_W^2 \cos^2 2\beta+ \lambda^2 v^2 \sin^2 2 \beta,$ while for $g_R^2=\frac{g_L^2}{2}$ we get
$M_{h_{tree}}^2 = \frac{3}{2} M_W^2 \cos^2 2\beta+ \lambda^2 v^2 \sin^2 2 \beta.$  Note that the second term in Eq. (\ref{limit1}) has the same origin as in NMSSM due to the presence of a gauge singlet scalar in the model.

Including the leading radiative corrections from the top and stop sector, the upper limit on the lightest CP even Higgs boson mass is:
\begin{eqnarray}
M_h^2 &=& (2 M_W^2 \cos^2 2 \beta+ \lambda^2 \sin^2 2 \beta) \D_1 + \D_2
\end{eqnarray}
where
\begin{eqnarray}
\D_1 &=& \left(1-\frac{3}{8 \pi^2} \frac{m_t^2}{v^2}t\right), \notag \\
\D_2 &=& \frac{3}{4 \pi^2} \frac{m_t^4}{v^2}\left[ \frac{1}{2} \tilde{X}_t+t+\frac{1}{16 \pi^2} \left(\frac{3}{2} \frac{m_t^2}{v^2}-32 \pi \alpha_3 \right) \left( \tilde{X}_t t+t^2 \right)   \right],
\label{eq:rad}
\end{eqnarray}
and $m_t$ is the running top quark mass, $v=\sqrt{v_1^2+v_2^2}\approx 174$ GeV, $\alpha_3$ is the running QCD coupling, $\tilde{X}_t$ is the left--right stop squark mixing parameter, and $t = {\text{log}} \frac{M_{S}^2}{M_t^2}$ with $M_t$ being the top quark pole mass and $M_S$ being the geometric mean of the two stop squark masses.

\begin{figure}[h!]\centering
\includegraphics[width=3.1in]{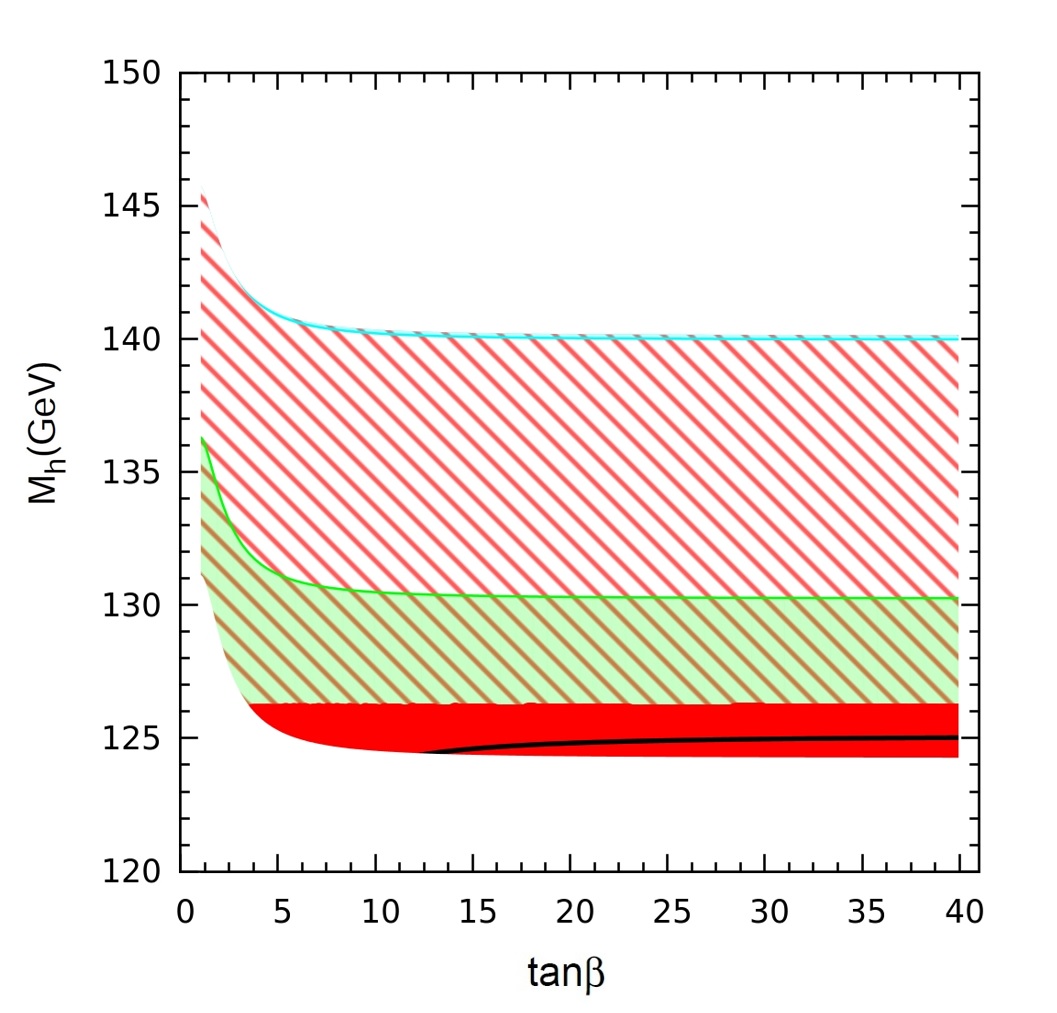}
\includegraphics[width=3.05in]{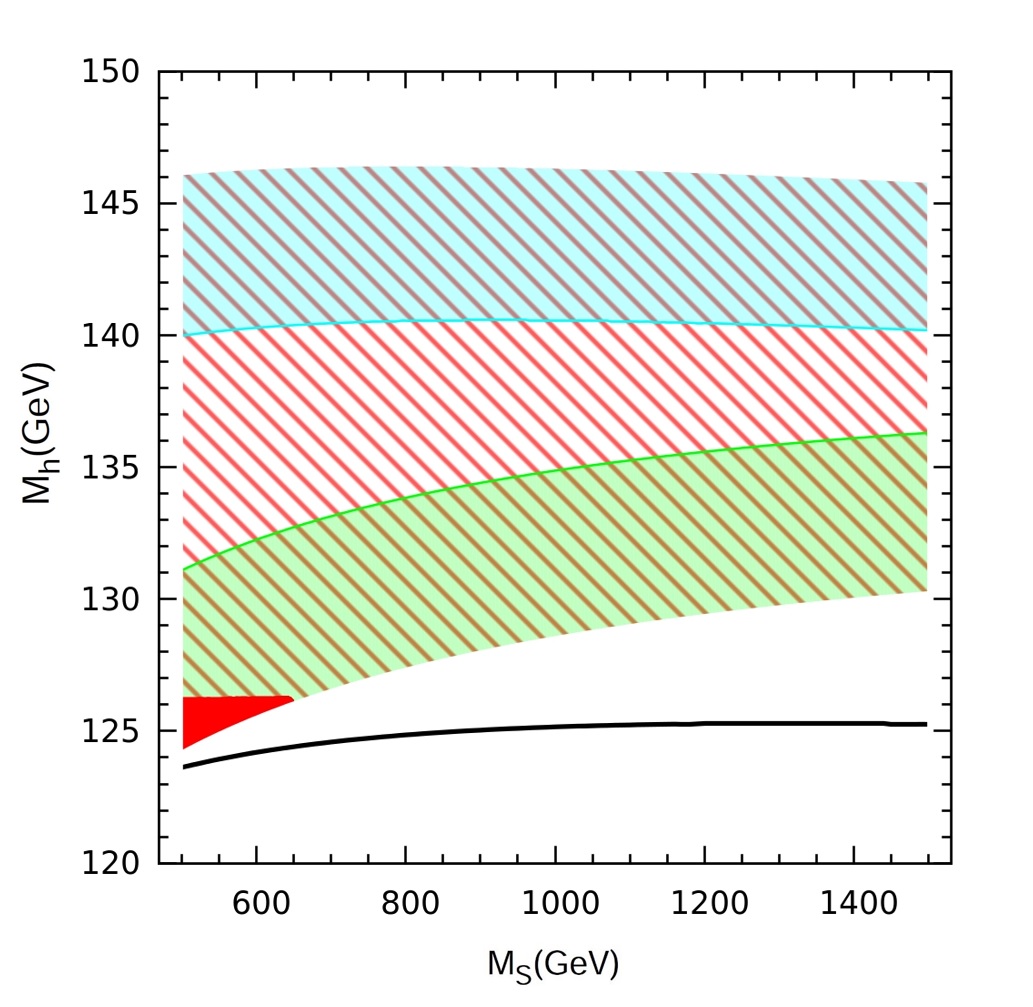}
\caption{{\sl(a) Variation of Higgs boson mass with $\tan \beta$, (b) Higgs boson mass as a function of $M_S$ for the case
with Higgs triplets, one bidoublet and a singlet.
The red region represents the band where 124 GeV $< M_h <$ 126 GeV. The light green region corresponds to 
$X_t=0$ while the blue upper region is for $X_t=6$.  The red shaded region is for all values of Higgs mass greater than 126 GeV and it is overlapped by the blue and the green regions.  The black solid line represents the MSSM upper limit for the Higgs mass.
}}
\label{fig:1d}
\end{figure}

The upper limit of the Higgs boson mass in this case is plotted in Fig.~\ref{fig:1d}(a) as a function of $\tan \beta$. The red region in the figure represents the band where the mass is between 124 GeV and 126 GeV. Anything below this has not been included as that will be ruled out by experiments. Any point above this can always be lowered by choosing a different set of parameters, as one must remember that we have chosen our parameter space so as to maximize the lightest Higgs boson mass. The light green region represents the area where the stop squark mixing is minimum, i.e., $X_t=0$ while the blue upper region is for maximal mixing where $X_t=6$. The red shaded region is for all values of Higgs mass greater than 126 GeV and it is overlapped by the blue and the green regions. Fig.~\ref{fig:1d}(b) represents the upper limit of the Higgs mass and as a function $M_S$ in Fig.~\ref{fig:1d}(b). Again the red band is where the Higgs boson mass is between 124 GeV and 126 GeV, green region is for $X_t=0$, blue region represents $X_t=6$ and shaded red region is for all values of Higgs mass greater than 126 GeV which is overlapped by the green and the blue regions. The black solid line in each case represents the MSSM upper limit for the Higgs mass. We can see that a Higgs mass of 124 GeV can be very easily achieved in this case for a very small mass of stop squark and even for minimal mixing between them.

The $2\times 2$ mass-squared matrix corresponding to the neutral left-handed triplet scalar Higgs fields in the original basis is given as
\begin{equation}
\begin{bmatrix}m_3^2 + \frac{g_L^2}{2} (v_1^2 - v_2^2) +
       2 g_V^2 (-v_R^2 +\ov v_R^2) + ( \lambda v_S + \mu_1)^2 & - \lambda (M^2 - \lambda v_R \ov v_R  +
      {\l^{\prime}} v_1 v_2 - \mu_S v_S-\kappa v_S^2)+\lambda A_\lambda v_S + \mu_1 B_1  \\ - \lambda (M^2 - \lambda v_R \ov v_R  +
      {\l^{\prime}} v_1 v_2 - \mu_S v_S-\kappa v_S^2)+\lambda A_\lambda v_S + \mu_1 B_1&
      m_4^2 - \frac{g_L^2}{2} (v_1^2 - v_2^2) +
       2 g_V^2 (v_R^2 -\ov v_R^2) + ( \lambda v_S + \mu_1)^2\end{bmatrix}
\label{eq:1dleft}
\end{equation}

We now look at the pseudo-scalar Higgs boson masses in this model. The structure of this sector is very similar to the scalar Higgs boson in the sense that the left-handed triplet fields decouple to form a $2\times 2$ matrix which is exactly the same as given in Eq.~(\ref{eq:1dleft}) while the imaginary components of the other neutral Higgs bosons form a $5\times 5$ matrix. We choose a basis given as
\begin{equation}
g_1 = \frac{v_1 \phi^0_{1} - v_2 \phi^0_{2}}{\sqrt{v_1^2+v_2^2}},~~~~
g_2 = \frac{v_R \d^{c^0} -\ov v_R \ov \d^{c^0}}{\sqrt{v_R^2+\ov v_R^2}},~~~~
h_1 = \frac{v_2 \phi^0_{1} + v_1 \phi^0_{2}}{\sqrt{v_1^2+v_2^2}},~~~~
h_2 = \frac{\ov v_R \d^{c^0} + v_R \ov \d^{c^0}}{\sqrt{v_R^2+\ov v_R^2}}.
\end{equation}
The Im($g_1$) and Im($g_2$) fields can be identified as the Goldstone bosons which are absorbed by the $Z_R$-boson and the $Z$-boson to make them massive. Integrating out these Goldstone states, the resulting $3\times 3$ matrix in the basis $({\text{Im}}(h_2),{\text{Im}}(h_1),{\text{Im}}(S))$ is given as
\begin{eqnarray}
M_{11} &=&  m_1^2 + m_2^2 + \l^2(v_R^2+\ov v_R^2+2 v_S^2) +2 \mu_1 (2 \l v_s +\mu_1), \notag \\
M_{12} &=& -\l {\l^{\prime}} \sqrt{(v_1^2+v_2^2)(v_R^2+\ov v_R^2)} , \notag \\
M_{13} &=& \l (\mu _S+2 \kappa v_S-A_ \l) \sqrt{v_R^2+\ov v_R^2}, \notag \\
M_{22} &=& 2 m_5^2+{\l^{\prime}}^2(v_1^2+v_2^2+2 v_S^2)+2 \mu(2 {\l^{\prime}} v_S+\mu), \notag \\
M_{23} &=& {\l^{\prime}} (2A_{{\l^{\prime}}}-\mu_S-2 \kappa v_S) \sqrt{v_1^2+v_2^2}, \notag \\
M_{33} &=& m_S^2 +\l^2 (v_R^2+\ov v_R^2)+{\l^{\prime}}^2(v_1^2+v_2^2)-\mu_S(2 B_S-\mu_S) \notag \\
&+& 2\kappa(M^2-\l v_R \ov v_R +\l ^{\prime} v_1 v_2+\mu_S v_S+\kappa v_S^2-3 A_\kappa v_S).
\label{eq:1dpseudo}
\end{eqnarray}

The charged Higgs boson sector has six singly-charged fields in this model. Their mass-squared matrix can be split into two block diagonal matrices. There is a $2\times 2$ matrix corresponding to the $\d^+$ and $\ov \d^-$ fields which in its original basis is given as
\begin{equation}
\begin{pmatrix} g_V^2(\ov v_R^2-v_R^2)+ m_3^2 + (\mu_1+\l v_S)^2 & - \lambda (M^2 - \lambda v_R \ov v_R  +
      {\l^{\prime}} v_1 v_2 - \mu_S v_S-\kappa v_S^2-A_\lambda v_S) + \mu_1 B_1 \\ - \lambda (M^2 - \lambda v_R \ov v_R  +
      {\l^{\prime}} v_1 v_2 - \mu_S v_S-\kappa v_S^2-A_\lambda v_S) + \mu_1 B_1 & g_V^2(v_R^2-\ov v_R^2)+ m_4^2 + (\mu_1+\l v_S)^2 \end{pmatrix}.
\label{eq:1dchl}
\end{equation}
The other $4\times4$ block has two Goldstone bosons which are absorbed by $W_R$ and $W$ gauge bosons to get mass. We choose a basis given as:
\begin{equation}
\s_1^+ = \frac{v_1 \phi_1^+ - v_2 \phi_2^{-^*} }{\sqrt{v_1^2+v_2^2}},~\s_2^+ = \frac{v_2 \phi_1^+ + v_1 \phi_2^{-^*} }{\sqrt{v_1^2+v_2^2}},~\s_3^+ = \frac{v_R \ov \d^{c^{+}} -\ov v_R \d^{c^{-^*}}}{\sqrt{v_R^2+\ov v_R^2}},~\s_4^+ = \frac{\ov v_R \ov \d^{c^{+}} + v_R \d^{c^{-^*}}}{\sqrt{v_R^2+\ov v_R^2}},
\end{equation}
where the Goldstone eigenstates can be identified as
\begin{equation}
g_1^+ = \s_1^+,~~~~ g_2^+ = \frac{\sqrt{2(v_1^2+v_2^2)(v_R^2+\ov v_R^2)}\s_4^+ + (v_2^2-v_1^2)\s_2^+}{\sqrt{(v_2^2-v_1^2)^2+2(v_1^2+v_2^2)(v_R^2+\ov v_R^2)}}.
\end{equation}
This gives us the actual basis for the physical singly-charged Higgs bosons to be
\begin{equation}
h_1^+ = \s_3^+,~~~~
h_2^+ = \frac{(v_2^2-v_1^2)\s_4^+ - \sqrt{2(v_1^2+v_2^2)(v_R^2+\ov v_R^2)}\s_2^+}{\sqrt{(v_2^2-v_1^2)^2+2(v_1^2+v_2^2)(v_R^2+\ov v_R^2)}},
\end{equation}
and the $2\times 2$ singly-charged Higgs boson mass-squared matrix elements are given as
\begin{eqnarray}
M_{11}&=& \frac{g_R^2 \left\{(v_1^2-v_2^2)(v_R^2-\ov v_R^2)+2(v_R^2+\ov v_R^2)^2 \right\} }{(v_R^2 +\ov v_R^2)} \notag \\
  &-& \frac{ 2(v_R^2+\ov v_R^2) \left\{ \l(-M^2+A_\l v_S +\l v_R \ov v_R-\l^{\prime}v_1 v_2+\mu_S v_S+\kappa v_S^2)+B_2 \mu_1\right\} }{v_R \ov v_R}, \notag \\
M_{12} &=& \frac{2 g_R^2 v_R \ov v_R \sqrt{
 v_1^4 + 2 v_1^2 (-v_2^2 + v_R^2 +\ov v_R^2) +
  v_2^2 [v_2^2 + 2 (v_R^2 +\ov v_R^2)]}}{v_R^2 + \ov v_R^2}, \notag \\
M_{22}&=&-\frac{g_R^2 (v_R^2 -\ov v_R^2) \left[v_1^4 + 2 v_1^2 (-v_2^2 + v_R^2 +\ov v_R^2) +  v_2^4 +2 v_2^2 (v_R^2 +\ov v_R^2)\right]}{(v_1^2 - v_2^2) (v_R^2 +\ov v_R^2)}.
  \label{eq:1dch}
\end{eqnarray}

Using the minimization conditions given in Eq.~(\ref{eq:min1d}), we eliminate $B_2$, $m_2$, $m_5$, $B$ and $C_\l$ in terms of the other parameters and numerically calculate the Higgs boson mass spectrum. We choose the parameters such that the lightest neutral scalar Higgs boson mass is 125 GeV after the radiative corrections and with a stop squark mass of 570 GeV and a stop squark mixing parameter $X_t$ = 4, as an example. The soft quadratic mass terms for the left-handed and right-handed triplets were chosen to be different since otherwise it leads to unphysical states with some negative eigenvalues for the Higgs boson mass-squared matrices. The numerical values of the chosen parameters and the masses obtained are given in Table~\ref{tab:one}. It is easy to identify the left-handed triplet Higgs boson eigenvalues since they decouple in each case as discussed earlier. In the table, $M_{H_1^\D}$ and $M_{H_2^\D}$ denote the mass-squared values for the left-handed triplet scalar Higgs bosons. Similarly $M_{A_1^\D}$, $M_{A_2^\D}$ and $M_{\D_1^+}$, $M_{\D_2^+}$ are the squared masses for the pseudo-scalar and the single-charged left-handed triplet Higgs bosons respectively.  This numerical result shows the self-consistency of the model.

\vspace*{0.1in}
\noindent {\large{\bf{Chargino and Neutralino masses}}}
\vspace*{0.1in}

The particle spectrum of this model is much richer compared to the Minimal Supersymmetric Standard Model and hence the study of the chargino and neutralino masses is crucial for determining the lightest supersymmetric particle, a candidate for dark matter in the universe. The higgsinos and the gauginos mix to form charginos and neutralinos. The chargino mass matrix in this case is written as
\begin{equation}
{\mathcal{L}}_{ch} = -\frac{1}{2}\begin{pmatrix}\widetilde{\d}^{c^{-}}&\widetilde{\ov \d}^-&\widetilde{\phi}_2^-&\widetilde{W}_R^-&\widetilde{W}_L^-\end{pmatrix} \begin{pmatrix}\mu_1+\l^* v_S &0&0&-\sqrt{2}g_R v_R&0 \\ 0&\mu_1+\l v_S&0&0&0 \\ 0&0&\mu+\l^{\prime} v_S&g_R v_2&g_L v_2 \\ \sqrt{2} g_R \ov v_R&0&g_R v_1&M_R&0 \\ 0&0&g_L v_1&0&M_L \end{pmatrix}\begin{pmatrix}\widetilde{\ov \d}^{c^+} \\ \widetilde{\d}^+ \\ \widetilde{\phi}_1^+ \\ \widetilde{W}_R^+ \\ \widetilde{W}_L^+ \end{pmatrix}.
\label{eq:1dchargino}
\end{equation}
where the gaugino soft mass terms are given as
\begin{equation}
\mathcal{L}_G = -\frac{1}{2}\left(M_3 \widetilde{g}\widetilde{g}+ M_R \widetilde{W}_R \widetilde{W}_R + M_L \widetilde{W}_L \widetilde{W}_L + M_1 \widetilde{B} \widetilde{B} + h.c.\right)
\label{eq:gluino}
\end{equation}
The neutralino mass matrix splits into two matrices with the left-handed triplet Higgsino fields $\widetilde \d^0$ and $\widetilde {\ov \d^0}$ decoupling to form a $2\times 2$ matrix given as
\begin{equation}
\begin{pmatrix} 0&\mu_1+\l v_S \\ \mu_1+\l v_S&0 \end{pmatrix}~.
\end{equation}
The mass matrix for the other neutral fields in the basis $\begin{pmatrix}\widetilde{\d}^{c^0}&\widetilde{\ov \d}^{c^0}&\widetilde{\phi}_1^0&\widetilde{\phi}_2^0&\widetilde{B}&\widetilde{W}_{R_3}&\widetilde{W}_{L_3}&\widetilde{S}\end{pmatrix}$ is given as
\begin{equation}
\begin{pmatrix} 0&\mu_1+\l^* v_S&0&0&-\sqrt{2}g_V v_R&\sqrt{2}g_R v_R&0&\l^* \ov v_R \\ \mu_1+\l^* v_S&0&0&0&\sqrt{2} g_V \ov v_R&-\sqrt{2}g_R \ov v_R&0&\l^* v_R   \\ 0&0&0&-\mu-\l^{\prime} v_S&0&-\frac{g_R v_1}{\sqrt{2}}&-\frac{g_L v_1}{\sqrt{2}}&-\l^{\prime} v_2 \\ 0&0&-\mu-\l^{\prime} v_S&0&0&\frac{g_R v_2}{\sqrt{2}}&\frac{g_L v_2}{\sqrt{2}}&-\l^{\prime} v_1 \\ -\sqrt{2} g_V v_R&\sqrt{2}g_V \ov v_R&0&0&M_1&0&0&0 \\ \sqrt{2} g_R v_R&-\sqrt{2}g_R \ov v_R&-\frac{g_R v_1}{\sqrt{2}}&\frac{g_R v_2}{\sqrt{2}}&0&M_R&0&0 \\ 0&0&-\frac{g_L v_1}{\sqrt{2}}&\frac{g_L v_2}{\sqrt{2}}&0&0&M_L&0 \\ \l ^* \ov v_R & \l^* v_R &-\l^{\prime} v_2&-\l^{\prime} v_1&0&0&0&\mu_S\end{pmatrix},
\end{equation}
where $M_R, M_L$ and $M_1$ are defined above. Parity invariance further demands $M_L=M_R^*$ and $M_1$ and $M_3$ are real.

The chargino and neutralino masses for this models are given in Table~\ref{tab:one}. Here again the states $\wdt \D_1^+$ and $\wdt \D_{1,2}^0$ refer to the chargino and neutralino states corresponding to the left-handed triplet Higgsinos.

\begin{center}
\begin{table}[h!]
\centering
\begin{tabular}{|C{2.8cm}|C{2.8cm}|C{2.8cm}|C{2.9cm}|C{3.1cm}|}
        \hline
{\bf{Scalar Higgs boson masses}} & {\bf{Pseudo-scalar Higgs boson masses}} & {\bf{Single charged Higgs boson masses}}& \bf{Chargino masses}&{\bf{Neutralino masses}}  \\ \hline
$M_{H_1}$=6.26 TeV, $M_{H_2}$=2.59 TeV, $M_{H_3}$=1.21 TeV, $M_{H_4}$=468 GeV, $M_{H_1^\D}$=4.51 TeV, $M_{H_2^\D}$=1.92 TeV  & $M_{A_1}$=4.53 TeV, $M_{A_2}$=3.34 TeV, $M_{A_3}$=514 GeV, $M_{A_1^\D}$=4.51 TeV, $M_{A_2^\D}$=1.92 TeV & $M_{H_1^+}$=4.77 TeV, $M_{H_2^+}$=513 GeV, $M_{\D_1^+}$=4.51 TeV, $M_{\D_2^+}$=1.92 TeV & $M_{\wdt \D_1^+}$= 2.65 TeV, $M_{\wdt \chi_1^+}$= 4.23 TeV, $M_{\wdt \chi_2^+}$= 2.38 TeV, $M_{\wdt \chi_3^+}$= 809 GeV, $M_{\wdt \chi_4^+}$= 348 GeV & $M_{\wdt \D_{1,2}^0}$= 2.65 TeV, $M_{\wdt \chi_1^0}$= 5.98 TeV, $M_{\wdt \chi_2^0}$= 4.85 TeV, $M_{\wdt \chi_3^0}$= 3.09 TeV, $M_{\wdt \chi_4^0}$= 2.00 TeV, $M_{\wdt \chi_5^0}$= 1.15 TeV, $M_{\wdt \chi_6^0}$= 885 GeV, $M_{\wdt \chi_7^0}$= 352 GeV, $M_{\wdt \chi_8^0}$= 346 GeV\\ \hline
    \end{tabular}
    \caption{{Higgs boson, chargino and neutralino masses for a sample point for case with triplets using the parameters given as: 
     $\l^{\prime}$=0.7, $\l$=-0.3, $v_1$=173.14 GeV, $v_2$=17.3 GeV, $v_R$=3 TeV, $\ov v_R$=3.1 TeV, $\mu_1$=3.1 TeV, $\mu_2$=3.1 TeV, $\mu$=-1.4 TeV, $\mu_S$=-700 GeV, $m_S^2$=9 TeV$^2$, $v_S$=1.5 TeV, $B_S$=2 TeV, $m_1^2=m_3^2$=1 TeV$^2$, $\kappa$=0.1, $A_\kappa$=1 TeV, $m_4^2$=9 TeV$^2$, $A_{\l}$=-4 TeV, $A_{\l^{\prime}}$=-1 TeV, $M_R$=800 GeV, $M_L$=800 GeV, $M_1$=400 GeV and $B_1$ is chosen to be equal to $B_2$ which was fixed using the minimzation conditions.}}
\label{tab:one}
\end{table}
\end{center}

The doubly charged Higgs boson sector of the model is discussed in more detail in Sec. \ref{2chhiggs} where we also cary
out the one-loop radiative corrections to its mass and show the consistency of the framework.  

\subsection{Symmetry breaking with a pair of Higgs triplets, a bidoublet and a heavy singlet} 
\label{heavysinglet}

We now look at the case where the single Higgs $S$ is heavy and can be integrated out from the low energy sector of the model to give the following superpotential:
\begin{eqnarray}
W&=&\mu_1{\text{Tr}}(\Delta \overline{\Delta})+\mu_2{\text{Tr}}(\Delta^{c}\overline{\Delta}^{c}) +\e \text{Tr}\left[\D^c \ov\D^c\right]^2 +\frac12 \mu {\text{Tr}}(\Phi^{T}\tau_{2}\Phi\tau_{2}).
\end{eqnarray}
Here $\e$ is proportional to $1/{M_S}$ with $M_S$ being the scale at which the singlet is integrated out. Note that $\e$ is a relevant operator which is kept in our analysis, although the field $S$ has been integrated out. Since $\e$ is very small, we only kept the $\e\text{Tr}(\Delta ^c \overline {\Delta}^c)^2$ term in the superpotential as other terms will have no significant effect to the lightest CP-even Higgs boson mass.

The $D$-term of the Higgs potential is exactly same as in Eq.~(\ref{eq:two}) but there will be different contributions to the  $F$-term and the soft supersymmetry breaking terms. They are given by:
\begin{eqnarray}
V_F&=&|\mu_1|^2{\text{Tr}}(\D^{\dagger} \D+{\ov\D}^{\dagger} \ov\D)+{\text{Tr}}\left[ \left|\mu_2^2{\D^c}+2 \e \D^c \ov\D^c \D^c \right|^2+\left|\mu_2^2 \ov\D^c+2 \e \ov\D^c \D^c \ov\D^c \right|^2\right] \notag \\
&+& |\mu|^2 {\text{Tr}}(\Phi^\dagger \Phi), \\
V_{Soft}&=& m_1^2 {\text{Tr}}(\Phi^\dagger \Phi)+\left[ B \mu {\text{Tr}}(\Phi^T \tau_2 \Phi \tau_2)+h.c.\right]+m_3^2{\text{Tr}}(\D^\dagger\D)+m_4^2{\text{Tr}}(\ov\D^\dagger\ov\D) \nonumber \\
&+&m_5^2{\text{Tr}}({\D^c}^\dagger\D^c)+m_6^2{\text{Tr}}({\ov{\D}^c}^\dagger\ov\D^c)+{\text{Tr}}(B_1 \mu _1\D \ov\D + h.c.)\nonumber \\
&+&{\text{Tr}}(B_2 \mu_2\D^c \ov{\D}^c+h.c.)+\left[ \e D_{\e}  {\text{Tr}} ( {\D^c} \ov\D^c )^2 + h.c. \right].
\label{eq:pot1c}
\end{eqnarray}

We use the same basis field redefinition as in Eq.~(\ref{eq:basis1d}). The minimization conditions are given as:
\begin{align}
0&= -4  B \mu {v_2}+ {v_1} \left( 4  {m_1}^2+ {g_L}^2 ( {v_1}^2- {v_2}^2)+ {g_R}^2 \left( {v_1}^2- {v_2}^2+2  {v_R}^2-2 \ov{v}_R^2\right)+4 \mu ^2\right), \nonumber \\
0& = -4 B \mu {v_1}+ {v_2} \left(4  {m_1}^2+ {g_L}^2 \left(- {v_1}^2+ {v_2}^2\right)+ {g_R}^2 \left(- {v_1}^2+ {v_2}^2-2  {v_R}^2+2  \ov{v}_R^2\right)+4 \mu ^2\right), \nonumber \\
0& = 2 B_2 \mu_2  {\ov v_R}+\left[2  {m_5}^2+2  {\mu_2}^2+ {g_R}^2 \left( {v_1}^2- {v_2}^2\right)\right]  {v}_R+2 \left( {g_R}^2+ {g_V}^2\right) {v}_R \left(- {\ov v_R}^2+ {v}_R^2\right) \nonumber  \\
&+ 4 \e \ov{v}_R\left[D_{\e} v_R \ov v_R +\mu_2 (3 v_R^2+\ov v_R^2)+2 \e v_R \ov v_R(2 v_R^2+\ov v_R^2) \right], \nonumber \\
0& =  2  B_2 \mu_2  {v_R}+\left[2  {m_6}^2+2  {\mu_2}^2+ {g_R}^2 \left(- {v_1}^2+ {v_2}^2\right)\right]  \ov{v}_R+2 \left( {g_R}^2  + {g_V}^2\right) \ov{v}_R \left(- {v_R}^2+ \ov{v}_R^2\right) \nonumber \\
&+ 4 \e v_R\left[D_{\e} v_R \ov v_R +\mu_2 (v_R^2+3\ov v_R^2)+2 \e v_R \ov v_R( v_R^2+2\ov v_R^2) \right].
\label{eq:min1c}
\end{align}

Calculating the neutral CP-even Higgs boson mass-squared matrix subject to these minimization conditions, the matrix elements can be obtained from Eq.~(\ref{eq:masmat1d}) by putting all the triplet and bidoublet couplings to the singlet Higgs to be zero with some extra terms in the $M_{33},M_{34},M_{44}$ elements. The relevant terms in the mass-squared matrix are:
\begin{align}
&M_{11} = \frac{(g_L^2+g_R^2)(v_1^2-v_2^2)^2}{2(v_1^2+v_2^2)}, \notag \\
& M_{13} = \frac{g_R^2 (v_R^2-\ov v_R^2)(v_1^2-v_2^2)}{\sqrt{(v_1^2+v_2^2)(v_R^2+\ov v_R^2)}}, \notag \\
& M_{14} = \frac{2g_R^2 v_R \ov{v}_R(v_1^2-v_2^2)}{\sqrt{(v_1^2+v_2^2)(v_R^2+\ov v_R^2)}}, \\
& M_{33} = \frac{2(g_R^2+g_V^2)v_R^3-B_2 \mu_2\ov{v}_R-2 \e \ov v_R \left[ \mu_2 (\ov{v}_R^2-3 v_R^2)-8 \e v_R^3 \ov v_R \right]}{v_R}, \notag \\
& M_{34}= B_2 \mu_2-2(g_R^2+g_V^2)v_R\ov{v}_R+\e \left[3 \mu_2(v_R^2+\ov v_R^2)+2 v_R \ov v_R(D_{\e}+4 \e(v_R^2+\ov v_R^2) \right], \notag \\
& M_{44} = \frac{2(g_R^2+g_V^2)\ov{v}_R^3-B_2 \mu_2v_R+2 \e v_R \left[ \mu_2 (3 \ov{v}_R^2- v_R^2)+8 \e \ov v_R^3 v_R \right]}{\ov v_R}.
\label{eq:masmat1c}
\end{align}
We calculate the contribution of the off-diagonal ($M_{13},M_{14}$) entries in the mass-squared matrix to the lightest eigenvalue using the seesaw formula. For simplicity we take the approximation $D_\e =0$ and we get the following result:
\begin{equation}
M_{h_{tree}}^2 = 2 M_W^2 {\cos^2 2\beta}\left[1-\frac{x}{2 \left( \frac{g_R^4 x}{g_R^2-{g^\prime}^2}+y \right)} \right]
\label{inter}
\end{equation}
where
\begin{align}
x &= B_2 \mu_2 (v_R^2 - \ov v_R^2)^2 +
 2 \e (v_R^2 +
    \ov v_R^2) \left[\mu_2 (v_R^4 - 10 v_R^2 \ov v_R^2 + \ov v_R^4) -
    24 \e v_R^3 \ov v_R^3\right], \nonumber \\
y &= 8 v_R \ov v_R \e ( B_2 \mu_2^2 (v_R^2 + \ov v_R^2) +
   \mu_2^2 [3 v_R^4 + 2 v_R^2 \ov v_R^2 + 3 \ov v_R^4) \e +
   2 \mu_2 \ov v_R (7 v_R^5 + 6 v_R^3 \ov v_R^2 + 7 v_R \ov v_R^4) \e^2 \nonumber \\
   & +
   v_R \ov v_R \e (3 m_8^2 (v_R^2 +\ov v_R^2) +
      16 v_R \ov v_R (v_R^4 + v_R^2 \ov v_R^2 + \ov v_R^4) \e^2], \notag
\end{align}
$\tan \beta = \frac{v_1}{v_2}$ and $g_R=g_L$.
This result shows that the lightest CP-even Higgs boson mass has an upper limit of $ \sqrt{2} M_W$ in this case which can be realized if $x=0$. If we consider $\ov v_R^2-v_R^2 \sim M_{SUSY}^2$ and $v_R, \ov v_R >> M_{SUSY}$, we get an upper limit of $M_Z$ for the lightest scalar Higgs boson mass. Eq. (\ref{inter}) interpolates between the two interesting cases of
$v_R \sim M_{SUSY}$ and $v_R \gg M_{SUSY}$.  Taking the least constraining of the limits we have for the largest allowed
$M_h$ at tree level,
\begin{equation}
M^2_{h_{tree}} = 2 M_W^2 \cos^2 2 \beta.
\end{equation}
Including the one and two loop corrections from the top quark and stop squark, we get:
\begin{eqnarray}
M_{h_{max}}^2 &=& (2 M_W^2 \cos^2 2 \beta) \D_1 + \D_2,
\end{eqnarray}
where $\D_1$ and $\D_2$ are defined in Eq.~(\ref{eq:rad}).

\begin{figure}[h!]\centering
\includegraphics[width=3.1in]{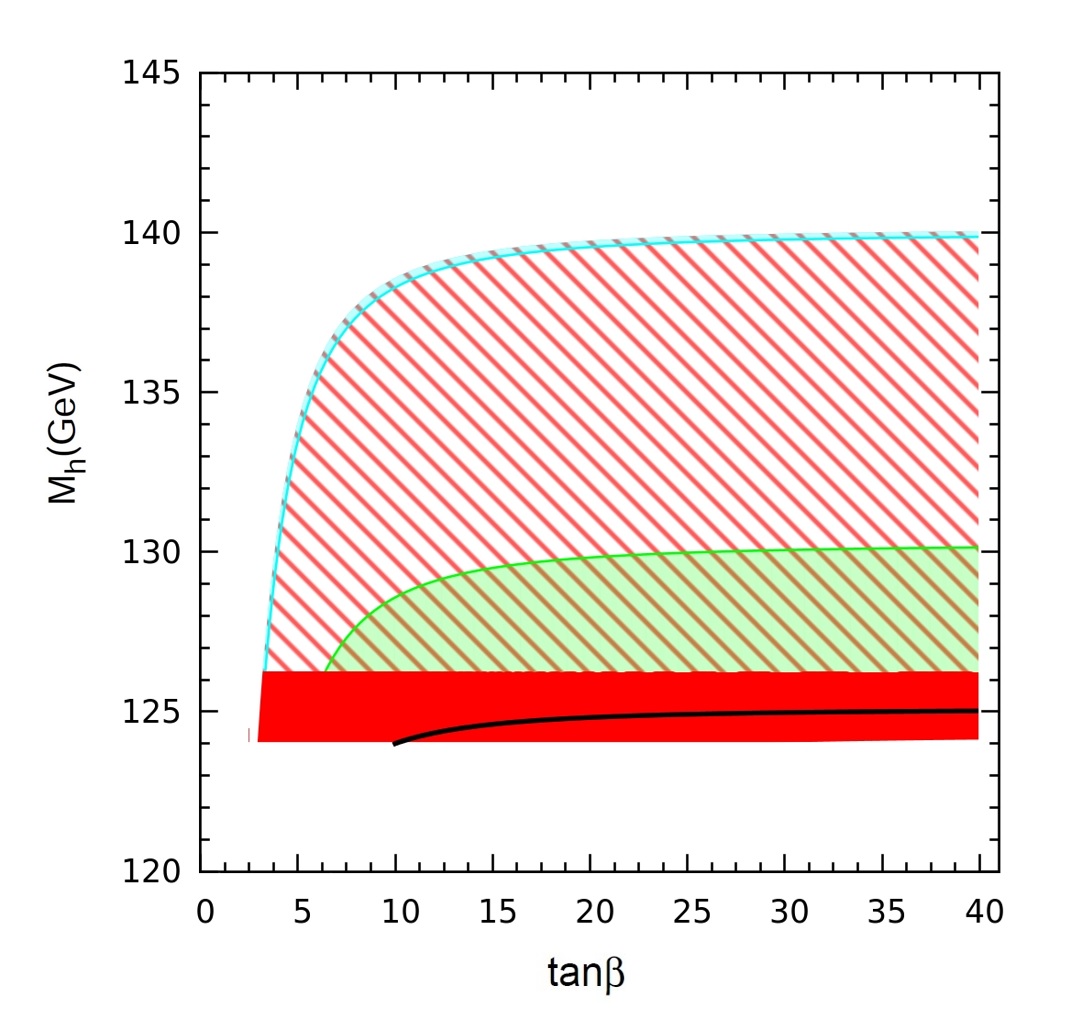}
\includegraphics[width=3.1in]{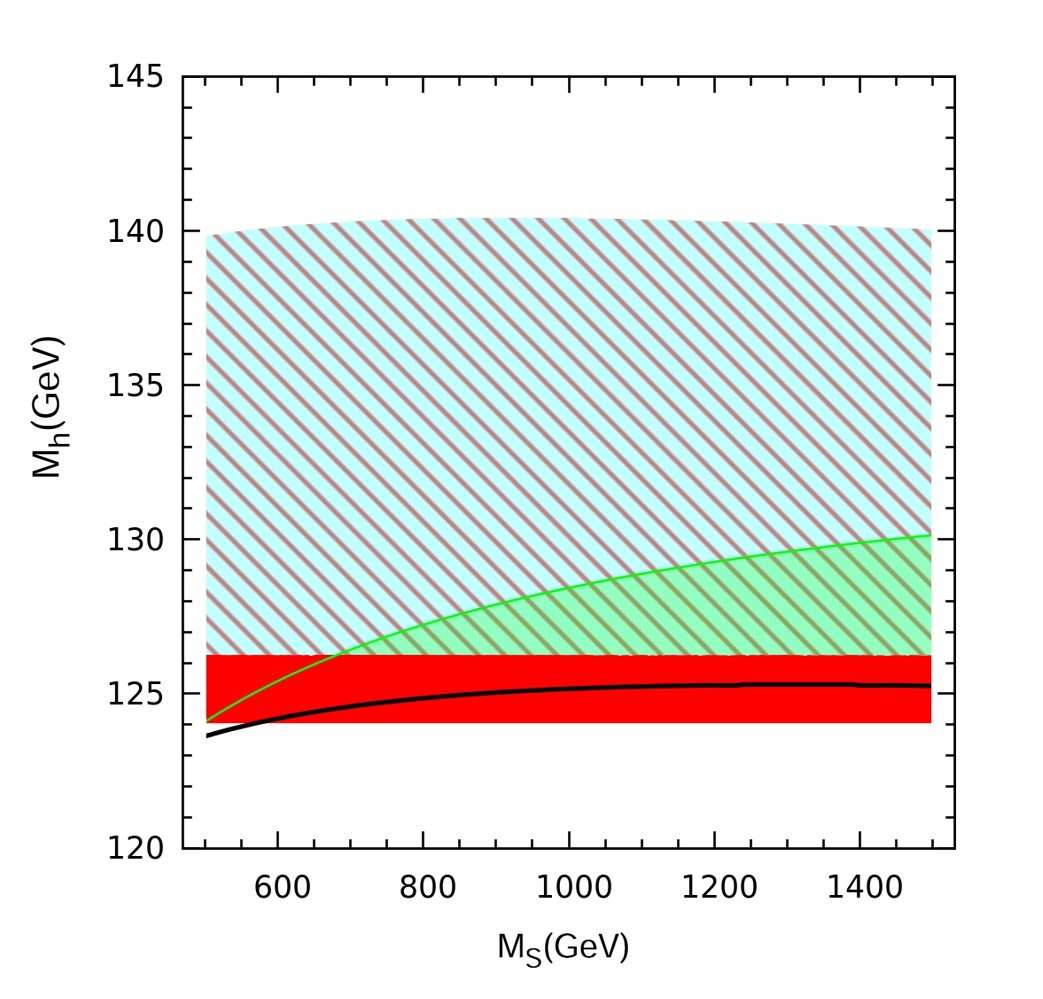}
\caption{{\sl(a) Variation of Higgs boson mass with $\tan \beta$, (b) Higgs boson mass as a function of $M_S$
for the case with Higgs triplets, one bidoublet and a heavy singlet.
Notation here is the same as in Fig. \ref{fig:1d}.}}
\label{fig:1c}
\end{figure}

The Higgs boson mass is plotted in Fig~\ref{fig:1c}(a) as a function of $\tan \beta$. The red region in the figure represents the band where the mass is between 124 GeV and 126 GeV. The light green region represents the area where the stop squark mixing is minimum i.e. $X_t=0$ while the blue upper region is for maximal mixing where $X_t=6$. The red shaded region is for all values of Higgs mass greater than 126 GeV and it is overlapped by the blue and the green regions. Fig.~\ref{fig:1c}(b) represents the upper limit of the Higgs mass and as a function $M_S$. Again the red band is where the Higgs boson mass is between 124 GeV and 126 GeV, green region is for $X_t=0$, blue region represents $X_t=6$ and shaded red region is for all values of Higgs mass greater than 126 GeV which is overlapped by the green and the blue regions. The black solid line in each case represents the MSSM upper limit for the Higgs mass.


The pseudo-scalar mass-squared matrix is again two $2\times2$ blocks which can be obtained by putting all the singlet couplings to zero in Eq.~(\ref{eq:1dleft}) and Eq.~(\ref{eq:1dpseudo}).

The charged Higgs boson mass-squared matrix is exactly the same as in Eq.~(\ref{eq:1dchl}) and Eq.~({\ref{eq:1dch}) with all the singlet couplings set to zero and in the limit where we take $D_{\e}\rightarrow 0$.

\vspace*{0.1in}
\noindent {\large{\bf{Chargino and Neutralino masses}}}
\vspace*{0.1in}

We now look at the chargino and neutralino sector in this case. The chargino basis is exactly the same as in the case discussed in {{section 3.1}}. The chargino mass matrix in this case is written as
\begin{equation}
M_{ch} = \begin{pmatrix}\mu_2+\e v_R \ov v_R &0&0&-\sqrt{2}g_R v_R&0 \\ 0&\mu_1&0&0&0 \\ 0&0&\mu&g_R v_2&g_L v_2 \\ \sqrt{2} g_R \ov v_R&0&g_R v_1&M_R&0 \\ 0&0&g_L v_1&0&M_L \end{pmatrix}.
\end{equation}
The neutralino mass matrix again splits into two matrices. The matrix in the basis $\begin{pmatrix}\widetilde{\d}^0&{\widetilde{\ov \d}^0} \end{pmatrix}$ is given as
\begin{equation}
\begin{pmatrix} 0&\mu_1 \\ \mu_1&0 \end{pmatrix},
\end{equation}
while the mass matrix in the basis $\begin{pmatrix}\widetilde{\d}^{c^0}&\widetilde{\ov \d}^{c^0}&\widetilde{\phi}_1^0&\widetilde{\phi}_2^0&\widetilde{B}&\widetilde{W}_{R_3}&\widetilde{W}_{L_3}\end{pmatrix}$ is given as
\begin{equation}
\begin{pmatrix} \e \ov v_R^2&\mu_2+\e v_R \ov v_R&0&0&-\sqrt{2}g_V v_R&\sqrt{2}g_R v_R&0 \\ \mu_2+\e v_R \ov v_R&\e v_R^2&0&0&\sqrt{2} g_V \ov v_R&-\sqrt{2}g_R \ov v_R&0  \\ 0&0&0&-\mu&0&-\frac{g_R v_1}{\sqrt{2}}&-\frac{g_L v_1}{\sqrt{2}} \\ 0&0&-\mu&0&0&\frac{g_R v_2}{\sqrt{2}}&\frac{g_L v_2}{\sqrt{2}} \\ -\sqrt{2} g_V v_R&\sqrt{2}g_V \ov v_R&0&0&M_1&0&0\\ \sqrt{2} g_R v_R&-\sqrt{2}g_R \ov v_R&-\frac{g_R v_1}{\sqrt{2}}&\frac{g_R v_2}{\sqrt{2}}&0&M_R&0 \\ 0&0&-\frac{g_L v_1}{\sqrt{2}}&\frac{g_L v_2}{\sqrt{2}}&0&0&M_L\end{pmatrix}.
\label{eq:1cneutalino}
\end{equation}

\subsection{Case with two pair of triplets and a bidoublet} \label{case1a}

This is a special case of the one discussed in {{Section~\ref{case1d}}}. We do not have the singlet Higgs and as a result it will be seen that the lightest Higgs boson mass upper limit becomes the same as MSSM.  We also show explicitly the
self-consistency of this model which requires $v_R$ and $M_{SUSY}$ to be of the same order.  

The most general superpotential relevant to our calculation is given by:
\begin{eqnarray}
W&=&\mu_1{\text{Tr}}(\Delta \overline{\Delta})+\mu_2{\text{Tr}}(\Delta^{c}\overline{\Delta}^{c}) +\frac12 \mu {\text{Tr}}(\Phi^{T}\tau_{2}\Phi\tau_{2}).
\end{eqnarray}

The $D$-term in the Higgs potential is exactly the same as given in Eq.~(\ref{eq:two}), the $F$-term can be obtained from Eq.~(\ref{eq:1df}) by putting all the singlet couplings to zero. The soft supersymmetry breaking terms are given by:
\begin{eqnarray}
V_{Soft}&=& m_1^2{\text{Tr}}({\D^c}^\dagger\D^c)+m_2^2{\text{Tr}}({\ov{\D}^c}^\dagger\ov\D^c)+m_3^2{\text{Tr}}(\D^\dagger\D)+m_4^2{\text{Tr}}(\ov\D^\dagger\ov\D) \nonumber \\
&+&m_5^2 {\text{Tr}}(\Phi^\dagger \Phi)+ \left[B \mu {\text{Tr}}(\Phi^T \tau_2 \Phi \tau_2)+h.c.\right]\nonumber \\
&+&\left[{B_1 \mu_1 \text{Tr}}(\D \ov\D) + h.c.\right] + \left[ {B_2 \mu_2\text{Tr}}(\D^c \ov{\D}^c)+h.c.\right].
\end{eqnarray}

We use this potential to calculate the Higgs boson mass-squared matrices for the charged, neutral CP-even and neutral CP-odd Higgs bosons. To easily identify the field corresponding to the lightest eigenvalue, we redefine the Higgs fields. This redefinition is the same as in Eq.~(\ref{eq:basis1d}).

The minimization conditions and the Higgs mass-squared in this case can again be obtained by putting all the singlet couplings to zero in the model of {{Section~\ref{case1d}}}.

Calculating the lightest eigenvalue for the CP-even Higgs boson mass-squared matrix we get:
\begin{equation}
M_{h_{tree}}^2 = \frac{g_L^4 ({g^\prime}^2 + g_R^2) (v_1^2 - v_2^2)^2}{2 [g_L^2 g_R^2 +
   {g^\prime}^2 (g_L^2 - g_R^2)] (v_1^2 + v_2^2)}.
\end{equation}
If we assume that the $SU(2)_R$ gauge coupling ($g_R$) is equal to the $SU(2)_L$ gauge coupling ($g_L$), $\tan\beta = \frac{v_1}{v_2}$ and $v^2 = v_1^2+v_2^2$, then
\begin{equation}
M_{h_{tree}}^2 = \frac{(g_L^2+{g^ \prime}^2)}{2} v^2 \cos^2{2\beta}.
\end{equation}
The mass of the $Z$ boson in this model is $\sqrt{\frac{g_L^2+{g^\prime}^2}{2}} v$. So we see that the tree-level lightest CP-even Higgs mass has an upper limit of $M_Z$. This is same as the case of MSSM.

The charged mass-squared matrix is the same as in Eq.~(\ref{eq:1dch}) while the pseudo-scalar mass-squared matrix is composed of two $2\times2$ block which can be obtained from Eq.~(\ref{eq:1dleft}) and Eq.~(\ref{eq:1dpseudo}) by putting all the singlet couplings to zero.

The chargino mass matrix in this case is a special limit of {{Section~\ref{case1d}}} obtained by neglecting all the singlet couplings while the neutralino mass matrix is obtained from Eq.~(\ref{eq:1cneutalino}) by putting $\e=0$.

\subsection{Case with two pair of triplets and two bidoublets} \label{case1b}

This case is a realistic model where, unlike previous cases, we can generate the CKM matrices for quarks directly. The calculation of the Higgs mass, though shows that the result for the upper limit on $M_h$ is exactly the same as the case with only one bidoublet. Due to the complexity of the calculations, we only discuss the neutral CP-even Higgs boson mass in this case and see that the largest $M_h$ is the same as with one bidoublet. The particle content of the Higgs sector will be exactly as in Eq.~(\ref{eq:triphig}) except in this case $a = 1,2$.

The superpotential of the model is given as:
\begin{eqnarray}
W&=&\mu_1{\text{Tr}}(\Delta \overline{\Delta})+\mu_2{\text{Tr}}(\Delta^{c}\overline{\Delta}^{c}) +\frac 12\mu_{ab} {\text{Tr}}(\Phi_a^{T}\tau_{2}\Phi_b\tau_{2}).
\end{eqnarray}

The relevant terms in the Higgs potential is given by:
\begin{eqnarray}
V_F&=&|\mu_1|^2{\text{Tr}}(\D^{\dagger} \D+{\ov\D}^{\dagger} \ov\D)+|\mu_2|^2{\text{Tr}}({\D^c}^{\dagger} \D^c+{\ov\D^c}^\dagger \ov\D^c) \notag \\
&+&\sum\limits_{a=1}^2 {\text{Tr}} |(\mu_{a1} \Phi_{1}+\mu_{a2} \Phi_2)|^2, \\
V_D&=&\frac{g_L^2}{8}\sum \limits_{a=1}^3 \left| {\text{Tr}}(2\D^\dagger \tau_a \D+2\ov\D^\dagger \tau_a \ov\D+(\Phi_1^\dagger \tau_a \Phi_1)+(\Phi_2^\dagger \tau_a \Phi_2)\right| ^2\nonumber \\
&+&\frac{g_R^2}{8}\sum \limits_{a=1}^3  \left| {\text{Tr}}(2{\D^c}^\dagger \tau_a \D^c+2\ov{\D^c}^\dagger \tau_a \ov\D^c+(\Phi_1^* \tau_a \Phi_1^T)+(\Phi_2^* \tau_a \Phi_2^T)\right|^2 \nonumber \\
&+& \frac{g_V^2}{2}\left| {\text{Tr}}(\D^\dagger \D- \ov\D^\dagger \ov\D-{\D^c}^\dagger \D^c+ \ov{\D^c}^\dagger \ov\D^c)\right| ^2, \\
V_{Soft}&=& m_{ab}^2 {\text{Tr}}(\Phi_a^\dagger \Phi_b)+\sum\limits_{a,b=1}^2{{B_{ab} \mu_{ab}}} \left[{\text{Tr}}(\Phi_a^T \tau_2 \Phi_b \tau_2)+h.c.\right]+m_3^2{\text{Tr}}(\D^\dagger\D)+m_4^2{\text{Tr}}(\ov\D^\dagger\ov\D) \nonumber \\
&+&m_5^2{\text{Tr}}({\D^c}^\dagger\D^c)+m_6^2{\text{Tr}}({\ov{\D}^c}^\dagger\ov\D^c)+ [B_1 \mu_1{\text{Tr}}(\D\ov\D)+h.c.] \notag \\
&+& [B_2 \mu_2{\text{Tr}}(\D^c\ov\D^c)+h.c.].
\label{eq:three}
\end{eqnarray}

We use this Higgs potential for this variation of the LRSUSY model and calculate the mass-squared matrix for the neutral CP-even Higgs boson. The vacuum structure for this model is given by:
\begin{equation}
\left<\D^c\right> = \begin{pmatrix}
0 & v_R \\ 0 & 0  \end{pmatrix},~
\left<\ov\D^c\right> = \begin{pmatrix}
0&0\\\ov{v}_R & 0  \end{pmatrix},~
\left<\Phi_1\right> ={\begin{pmatrix}
0 & v_{d_1} \\ v_{u_1} & 0 \end{pmatrix}},~
\left<\Phi\right> ={\begin{pmatrix}
0 & v_{{d_2}} \\ v_{{u_2}} & 0 \end{pmatrix}}.
\end{equation}
The left-handed triplet fields $\D$ and $\ov\D$ do not get any VEV. We do a field redefinition with the $\phi_{11}^0, \phi_{21}^0, \phi_{12}^0, \phi_{22}^0$ fields so that only one of the new fields get a non-zero vacuum expectation value. The transformation we use is given by:
\begin{eqnarray}
\rho_1&=&\frac{v_{{u_1}} \phi_{11}^0 +v_{{d_1}} \phi_{21}^0+v_{{u_2}}\phi_{12}^0+v_{{d_2}} \phi_{22}^0}{\sqrt{v_{{u_1}}^2+v_{{d_1}}^2+v_{{u_2}}^2+v_{{d_2}}^2}},~~\rho_2=\frac{v_{{d_1}}\phi_{11}^0-v_{{u_1}} \phi_{21}^0}{\sqrt{v_{{u_1}}^2+v_{{d_1}}^2}},~~\rho_3=\frac{v_{{d_2}}\phi_{21}^0-v_{{u_2}} \phi_{22}^0}{\sqrt{v_{{u_2}}^2+v_{{d_2}}^2}}, \nonumber \\
\rho_4&=&\frac{v_{{u_1}}(v_{{u_2}}^2+v_{{d_2}}^2)\phi_{11}^0+v_{{d_1}}(v_{{u_2}}^2+v_{{d_2}}^2)\phi_{21}^0 -v_{{u_2}}(v_{{u_1}}^2+v_{{d_1}}^2)\phi_{12}^0-v_{{d_2}}(v_{{u_1}}^2+v_{{d_1}}^2)\phi_{22}^0}{\sqrt{(v_{{u_1}}^2+v_{{d_1}}^2)(v_{{u_2}}^2+v_{{d_2}}^2)(v_{{u_1}}^2+v_{{d_1}}^2+v_{{u_2}}^2+v_{{d_2}}^2)}}.\notag
\end{eqnarray}
The $\rho_1$ field gets a VEV of $\sqrt{v_{{u_1}}^2+v_{{d_1}}^2+v_{{u_2}}^2+v_{{d_2}}^2}$, the other fields do not get any VEV. The $\D$ and $\ov\D$ fields decouple and we get a $6\times6$ mass-square matrix in the basis $({\text{Re}}\rho_1,{\text{Re}}\rho_2,{\text{Re}}\rho_3,{\text{Re}}\rho_4,{\text{Re}}{\d^c}^0,{\text{Re}}{\ov\d^c}^0)$. The minimization conditions for this case are given in the Appendix. The matrix elements for this case are not quoted here as they are lengthy and this case is not very interesting in terms of the final result which comes out to be exactly as {{section~\ref{case1a}}}.

Using the minimization conditions and the assumption that the right-handed symmetry breaking scale is much above the electroweak scale, we get the lightest eigenvalue to be:
\begin{equation}
M_{h_{tree}}^2 = \frac{\left(g_L^2+g^\prime\right)  (v_{{u_2}}^2 - v_{{d_2}}^2 + v_{{u_1}}^2 - v_{{d_1}}^2)^2}{ 2(v_{{u_2}}^2 + v_{{d_2}}^2 +
   v_{{u_1}}^2 + v_{{d_1}}^2)} = M_Z^2 \cos^2 2\beta
\end{equation}
where $\tan\beta = \frac{\sqrt{(v_{{u_1}}^2+v_{{u_2}}^2)}}{\sqrt{(v_{{d_1}}^2+v_{{d_2}}^2)}}$ and $v^2 = \sqrt{v_{{u_1}}^2+v_{{d_1}}^2+v_{{u_2}}^2+v_{{d_2}}^2}$. We have made the assumption that $g_R=g_L$.

This result is the same as the previous case with one bidoublet and gives an upper-limit for the tree-level mass of lightest CP-even neutral Higgs boson to be $M_Z$.

\section{Inverse seesaw model} \label{iseesaw}

The Higgs spectrum of this model is given in Eq.~(\ref{eq:2dh}). The most general superpotential terms needed for calculation of the Higgs boson mass are given as:
\begin{equation}
W = i \mu_1 H_L^T \tau_2 \ov H_L + i \mu_1 H_R^T \tau_2 \ov H_R + \lambda H_L^T \tau_2 \Phi \tau_2 H_R + \lambda\ov H_L^T \tau_2 \Phi \tau_2 \ov H_R + \mu {\text{Tr}}\left[\Phi \tau_2 \Phi^T \tau_2\right].
\end{equation}
The relevant Higgs potential in this case is given as:
\begin{eqnarray}
V_F &=& \text{Tr} \left[ \left| i \mu_1 \tau_2 \ov H_L + \lambda \tau_2 \Phi \tau_2 H_R \right| ^2 +\left| i \mu_1 \tau_2 \ov H_R + \lambda \tau_2 \Phi ^T \tau_2 H_L \right| ^2 \right. \nonumber \\
&+&~~~~~\left| -i \mu_1 \tau_2 H_L + \lambda \tau_2 \Phi \tau_2 \ov H_R \right|^2 + \left|- i \mu_1 \tau_2 H_R + \lambda \tau_2 \Phi ^T \tau_2 \ov H_L \right|^2 \nonumber \\
&+&~~~~~\left. \left| \lambda H_R H_L^T+ \lambda \ov H_R \ov H_L^T +2 \mu \phi ^T \right|^2 \right] , \\
V_D &=&\frac{g_L^2}{8}\sum \limits_{a=1}^3 \left| H_L^\dagger \tau_a H_L+\ov H_L^\dagger \tau_a \ov H_L+{\text{Tr}}(\Phi^\dagger \tau_a \Phi)\right| ^2 \nonumber \\ &+&\frac{g_R^2}{8}\sum \limits_{a=1}^3 \left| H_R^\dagger \tau_a H_R+\ov H_R^\dagger \tau_a \ov H_R+{\text{Tr}}(\Phi^* \tau_a \Phi^T)\right| ^2\nonumber \\
&+& \frac{g_V^2}{8}\left| H_R^\dagger H_R-\ov H_R^\dagger \ov H_R-H_L^\dagger H_L+\ov H_L^\dagger \ov H_L \right| ^2, \\
V_{Soft}&=& \text{Tr} \left[ m_1^2 H_L^\dagger H_L+m_2^2 H_R^ \dagger H_R+ m_3^2 \ov H_L^\dagger \ov H_L + m_4^2 \ov H_R^\dagger \ov H_R + m_5^2 \Phi ^ \dagger \Phi \right. \nonumber \\
&+&~~~ \left( \lambda A_{\lambda} H_L^T \tau_2 \Phi \tau_2 H_R + \lambda A_{\lambda} \ov H_L^T \tau_2 \Phi \tau_2 \ov H_R + h.c. \right)+ \left( B \mu \Phi^T \tau_2 \Phi \tau_2 + h.c. \right)   \nonumber \\
&+&~~~ \left. \left( i B_1 \mu_1 H_L^T \tau_2 \ov H_L + i B_1 \mu_1 H_R^T \tau_2 \ov H_R + h.c \right) \right].
\end{eqnarray}
Parity conservation would require $m_1^2=m_2^2$ and $m_3^2=m_4^2$ but as in the previous case, we allow for soft breaking
of parity by the bilinear terms and choose these parameters to be different.

The vacuum expectation values of the Higgs fields are given as:
\begin{eqnarray}
\left<H_L\right> &=& \begin{pmatrix}
v_L \\ 0  \end{pmatrix},
\left<H_R\right> = \begin{pmatrix}
0 \\ v_R   \end{pmatrix},
\left<\ov H_L\right> = \begin{pmatrix}
0 \\ \ov v_L   \end{pmatrix},\nonumber \\
\left<\ov H_R\right> &=& \begin{pmatrix}
\ov v_R \\ 0  \end{pmatrix},
\left<\Phi\right> ={\begin{pmatrix}
0 & v_2 \\ v_1 & 0 \end{pmatrix}}.
\end{eqnarray}

We again choose a rotated basis given as
\begin{eqnarray}
\rho_1&=&\frac{v_{{u_1}} H_L^0 +v_{{d_1}} \ov H_L^0+v_{{u_2}}\phi_{1}^0+v_{{d_2}} \phi_{2}^0}{\sqrt{v_{{u_1}}^2+v_{{d_1}}^2+v_{{u_2}}^2+v_{{d_2}}^2}},~~\rho_2=\frac{v_{{d_1}}H_L^0-v_{{u_1}} \ov H_L^0}{\sqrt{v_{{u_1}}^2+v_{{d_1}}^2}},\notag \\
\rho_3&=&\frac{v_1 v_L H_L^0+v_1 \ov v_L \ov H_L^0-(v_L^2+\ov v_L^2) \phi_1^0}{\sqrt{(v_L^2+\ov v_L^2+v_1^2)(v_L^2+\ov v_L^2)}},\notag \\
\rho_4&=&\frac{v_2 v_L H_L^0+v_2 \ov v_L \ov H_L^0+v_1 v_2 \phi_1^0-(v_1^2+v_L^2+\ov v_L^2)\phi_2^0}{\sqrt{(v_L^2+\ov v_L^2+v_1^2+v_2^2)(v_L^2+\ov v_L^2+v_1^2)}},
\end{eqnarray}
such that only $\rho_1$ gets a non-zero vacuum expectation value at the electroweak symmetry breaking scale. The right-handed doublets get VEVs of order the right-handed symmetry breaking scale. The minimization conditions in this case are given as:
\begin{align}
0= & 2 m_5^2 v_1+\frac{v_1}{2}\left[ g_L^2 \left( v_1^2-v_2^2-v_L^2+\ov v_L^2 \right) +g_R^2
\left( v_1^2-v_2^2-v_R^2+\ov v_R^2 \right) + 4 \lambda^2 \left( v_L^2+v_R^2 \right) \right] \notag \\ & -2 \lambda A_{\lambda} v_L v_R+2 \lambda \mu_1 \left( v_L \ov v_R-v_R \ov v_L \right) + 4 \mu \left( \lambda \ov v_L \ov v_R-B v_2+2 \mu v_1 \right), \notag \\
0= & 2 m_5^2 v_2+\frac{v_2}{2}\left[ g_L^2 \left( -v_1^2+v_2^2+v_L^2-\ov v_L^2 \right) - g_R^2
\left( -v_1^2+v_2^2+v_R^2-\ov v_R^2 \right) + 4 \lambda^2 \left( \ov v_L^2+\ov v_R^2 \right) \right] \notag \\ & -2 \lambda A_{\lambda} \ov v_L \ov v_R-2 \lambda \mu_1 \left( v_L \ov v_R-v_R \ov v_L \right) + 4 \mu \left( \lambda v_L v_R-B v_1+2 \mu v_2 \right), \notag \\
0=& 2 m_1^2 v_L+\frac{v_L}{2}\left[ g_L^2 \left( -v_1^2+v_2^2+v_L^2-\ov v_L^2 \right) + g_V^2
\left( v_L^2-\ov v_L^2-v_R^2+\ov v_R^2 \right) + 4 \lambda^2 \left(  v_1^2+ v_R^2 \right) \right] \notag \\ & -2 \lambda A_{\lambda}  v_1  v_R+2 \lambda \mu_1 \ov v_R \left( v_1 -v_2 \right) + 2 \mu_1^2 v_L+2 B_1 \mu_1 \ov v_L +  4 \mu \lambda v_2 v_R, \notag \\
0=& 2 m_2^2 v_R+\frac{v_R}{2}\left[ g_R^2 \left( -v_1^2+v_2^2+v_R^2-\ov v_R^2 \right) + g_V^2
\left( -v_L^2+\ov v_L^2+v_R^2-\ov v_R^2 \right) + 4 \lambda^2 \left(  v_1^2+ v_L^2 \right) \right] \notag \\ & -2 \lambda A_{\lambda}  v_1  v_L-2 \lambda \mu_1 \ov v_L \left( v_1 -v_2 \right) + 2 \mu_1^2 v_R-2 B_1 \mu_1 \ov v_R +  4 \mu \lambda v_2 v_L, \notag \\
0=& 2 m_3^2 \ov v_L+\frac{\ov v_L}{2}\left[ g_L^2 \left( v_1^2-v_2^2-v_L^2+\ov v_L^2 \right) - g_V^2 \left( v_L^2-\ov v_L^2-v_R^2+\ov v_R^2 \right) + 4 \lambda^2 \left(  v_2^2+\ov v_R^2 \right) \right] \notag \\ & -2 \lambda A_{\lambda}  v_2 \ov  v_R-2 \lambda \mu_1 v_R \left( v_1 -v_2 \right) + 2 \mu_1^2 \ov v_L+2 B_1 \mu_1 v_L +  4 \mu \lambda v_1 \ov v_R, \notag \\
0=& 2 m_4^2 \ov v_R+\frac{\ov v_R}{2}\left[ g_R^2 \left( v_1^2-v_2^2-v_R^2+\ov v_R^2 \right) + g_V^2 \left( v_L^2-\ov v_L^2-v_R^2+\ov v_R^2 \right) + 4 \lambda^2 \left(  v_2^2+\ov v_L^2 \right) \right] \notag \\ & -2 \lambda A_{\lambda}  v_2 \ov  v_L+2 \lambda \mu_1 v_L \left( v_1 -v_2 \right) + 2 \mu_1^2 \ov v_R-2 B_1 \mu_1 v_R +  4 \mu \lambda v_1 \ov v_L.
\label{eq:iseesawmin}
\end{align}

The relevant mass-matrix elements in this case are given as:
\begin{eqnarray}
M_{11}&=&  \frac{g_R^2 \left(v_1^2-v_2^2\right)^2+g_V^2 \left( v_L^2-\ov v_L ^2 \right)^2+g_L^2 \left(v_1^2-v_2^2-v_L^2+\ov{v}_L^2\right)^2+8 \lambda ^2 \left(v_1^2 v_L^2 +v_2^2
\ov{v}_L^2 \right) }{2 \left(v_1^2+v_2^2+v_L^2+\ov{v}_L^2\right)},  \nonumber \\
M_{12}&=& \frac{v_L \ov{v}_L \left(g_V^2 \left(v_L^2-\ov{v}_L^2\right)+g_L^2 \left(-v_1^2+v_2^2+v_L^2-\ov{v}_L^2\right)+2
\left(v_1^2-v_2^2\right) \lambda ^2\right)}{\sqrt{v_L^2+\ov{v}_L^2} \sqrt{v_1^2+v_2^2+v_L^2+\ov{v}_L^2}}, \nonumber \\
M_{13}&=&\left[v_1 \left\{g_V^2 (v_L^2-\ov{v}_L^2)^2+2 g_L^2 v_L^2
\left(-v_1^2+v_2^2+v_L^2-\ov{v}_L^2\right)\right. \right. \nonumber \\
&-& g_R^2 \left(v_1^2-v_2^2\right) \left(v_L^2+\ov{v}_L^2\right)
+\left. \left. 4 \lambda ^2 (v_1^2 v_L^2 - v_L^4 + v_2^2 \ov{v}_L^2 - v_L^2 \ov{v}_L^2)\right\}\right]\mbox {\Large$ /$}\notag \\
&~& ~\left(2 \sqrt{\left(v_L^2+\ov{v}_L^2\right) \left(v_1^2+v_L^2+\ov{v}_L^2\right)} \sqrt{v_1^2+v_2^2+v_L^2+\ov{v}_L^2}\right),\notag \\
M_{14}&=&\left[v_2 \left\{g_V^2 (v_L^2-\ov{v}_L^2)^2+2 g_L^2 (v_1^2+\ov v_L^2)
\left(v_1^2-v_2^2-v_L^2+\ov{v}_L^2\right)\right. \right. \nonumber \\
&+& \left. \left. g_R^2 \left(v_1^2-v_2^2\right) \left(2 v_1^2+v_L^2+\ov{v}_L^2\right)+4 \lambda ^2 \left( 2 v_1^2 v_L^2 -v_1^2 \ov{v}_L^2 + v_2^2 \ov{v}_L^2 -
v_L^2 \ov{v}_L^2 - \ov{v}_L^4 \right) \right\}\right]\mbox {\Large$ /$} \nonumber \\
&~&~ \left(2(v_1^2+v_2^2+v_L^2+\ov{v}_L^2)
\sqrt{\left(v_1^2+v_L^2+\ov{v}_L^2\right)}\right), \notag \\
M_{15} &=& \left[g_R^2 \left(-v_1^2+v_2^2\right) v_R+g_V^2 \left(-v_L^2+\ov{v}_L^2\right) v_R+4 \lambda  \left\{ -A_\lambda v_1 v_L+\mu_1 (-v_1+v_2) \ov{v}_L \right. \right. \notag \\
&+& \left. \left. \lambda v_1^2 v_R  + \lambda v_L^2 v_R +2 \mu v_2 v_L  \right\}\right]/\left(2 \sqrt{v_1^2+v_2^2+v_L^2+\ov{v}_L^2}\right),\notag \\
M_{16} &=&\left[g_R^2 \left(v_1^2-v_2^2\right) \ov{v}_R+g_V^2 \left(v_L^2-\ov{v}_L^2\right) \ov{v}_R+4
\lambda  \left\{\mu_1 (v_1-v_2) v_L-A_\lambda v_2 \ov{v}_L+\lambda v_2^2 \ov{v}_R  \right. \right. \notag \\
&+&\left. \left. \lambda \ov{v}_L^2
\ov{v}_R +2 \mu v_1 \ov{v}_L  \right\}\right]/\left(2 \sqrt{v_1^2+v_2^2+v_L^2+\ov{v}_L^2}\right),\notag \\
M_{55} &=& \mu_1^2+m_2^2-\frac14 \left[ g_R^2(v_1^2-v_2^2-3 v_R^2+\ov v_R^2)+g_V^2(v_L^2-\ov v_L^2-3 v_R^2+\ov v_R^2)\right] +\l^2(v_1^2+v_L^2),\notag \\
M_{56}&=& -\frac{(g_R^2+g_V^2) v_R \ov v_R+2B_1\mu_1}{2}, \notag \\
M_{66}&=&\mu_1^2+m_4^2+\frac14 \left[ g_R^2(v_1^2-v_2^2- v_R^2+3 \ov v_R^2)+g_V^2(v_L^2-\ov v_L^2- v_R^2+3\ov v_R^2)\right] +\l^2(v_2^2+\ov v_L^2).~~~~~~~~~~
\end{eqnarray}

All the other elements in the mass matrix are of order SUSY breaking scale squared or the right-handed symmetry breaking scale squared. The only matrix elements that can provide significant contributions to the lightest eigenvalue comes from $M_{15}$ and $M_{16}$. We focus on the $3\times3$ sector formed by $M_{11},M_{15},M_{16},M_{55},M_{56},M_{66}$. We choose some of the parameters such that the $M_{15}$ and $M_{16}$ terms become zero and check that we have enough freedom to consistently keep the other eigenvalues of the matrix to be positive. The smallest eigenvalue in this case for the lightest CP-even Higgs boson
is given by:
\footnotesize
\begin{equation}
M^2_{h_{tree}} =\frac{g_R^2 \left(v_1^2-v_2^2\right)^2+g_V^2 \left( v_L^2-\ov v_L ^2 \right)^2+g_L^2 \left(v_1^2-v_2^2-v_L^2+\ov{v}_L^2\right)^2+8 \lambda ^2 \left(v_1^2 v_L^2 +v_2^2
\ov{v}_L^2 \right) }{2 \left(v_1^2+v_2^2+v_L^2+\ov{v}_L^2\right)}.
\end{equation}
\normalsize
We define $ v_1 = v \sin \beta \cos \phi$, $ v_2= v \cos \beta \sin \psi$, $v_L = v \cos \beta \cos \psi$, $\ov v_L = v \sin \beta \sin \phi$ and $ g_R=g_L$. Maximizing this  expression with respect to $\phi$ and $\psi$ gives the Higgs boson mass including the one and two loop corrections from the top and stop sector as:
\begin{eqnarray}
M_{h_{max}}^2 &=& \left(2 M_W^2  \sin^4 \beta +\frac{M_W^4}{2 M_W^2 - M_Z^2} \cos^4 \beta -\frac{M_W^2}{2} \sin^2 2\beta+ \lambda^2 v^2 \sin^2 2 \beta\right) \D_1 \notag \\
&+& \D_2,
\end{eqnarray}
where $\D_1$ and $\D_2$ are defined in Eq.~(\ref{eq:rad}). The coefficient of the $\D_1$ term is the tree-level lightest Higgs boson mass.

\begin{figure}[h!]\centering
\includegraphics[width=3.1in]{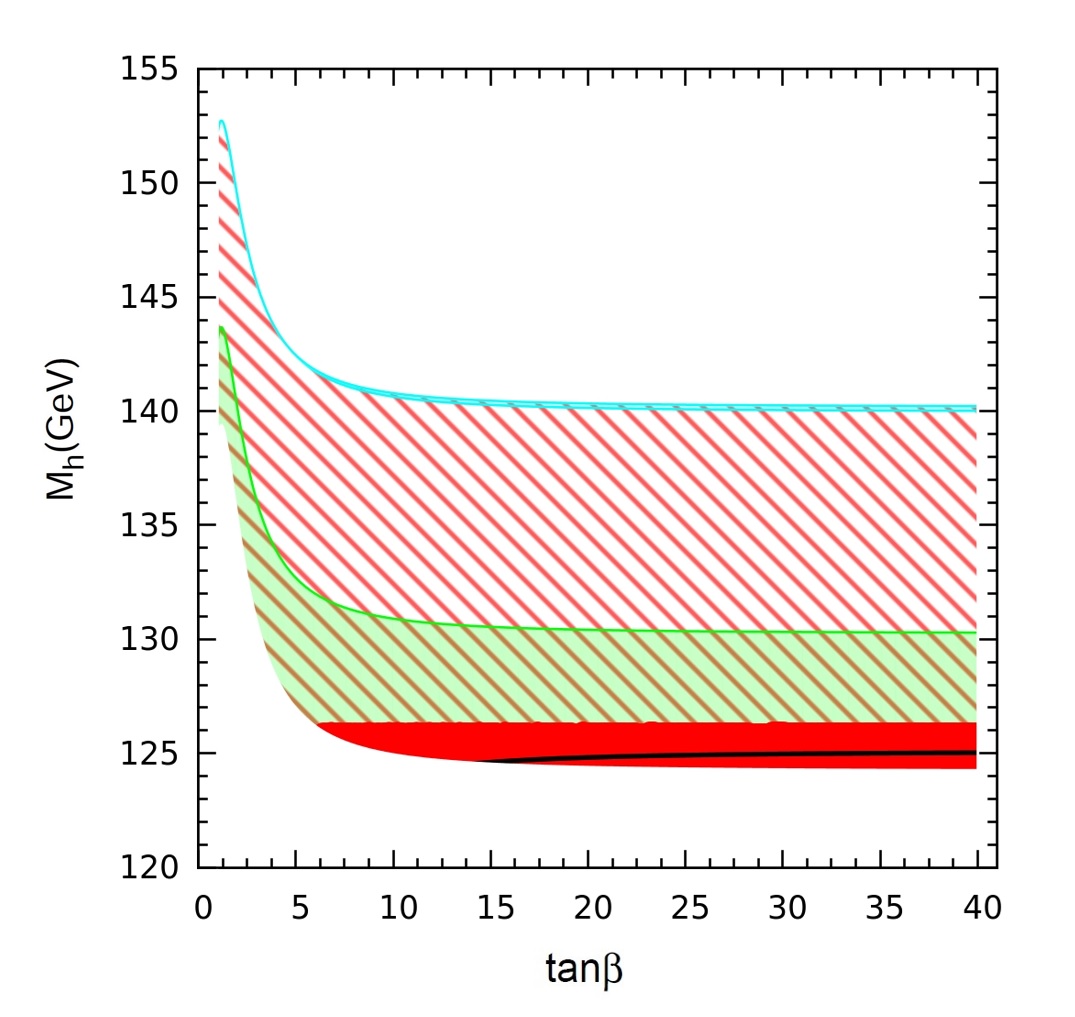}
\includegraphics[width=3.1in]{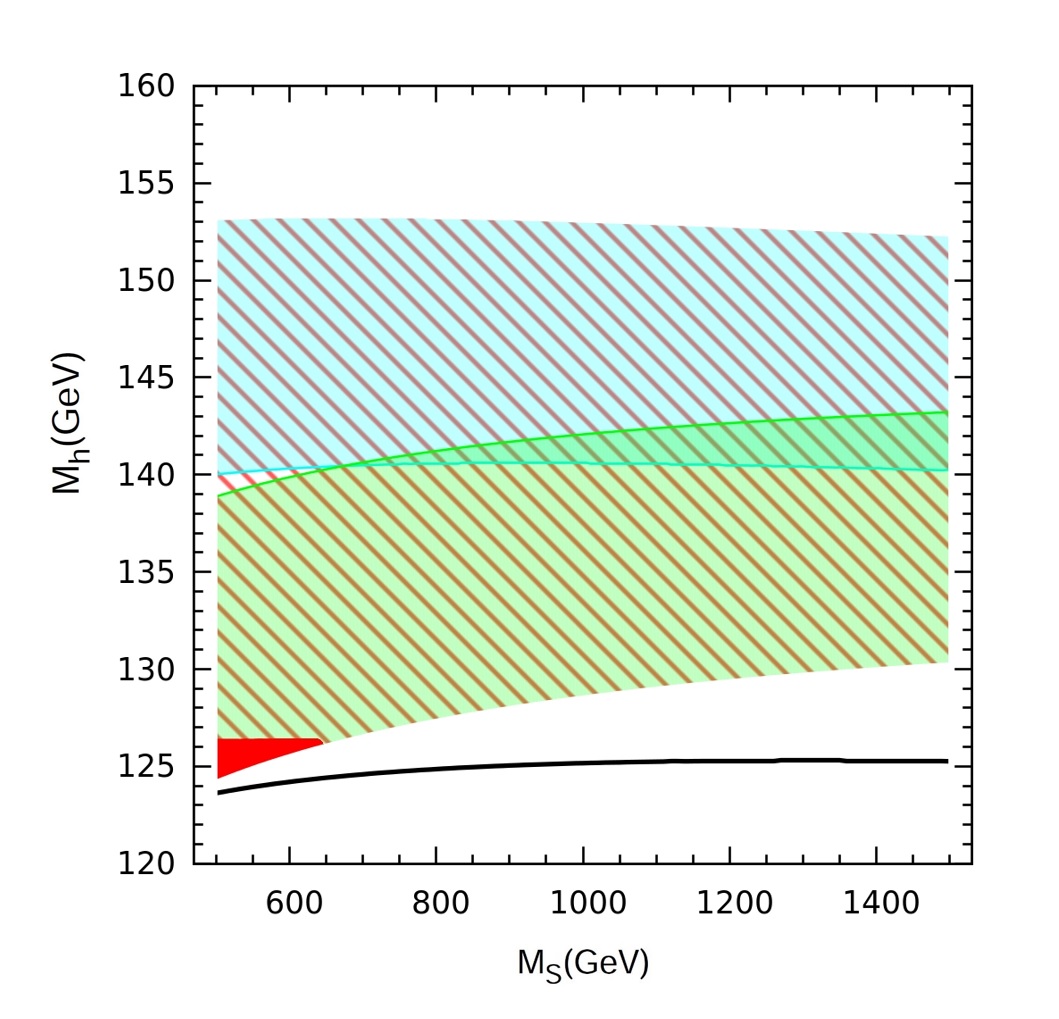}
\caption{{\sl(a) Variation of Higgs boson mass with $\tan \beta$, (b) Higgs boson mass as a function of $M_S$ in the inverse seesaw model.  Notation same as in Fig. \ref{fig:1d}}}
\label{fig:2d}
\end{figure}
The Higgs boson mass is plotted in Fig~\ref{fig:2d}(a) as a function of $\tan \beta$. The red region in the figure represents the band where the mass is between 124 GeV and 126 GeV. The light green region represents the area where the stop squark mixing is minimum, i.e., $X_t=0$ while the blue upper region is for maximal mixing where $X_t=6$. The red shaded region is for all values of Higgs mass greater than 126 GeV and it is overlapped by the blue and the green regions. Fig.~\ref{fig:2d}(b) represents the upper limit of the Higgs mass and as a function $M_S$. Again the red band is where the Higgs boson mass is between 124 GeV and 126 GeV, green region is for $X_t=0$, blue region represents $X_t=6$ and shaded red region is for all values of Higgs mass greater than 126 GeV which is overlapped by the green and the blue regions. The black solid line in each case represents the MSSM upper limit for the Higgs mass.

The pseudo-scalar mass-squared matrix in this case is a $4\times4$ matrix after eliminating the two Goldstone states which are absorbed by the $Z_R$ and $Z$ bosons to get mass. We choose a transformation given as
\begin{align}
& \rho_1 = \frac{v_L \ov H_L^0 + \ov v_L H_L^0}{\sqrt{v_L^2+\ov v_L^2}},~~~~ \rho_2 = \frac{v_R \ov H_R^0 + \ov v_R H_R^0}{\sqrt{v_R^2+\ov v_R^2}},~~~~ \rho_3 = \frac{v_1 v_L H_R^0+v_1 v_R H_L^0+v_L v_R \phi_1^0}{\sqrt{v_1^2 v_L^2+v_1^2 v_R^2+v_L^2 v_R^2}}, \notag \\
&\rho_4 = \frac{v_2 \ov v_L \ov H_R^0+v_2 \ov v_R \ov H_L^0+\ov v_L \ov v_R \phi_2^0}{\sqrt{v_2^2 \ov v_L^2+v_2^2 \ov  v_R^2+\ov v_L^2 \ov v_R^2}}.
\end{align}
 The matrix elements in the basis $(\text{Im}\rho_1, \text{Im}\rho_2,\text{Im}\rho_3,\text{Im}\rho_4)$ are given as:
\begin{eqnarray}
M_{11} &=& -\frac{(v_L^2 + \ov v_L^2) \left[B_1 \mu_1 v_L \ov v_L -
\mu_1 (v_1 \ov v_L v_R + v_2 v_L \ov v_R) \lambda +
2 \mu (\lambda v_2 v_L v_R  + \lambda v_1 \ov v_L \ov v_R -B v_1 v_2)\right]}{v_L^2 \ov v_L^2}, \notag \\
M_{12} &=& -\frac{2\mu \left(-B v_1 v_2 +\l v_2 v_L v_R+ \l v_1 \ov v_L \ov v_R\right)\sqrt{(v_L^2 + \ov v_L^2)(v_R^2 +\ov v_R^2)} }{v_L \ov v_L v_R\ov v_R}, \notag \\
M_{13} &=& -\frac{(\l \mu_1 \ov v_L v_R + 2 B \mu v_2 - 2 \l \mu \ov v_L \ov v_R) \sqrt{(v_L^2+\ov v_L^2)[v_L^2 v_R^2+v_1^2( v_L^2+ v_R^2)]}}{v_L^2 \ov v_L v_R}, \notag \\
M_{14} &=& -\frac{(\l \mu_1 \ov v_R v_L + 2 B \mu v_1 - 2 \l \mu v_L v_R) \sqrt{(v_L^2+\ov v_L^2)[\ov v_L^2 \ov v_R^2+v_2^2(\ov v_L^2+ \ov v_R^2)}}{v_L \ov v_L^2 \ov v_R}, \notag \\
M_{22} &=& \frac{(v_R^2 + \ov v_R^2) \left[B_1 \mu_1 v_R \ov v_R -
\mu_1 (v_2 \ov v_L v_R + v_1 v_L \ov v_R) \lambda -
2 \mu (\lambda v_2 v_L v_R  + \lambda v_1 \ov v_L \ov v_R -B v_1 v_2)\right]}{v_R^2 \ov v_R^2}, \notag \\
M_{23} &=& \frac{(\l \mu_1 \ov v_R v_L - 2 B \mu v_2 + 2 \l \mu \ov v_L \ov v_R) \sqrt{(v_R^2+\ov v_R^2)[v_L^2 v_R^2+v_1^2( v_L^2+ v_R^2)]}}{v_L \ov v_R v_R^2}, \notag \\
M_{24} &=& \frac{(\l \mu_1 \ov v_L v_R - 2 B \mu v_1 + 2 \l \mu v_L v_R) \sqrt{(v_R^2+\ov v_R^2) [\ov v_L^2 \ov v_R^2+v_2^2(\ov v_L^2+ \ov v_R^2)]}}{v_R \ov v_L \ov v_R^2}, \notag \\
M_{33} &=& \frac{[v_L^2 v_R^2 + v_1^2 (v_L^2 + v_R^2)] \left[ \l A_{\l} v_L v_R + \l \mu_1(\ov v_L v_R - v_L \ov v_R) + 2 B \mu v_2 -2 \l \mu \ov v_L \ov v_R \right]}{v_1 v_L^2 v_R^2}, \notag \\
M_{34}&=& \frac{2 \mu B \sqrt{v_L^2 v_R^2+v_1^2(v_L^2+v_R^2)}\sqrt{\ov v_L^2 \ov v_R^2+v_2^2(\ov v_L^2+\ov v_R^2)}}{v_L v_R \ov v_L \ov v_R}, \notag \\
M_{44} &=& \frac{[\ov v_L^2 \ov v_R^2 + v_2^2 (\ov v_L^2 +\ov  v_R^2)] \left[ \l A_{\l} \ov v_L \ov v_R - \l \mu_1(\ov v_L v_R - v_L \ov v_R) + 2 B \mu v_1 -2 \l \mu v_L v_R \right]}{v_2 \ov v_L^2 \ov v_R^2}.
\end{eqnarray}

The charged Higgs boson mass-squared matrix is a $6\times6$ matrix of which there are two zero mass eigenstates which are the Goldstone bosons required to give mass to the $W_R$ and the $W$ bosons. The elements of the $6\times6$ matrix in the original basis are given as:
\begin{eqnarray}
M_{11}&=&m_1^2 +\mu_1^2+\frac14 \left[ g_L^2(v_1^2-v_2^2+v_L^2+\ov v_L^2)+g_V^2(v_L^2-\ov v_L^2-v_R^2+\ov v_R^2)\right]+\l^2(v_2^2+v_R^2),\notag \\
M_{12}&=&-B_1 \mu_1+\frac{g_L^2 v_L \ov v_L}{2},\notag \\
M_{13}&=&\l(\l \ov v_L \ov v_R-A_\l v_2+2 \mu v_1), \notag \\
M_{14} &=&\l[\l v_R \ov v_L+\mu_1(v_1-v_2)],\notag \\
M_{15}&=&\frac12 g_L^2 v_1 v_L+\l( A_\l v_R-\mu_1 \ov v_R-\l v_1 v_L),\notag \\
M_{16}&=&\frac12 g_L^2 v_2 v_L-\l(\mu_1 \ov v_R-\l v_2 v_L-2\mu v_R),\notag \\
M_{22}&=&m_3^2 +\mu_1^2+\frac14 \left[ g_L^2(-v_1^2+v_2^2+v_L^2+\ov v_L^2)+g_V^2(-v_L^2+\ov v_L^2+v_R^2-\ov v_R^2)\right]+\l^2(v_1^2+\ov v_R^2),\notag \\
M_{23}&=&\l[\l v_L \ov v_R-\mu_1(v_1-v_2)],\notag \\
M_{24}&=&\l(\l v_L v_R-A_\l v_1+2 \mu v_2), \notag \\
M_{25}&=&\frac12 g_L^2 v_1 \ov v_L-\l(\mu_1 v_R+\l v_1 \ov v_L-2\mu \ov v_R),\notag \\
M_{26}&=&\frac12 g_L^2 v_2 \ov v_L+\l( A_\l \ov v_R-\mu_1 v_R-\l v_2 \ov v_L),\notag \\
M_{33}&=&m_2^2 +\mu_1^2+\frac14 \left[ g_R^2(v_1^2-v_2^2+v_R^2+\ov v_R^2)+g_V^2(-v_L^2+\ov v_L^2+v_R^2-\ov v_R^2)\right]+\l^2(v_2^2+v_L^2),\notag \\
M_{34}&=&B_1 \mu_1+\frac{g_R^2 v_R \ov v_R}{2},\notag \\
M_{35}&=&\frac12 g_R^2 v_2 v_R+\l(\mu_1 \ov v_L-\l v_2 v_R+2\mu v_L),\notag \\
M_{36}&=&\frac12 g_R^2 v_1 v_R+\l(\mu_1 \ov v_L+\l v_1 v_R+A_\l v_L),\notag \\
M_{44}&=&m_4^2 +\mu_1^2+\frac14 \left[ g_R^2(-v_1^2+v_2^2+v_R^2+\ov v_R^2)+g_V^2(v_L^2-\ov v_L^2-v_R^2+\ov v_R^2)\right]+\l^2(v_1^2+ \ov v_L^2),\notag \\
M_{45}&=&\frac12 g_R^2 v_2 \ov v_R+\l(\mu_1 v_L-\l v_2 \ov v_R+A_\l \ov v_L),\notag \\
M_{46}&=&\frac12 g_R^2 v_1 \ov v_R+\l(\mu_1 v_L-\l v_1 v_R+2\mu \ov v_L),\notag \\
M_{55}&=&m_5^2 +4 \mu^2+\frac14 \left[ g_L^2(v_1^2+v_2^2+v_L^2-\ov v_L^2)+g_R^2(v_1^2+v_2^2 -v_R^2+\ov v_R^2)\right]+\l^2(\ov v_L^2+v_R^2),\notag \\
M_{56}&=&\frac{(g_L^2+g_R^2)v_1 v_2+4 B \mu}{2}, \notag \\
M_{66}&=&m_5^2 +4 \mu^2+\frac14 \left[ g_L^2(v_1^2+v_2^2-v_L^2+\ov v_L^2)+g_R^2(v_1^2+v_2^2 +v_R^2-\ov v_R^2)\right]+\l^2( v_L^2+\ov v_R^2).
\end{eqnarray}

Using the minimization conditions given in Eq.~(\ref{eq:iseesawmin}), we eliminate $m_1$, $m_2$, $m_3$, $B$, $B_1$ and $m_5$. We then numerically calculate the pseudo-scalar and charged Higgs boson masses choosing the remaining parameters such the the lightest Higgs mass is 125 GeV after radiative corrections. We choose a stop squark mass of 500 GeV and the mixng parameter $X_t$=4. The numerical values of the Higgs masses are given in Table~\ref{tab:two} for this choice of parameters.

\vspace*{0.1in}
\noindent
{\large{\bf{Chargino and Neutralino masses}}}
\vspace*{0.1in}

The chargino mass terms in this case is written as
\begin{equation}
{\mathcal{L}}_{chargino} = -\frac{1}{2}\begin{pmatrix}\widetilde{H}_R^+&\widetilde{\ov H}_L^+&\widetilde{\phi}_1^+&\widetilde{W}_R^+ & \widetilde{W}_L^+\end{pmatrix} \begin{pmatrix}\mu_1&-\l v_2&\l v_L&g_R v_R&0 \\-\l v_1&-\mu_1&\l \ov v_R&0&g_L \ov v_L \\\l \ov v_L& \l v_R&2 \mu&g_R v_1&g_L v_1 \\ g_R \ov v_R&0&g_R v_2&M_R&0\\0&g_L v_L&g_L v_2&0&M_L \end{pmatrix} \begin{pmatrix}\widetilde{\ov H}_R^- \\ \widetilde{H}_L^-\\ \widetilde{\phi}_2^- \\ \widetilde{W}_R^- \\ \widetilde{W}_L^- \end{pmatrix},
\end{equation}
and the neutralino mass matrix in the basis $\begin{pmatrix}\widetilde{H}_R^0&\widetilde{H}_L^0&\widetilde{\ov H}_R^0&\widetilde{\ov H}_L^0&\widetilde{\phi}_1^0&\widetilde{\phi}_2^0&\widetilde{B}&\widetilde{W}_{R_3}&\widetilde{W}_{L_3}\end{pmatrix}$ is given as
\begin{equation}
M_{n} = \begin{pmatrix} 0&-\l v_1&-\mu_1&0&-\l v_L&0&\frac{g_V v_R}{\sqrt{2}}&-\frac{g_R v_R}{\sqrt{2}}&0 \\ -\l v_1&0&0&\mu_1&-\l v_R&0&-\frac{g_V v_L}{\sqrt{2}}&0&\frac{g_L v_L}{\sqrt{2}} \\ -\mu_1&0&0&-\l v_2&0&-\l \ov v_L&-\frac{g_V \ov v_R}{\sqrt{2}}&\frac{g_R \ov v_R}{\sqrt{2}}&0 \\0&\mu_1&-\l v_2&0&0&-\l \ov v_R &\frac{g_V \ov v_L}{\sqrt{2}}&0&-\frac{g_L \ov v_L}{\sqrt{2}} \\-\l v_L &-\l v_R&0&0&0&-2 \mu&0&-\frac{g_R v_1}{\sqrt{2}}&-\frac{g_L v_1}{\sqrt{2}} \\ 0&0&-\l \ov v_L& -\l \ov v_R&-2 \mu&0&0&\frac{g_R v_2}{\sqrt{2}}&\frac{g_L v_2}{\sqrt{2}} \\ \frac{g_V v_R}{\sqrt{2}}&-\frac{g_V v_L}{\sqrt{2}}&-\frac{g_V \ov v_R}{\sqrt{2}}&\frac{g_V \ov v_L}{\sqrt{2}}&0&0&M_1&0&0 \\ -\frac{g_R v_R}{\sqrt{2}}&0&\frac{g_R \ov v_R}{\sqrt{2}}&0&-\frac{g_R v_1}{\sqrt{2}}&\frac{g_R v_2}{\sqrt{2}}&0&M_R&0 \\ 0&\frac{g_L v_L}{\sqrt{2}}&0&-\frac{g_L \ov v_L}{\sqrt{2}}&-\frac{g_L v_1}{\sqrt{2}}&\frac{g_L v_2}{\sqrt{2}}&0&0&M_L\end{pmatrix},
\end{equation}

The chargino and neutralino masses are given in Table.~\ref{tab:two}.

\begin{center}
\begin{table}[h!]
\centering
\begin{tabular}{|C{2.7cm}|C{2.7cm}|C{2.8cm}|C{2.8cm}|C{2.9cm}|}  \hline
{\bf{Scalar Higgs boson masses}} & {\bf{Pseudo-scalar Higgs boson masses}} & {\bf{Single charged Higgs boson masses}}& \bf{Chargino masses}&{\bf{Neutralino masses}}  \\ \hline
$M_{H_1}$=5.80 TeV, $M_{H_2}$=5.43 TeV, $M_{H_3}$=3.08 TeV, $M_{H_4}$=694 GeV, $M_{H_5}$=436 GeV & $M_{A_1}$=29.6 TeV, $M_{A_2}$=4.67 TeV, $M_{A_3}$=2.80 TeV, $M_{A_4}$=478 GeV & $M_{H_1^+}$=5.80 TeV, $M_{H_2^+}$=5.21 TeV, $M_{H_3^+}$=3.08 TeV, $M_{H_4^+}$=454 GeV & $M_{\wdt \chi_1^+}$=5.80 TeV, $M_{\wdt \chi_2^+}$=3.87 TeV, $M_{\wdt \chi_3^+}$=2.86 TeV, $M_{\wdt \chi_4^+}$=1.88 TeV, $M_{\wdt \chi_5^+}$=800 GeV & $M_{\wdt \chi_{1,2}^0}$=5.80 TeV, $M_{\wdt \chi_3^0}$=4.31 TeV, $M_{\wdt \chi_4^0}$=2.90 TeV, $M_{\wdt \chi_{5,6}^0}$=2.86 TeV, $M_{\wdt \chi_7^0}$=2.09 TeV, $M_{\wdt \chi_8^0}$=800 GeV, $M_{\wdt \chi_9^0}$=526 GeV \\ \hline
    \end{tabular}
    \caption{{Higgs boson, chargino and neutralino masses for inverse seesaw model with a sample point given as:
    $\l$=0.36, $v_1$=165.8 GeV, $v_2$=8 GeV, $v_L$=10 GeV, $\ov v_L$=51 GeV, $v_R$=3 TeV, $\ov v_R$=4 TeV, $\mu_1$=-2.68 TeV, $\mu$=-2.8 TeV, $m_4^2$=$-700^2~\text{GeV}^2$, $A_\l$=700 GeV, $M_R$=800 GeV, $M_L$=800 GeV, $M_1$=400 GeV.}}
\label{tab:two}
\end{table}
\end{center}

\section{Universal Seesaw model} \label{useesaw}

\subsection{Case with a Singlet} \label{3.1}

The particle spectrum for this case is given in Eq.~(\ref{eq:2ch}) with an additional singlet Higgs field $S$. The superpotential is given as:
\begin{eqnarray}
W&=& S(i \lambda H_L^T \tau_2 \overline{H}_L + i\lambda^c H_R^T \tau_2\overline{H}_R-M^2),
\end{eqnarray}
where $\l^c=\l^*$ and $M^2$ is real from parity invariance.

The $D$-terms, $F$-terms and the soft supersymmetry breaking terms are given as:
\begin{eqnarray}
V_F&=& \left| \lambda
   {\text{Tr}} [i H_L^T \tau_2 \ov H_L + i H_R^T \tau_2 \ov H_R] -
     M^2 \right|^2\nonumber \\
     &+& |\lambda S|^2 {\text{Tr}}[H_L^{\dagger} H_L+{\ov H_L}^{\dagger} \ov H_L+H_R^{\dagger} H_R+{\ov H_R}^\dagger \ov H_R], \\
\label{eq:2cdterm}
V_D&=&\frac{g_L^2}{8}\sum \limits_{a=1}^3 |H_L^\dagger \tau_a H_L+\ov H_L^\dagger \tau_a \ov H_L|^2\nonumber \\
&+&\frac{g_R^2}{8}\sum \limits_{a=1}^3  |H_R^\dagger \tau_a H_R+\ov H_R^\dagger \tau_a \ov H_R|^2 \nonumber \\
&+& \frac{g_V^2}{8}|-H_L^\dagger H_L+\ov H_L^\dagger \ov H_L+H_R^\dagger H_R-\ov H_R^\dagger \ov H_R|^2 , \\
V_{Soft}&=&  m_3^2(H_L^{\dagger} H_L)+m_4^2({H_R}^{\dagger} H_R)+m_5^2({\ov H_L}^{\dagger} \ov H_L)+m_6^2({\ov H_R}^\dagger \ov H_R)+m_S^2 |S|^2\nonumber \\
&+&  \left[ \lambda A_\lambda S( H_L^T \tau_2 \overline{H}_L +  H_R^T \tau_2\overline{H}) +h.c. \right] +(\lambda C_{\lambda} M^2 S +h.c.) .
\end{eqnarray}

We choose a rotated basis which is exactly the same as in Eq.~(\ref{eq:basis1d}) with $\phi_1 \rightarrow H_L, \phi_2 \rightarrow \ov H_L,{\d^c}^0 \rightarrow H_R,{\ov \d^c}^0 \rightarrow \ov H_R, v_1 \rightarrow v_L,  v_2 \rightarrow \ov v_L$. The minimization conditions are slightly modified form of Eq.~(\ref{eq:min1d}) and are given by:
\begin{align}
0=&  v_L [4 m_3^2 + g_L^2 (-\ov v_L^2 + v_L^2) +
     g_V^2 (-\ov v_L^2 + v_L^2 -  v_R^2 +  \ov v_R^2)]
      + 4 \lambda A_\lambda  \ov v_L v_S+ 4 \lambda^2 v_L v_S^2 \nonumber \\ &+
  4 \lambda \ov v_L (-M^2 +\lambda v_L \ov v_L -\lambda v_R \ov v_R), \notag \\
0=& \ov v_L [4 m_5^2 + g_L^2 (-v_L^2 + \ov v_L^2) +
     g_V^2 (-v_L^2 + \ov v_L^2 +  v_R^2 -  \ov v_R^2)]
      + 4 \lambda A_\lambda  v_L v_S +4 \lambda^2\ov v_L v_S^2 \notag \\
     &+  4 \lambda v_L (-M^2 +\lambda v_L \ov v_L -\lambda v_R \ov v_R),  \nonumber \\
0=&  4 m_4^2 v_R - g_V^2 v_R (-v_L^2 + \ov v_L^2 +  v_R^2 -  \ov v_R^2) + g_R^2 v_R (v_R^2 -\ov v_R^2)\nonumber \\
 & - 4 \lambda A_\lambda \ov v_R v_S  +4 \lambda \ov v_R  (M^2 - \lambda v_L \ov v_L) +4 \lambda ^2 v_R (\ov v_R^2 + v_S^2), \nonumber \\
0=&  4 m_6^2 \ov v_R + g_V^2 \ov v_R (v_L^2 - \ov v_L^2 -  v_R^2 +  \ov v_R^2) +
 g_R^2 \ov v_R (\ov v_R^2 - v_R^2)\nonumber \\
 & - 4 \lambda A_\lambda v_R v_S  +4 \lambda v_R  (M^2 - \lambda v_L \ov v_L) +4 \lambda ^2 \ov v_R ( v_R^2 + v_S^2), \nonumber \\
0=&      2 m_S^2 v_S +2 C_\lambda M^2 \lambda +2 \lambda A_\lambda (v_L \ov v_L-v_R \ov v_R)+\lambda^2 (v_L^2 + \ov v_L^2 + v_R^2 +
      \ov v_R^2) v_S.
\label{eq:min2c}
\end{align}

Using this minimization and the basis $({\text{Re}}\rho_1,{\text{Re}}\rho_2,{\text{Re}}{H_R}^0,{\text{Re}}{\ov H_R}^0)$, the relevant mass-squared matrix elements are given by:
\begin{eqnarray} \label{eq:masmat2c}
M_{11}& =&  \frac{g_L^2 (v_L^2 - \ov v_L^2)^2 + g_V^2 (v_L^2 - \ov v_L^2)^2 +
   8 v_L^2 \ov v_L^2 \lambda^2}{2 (v_L^2 + \ov v_L^2)},\notag \\
   M_{12}& =& \frac{ v_L \ov v_L (v_L^2 - \ov v_L^2) (g_L^2 + g_V^2 - 2 \lambda^2)}{ (v_L^2 + \ov v_L^2)},\notag \\
M_{13}& =& \frac{-g_V^2 (v_L^2 - \ov v_L^2) (v_R^2 -\ov v_R^2) -
 8 \lambda^2 v_L \ov v_L v_R \ov v_R}{\sqrt{(v_L^2 + \ov v_L^2)(v_R^2 +\ov v_R^2)}},  \notag \\
 M_{14}& =& \frac{ -g_V^2 (v_L^2 - \ov v_L^2)v_R \ov v_R +
  2 \lambda^2 v_L \ov v_L  (v_R^2 -\ov v_R^2)}{\sqrt{(v_L^2 + \ov v_L^2)(v_R^2 +\ov v_R^2)}}, \notag \\
  M_{15} &=& \frac{\lambda [2 A_\lambda v_L \ov v_L  +
   2 (v_L^2 + \ov v_L^2) v_S \lambda]}{\sqrt{v_L^2 + \ov v_L^2}}, \notag \\
   M_{55} &=& m_S^2 + (v_L^2 + \ov v_L^2 + v_R^2 +\ov v_R^2) \lambda^2.
\end{eqnarray}

The other terms in the mass matrix are given in the appendix. We choose the ratio between $\ov v_R$ and $v_R$ such that the matrix element $M_{13}$ vanishes and we choose the value of $A_ \lambda$ such that $M_{15}$ becomes zero. Then we calculate the correction from the off-diagonal elements to the lightest eigenvalue of this mass-squared matrix. In the limit where the soft-supersymmetry breaking parameter $m_6$ is significantly larger $v_R$, we can show that this correction vanishes. We use the definitions of $\tan\beta = \frac{\ov v_L}{v_L}$ and $v^2 = v_L^2+\ov v_L^2$. Including the loop corrections from the top and stop sector, the Higgs boson mass is:
\begin{eqnarray}
M_h^2 &=& \left(\frac{M_W^4}{2M_W^2-M_Z^2}\cos^22\beta+\lambda^2 v^2 \sin^2 2 \beta\right) \D_1 + \D_2,
\end{eqnarray}
where $\D_1$ and $\D_2$ are defined in Eq.~(\ref{eq:rad}). As before the coefficient of the $\D_1$ term is the tree-level Higgs boson mass upper limit.

The Higgs boson mass is plotted in Fig~\ref{fig:2c}(a) as a function of $\tan \beta$. The red region in the figure represents the band where the mass is between 124 GeV and 126 GeV. The light green region represents the area where the stop squark mixing is minimum, i.e., $X_t=0$ while the blue upper region is for maximal mixing where $X_t=6$. The red shaded region is for all values of Higgs mass greater than 126 GeV and it is overlapped by the blue and the green regions. Fig.~\ref{fig:2c}(b) represents the upper limit of the Higgs mass and as a function $M_S$. Again the red band is where the Higgs boson mass is between 124 GeV and 126 GeV, green region is for $X_t=0$, blue region represents $X_t=6$ and shaded red region is for all values of Higgs mass greater than 126 GeV which is overlapped by the green and the blue regions. The black solid line in each case represents the MSSM upper limit for the Higgs mass.

\begin{figure}[h!]\centering
\includegraphics[width=3.0in]{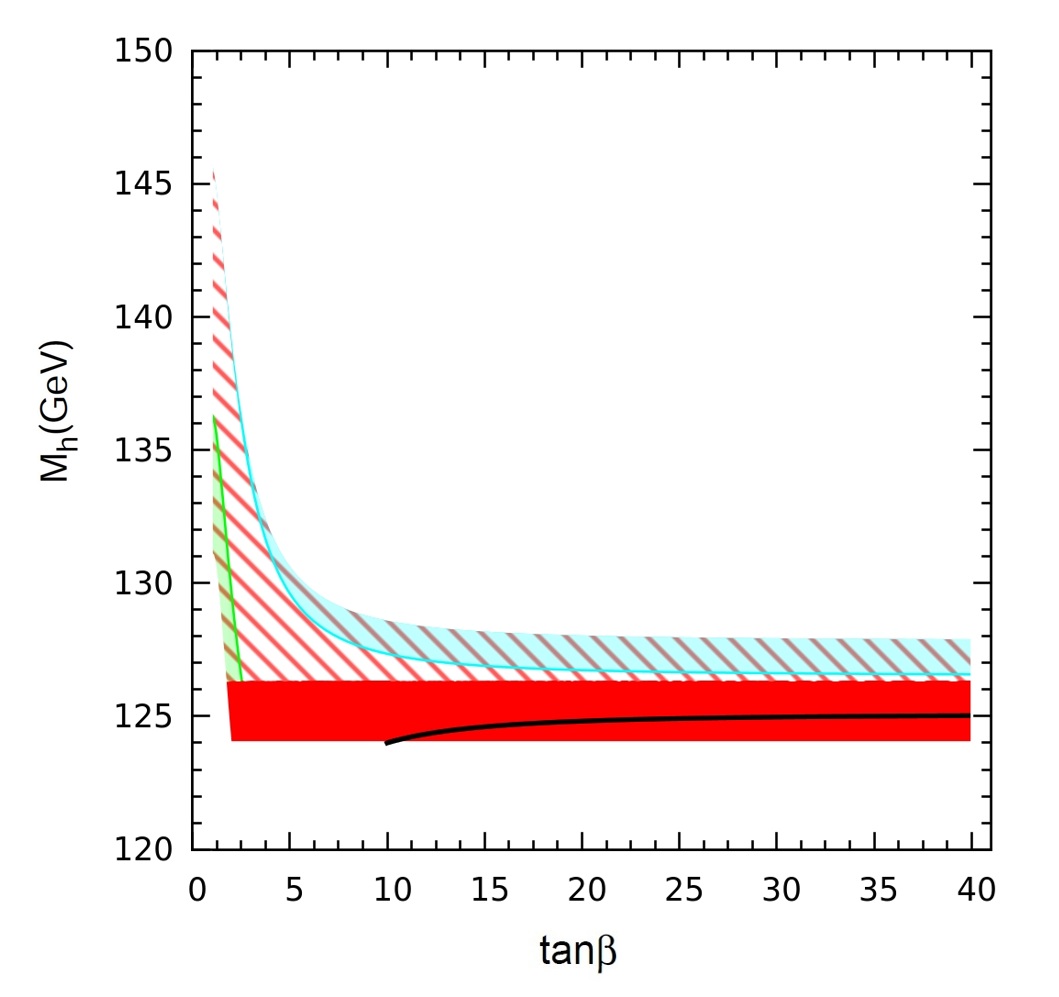}
\includegraphics[width=3.1in]{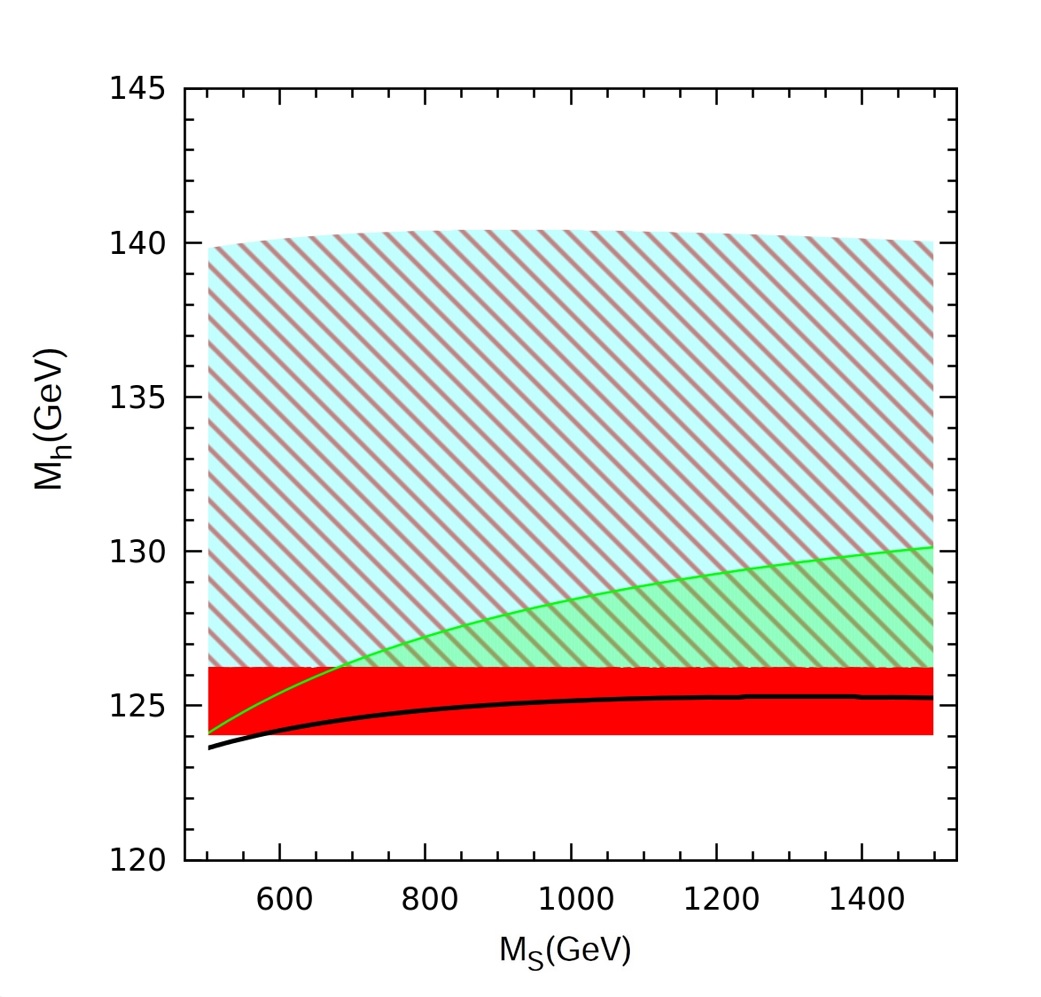}
\caption{{\sl(a) Variation of Higgs boson mass with $\tan \beta$, (b) Higgs boson mass as a function of $M_S$ in the universal seesaw model.  Notation same as in Fig. \ref{fig:1d}.}}
\label{fig:2c}
\end{figure}

To obtain the pseudo scalar mass-squared matrix , we make the following transformation
\begin{align}
\rho_1 = \frac{v_R \ov H_R^0 + \ov v_R H_R^0}{\sqrt{v_R^2+\ov v_R^2}},~~\rho_2 = \frac{v_L \ov H_L^0 + \ov v_L H_L^0}{\sqrt{v_L^2+\ov v_L^2}}, ~~g_1 = \frac{v_R H_R^0 - \ov v_R \ov H_R^0}{\sqrt{v_R^2+\ov v_R^2}},~~ g_2 = \frac{v_L H_L^0 - \ov v_L \ov H_L^0}{\sqrt{v_L^2+\ov v_L^2}}.
\end{align}
The imaginary components of $g_1$ and $g_2$ are identified as the Goldstone states, and the $3\times3$ pseudo-scalar matrix elements in the basis $(\text{Im}\rho_1, \text{Im}\rho_2,\text{Im}S)$ are given as:
\begin{eqnarray}
M_{11} &=& m_4^2+ m_6^2+\l^2(v_R^2+\ov v_R^2+ v_S^2), \notag \\
M_{12} &=& -\l^2  \sqrt{(v_R^2+\ov v_R^2)(v_L^2+\ov v_L^2)} \notag \\
M_{13} &=& \l A_{\l}\sqrt{v_R^2+\ov v_R^2} \notag \\
M_{22} &=& m_3^2+m_5^2+\l^2 (v_L^2+\ov v_L^2+2  v_S^2) \notag \\
M_{23} &=& -\l A_{\l}\sqrt{v_L^2+\ov v_L^2} \notag \\
M_{33} &=& m_S^2 + \l^2(v_L^2+\ov v_L^2+v_R^2+\ov v_R^2)
\end{eqnarray}

The charged Higgs boson matrix is obtained by identifying the Goldstone boson states to be:
\begin{equation}
g_1^+ = \frac{v_R H_R^+-\ov v_R {\ov H_R^-}^*}{\sqrt{v_R^2+\ov v_R^2}},~~~~
g_2^+ = \frac{v_L {H_L^-}^*-\ov v_L\ov H_L^+}{\sqrt{v_L^2+\ov v_L^2}},
\label{eq:useechgold}
\end{equation}
and the physical charged Higgs boson mass eigenstates as:
\begin{equation}
h_1^+ = \frac{\ov v_R H_R^++v_R {\ov H_R^-}^*}{\sqrt{v_R^2+\ov v_R^2}},~~~~
h_2^+ = \frac{\ov v_L {H_L^-}^*+ v_L\ov H_L^+}{\sqrt{v_L^2+\ov v_L^2}}.
\end{equation}
The two eigenvalues of the charged Higgs boson mass-squared matrix in this case are given by:
\begin{eqnarray}
M^2_{h^+_1} &=& m_4^2+m_6^2+\frac{1}{2} g_R^2(v_R^2+\ov v_R^2)+2 \l^2 v_S^2, \notag \\
M^2_{h^+_2} &=& m_3^2+m_5^2+\frac{1}{2} g_L^2(v_L^2+\ov v_L^2)+2 \l^2 v_S^2.
\end{eqnarray}

We again use the minimization conditions given in Eq.~(\ref{eq:min2c}) to eliminate $m_3$, $m_4$, $m_5$, $m_6$ and $C_\l$. We take a stop squark mass of 600 GeV and $X_t$ = 6. Using all these constraints on the aforementioned parameters and making sure that the lightest CP-even Higgs boson mass is 125 GeV, we numerically calculate the masses of the charged and pseudo-scalar Higgs boson for a sample point. The results are given in Table~\ref{tab:three} for this choice of parameters.

\vspace*{0.1in}
\noindent
{\large{\bf{Chargino and Neutralino masses}}}
\vspace*{0.1in}

The chargino mass terms in this case is written as
\begin{equation}
{\mathcal{L}}_{chargino} = -\frac{1}{2}\begin{pmatrix}\widetilde{H}_R^+&\widetilde{\ov H}_L^+&\widetilde{W}_R^+ & \widetilde{W}_L^+\end{pmatrix}\begin{pmatrix}\l^* v_S&0&g_R v_R&0 \\ 0&- \l v_S&0&g_L \ov v_L \\ g_R \ov v_R&0&M_R&0\\0&g_L v_L&0&M_L \end{pmatrix} \begin{pmatrix}\widetilde{\ov H}_R^-\\ \widetilde{H}_L^- \\ \widetilde{W}_R^- \\ \widetilde{W}_L^- \end{pmatrix},
\label{eq:2cch}
\end{equation}
and the neutralino mass matrix in the basis $\begin{pmatrix}\widetilde{H}_R^0&\widetilde{H}_L^0&\widetilde{\ov H}_R^0&\widetilde{\ov H}_L^0&\widetilde{B}&\widetilde{W}_{R_3}&\widetilde{W}_{L_3}&\widetilde{S}\end{pmatrix}$ is given as
\begin{equation}
M_{n} = \begin{pmatrix} 0&0&-\l^* v_S&0&\frac{g_V v_R}{\sqrt{2}}&-\frac{g_R v_R}{\sqrt{2}}&0&-\l^* \ov v_R \\ 0&0&0&\l v_S&-\frac{g_V v_L}{\sqrt{2}}&0&\frac{g_L v_L}{\sqrt{2}}&\l \ov v_L \\ -\l^* v_S &0&0&0&-\frac{g_V \ov v_R}{\sqrt{2}}&\frac{g_R \ov v_R}{\sqrt{2}}&0&-\l^* v_R \\ 0&\l v_S&0&0&\frac{g_V \ov v_L}{\sqrt{2}}&0&-\frac{g_L \ov v_L}{\sqrt{2}}&\l v_L \\ \frac{g_V v_R}{\sqrt{2}}&-\frac{g_V v_L}{\sqrt{2}}&-\frac{g_V \ov v_R}{\sqrt{2}}&\frac{g_V \ov v_L}{\sqrt{2}}&M_1&0&0&0 \\ -\frac{g_R v_R}{\sqrt{2}}&0&\frac{g_R \ov v_R}{\sqrt{2}}&0&0&M_R&0&0 \\ 0&\frac{g_L v_L}{\sqrt{2}}&0&-\frac{g_L \ov v_L}{\sqrt{2}}&0&0&M_L&0 \\ -\l^* \ov v_R&\l \ov v_L&-\l^* v_R&\l v_L&0&0&0&0\end{pmatrix},
\end{equation}
where $\widetilde{W}_R,\widetilde{W}_L$ and $\widetilde{B}$ are the superpartners of the right-handed gauge bosons, left-handed gauge bosons and the $U(1)_{B-L}$ gauge boson and $M_R, M_L$ and $M_1$ are their soft masses respectively as given in Eq.~{\ref{eq:gluino}. The numerical values of the masses for the chosen sample point are given in Table~{\ref{tab:three}.

\begin{center}
\begin{table}[h!]
\centering
\begin{tabular}{|C{2.7cm}|C{2.7cm}|C{2.8cm}|C{2.9cm}|C{2.9cm}|} \hline
{\bf{Scalar Higgs boson masses}} & {\bf{Pseudo-scalar Higgs boson masses}} & {\bf{Single charged Higgs boson masses}}& \bf{Chargino masses}&{\bf{Neutralino masses}}  \\ \hline
$M_{H_1}$=4.53 TeV, $M_{H_2}$=2.47 TeV, $M_{H_3}$=1.84 TeV, $M_{H_4}$=636 GeV & $M_{A_1}$=4.49 TeV, $M_{A_2}$=1.92 TeV, $M_{A_3}$=636 GeV & $M_{H_1^+}$=1.98 TeV, $M_{H_2^+}$=641 GeV & $M_{\wdt \chi_1^+}$=2.55 TeV, $M_{\wdt \chi_2^+}$=1.47 TeV, $M_{\wdt \chi_3^+}$=809 GeV, $M_{\wdt \chi_4^+}$=274 GeV & $M_{\wdt \chi_1^0}$=2.96 TeV, $M_{\wdt \chi_2^0}$=2.13 TeV, $M_{\wdt \chi_{3}^0}$=2.02 TeV, $M_{\wdt \chi_4^0}$=1.85 TeV, $M_{\wdt \chi_4^0}$=809 GeV, $M_{\wdt \chi_6^0}$=543 GeV, $M_{\wdt \chi_7^0}$=281 GeV, $M_{\wdt \chi_8^0}$=266 GeV \\ \hline
    \end{tabular}
    \caption{{Higgs boson, chargino and neutralino masses for Universal seesaw model with a singlet Higgs boson field using a sample point with parameters given as: 
    $\l$=0.46, $v_L$=7.4 GeV, $\ov v_L$=173.85 GeV, $v_R$=3 TeV, $\ov v_R$=3.1 TeV, $M^2$=-$2.2^2~ \text{TeV}^2$, $v_S$=600 GeV, $m_S^2$=16 TeV$^2$, $A_{\l}$=-1 TeV, $M_R$=800 GeV, $M_L$=800 GeV, $M_1$=400 GeV.}}
\label{tab:three}
\end{table}
\end{center}

\subsection{Case without singlet}\label{3.2}

The most general superpotential involving the Higgs fields in this case is given by:
\begin{eqnarray}
W&=&i \mu_1 H_L^T \tau_2 \overline{H}_L+i \mu_2 H_R^T \tau_2\overline{H}_R.
\end{eqnarray}
The $D$-terms in the superpotential is the same as in Eq.~(\ref{eq:2cdterm}). The $F$-terms and the soft supersymmetry breaking terms in the Higgs potential are given by:
\begin{eqnarray}
V_F&=&\mu_1^2(H_L^{\dagger} H_L+\ov H_L^{\dagger} \ov H_L)+\mu_2^2(H_R^{\dagger} H_R+\ov H_R^\dagger \ov H_R), \\
V_{Soft}&=& B_1 \mu_1(i H_L^T \tau_2 \overline{H}_L+h.c.)+B_2 \mu_2(i H_R \tau_2\overline{H}_R+h.c) \nonumber \\&+& m_3^2(H_L^{\dagger} H_L)+m_4^2({\ov H_L}^{\dagger} \ov H_L)+m_5^2({H_R}^{\dagger} H_R)+m_6^2({\ov H_R}^\dagger \ov H_R).
\label{eq:pot2b}
\end{eqnarray}

The vacuum structure in this case is given as:
\begin{equation}
\left<H_L\right> = \begin{pmatrix}
v_L \\ 0  \end{pmatrix},
\left<H_R\right> = \begin{pmatrix}
0 \\ v_R   \end{pmatrix},
\left<\ov H_L\right> = \begin{pmatrix}
0 \\ \ov v_L   \end{pmatrix},
\left<\ov H_R\right> = \begin{pmatrix}
\ov v_R \\ 0  \end{pmatrix}.
\end{equation}
We take a rotated basis given by:
\begin{equation}  \label{eq:basis2b}
\rho_1=\frac{v_L {H_L}^0 +\ov v_L {\ov {H}_L}^0}{\sqrt{v_L^2+\ov v_L^2}},~~~~~~~~~~\rho_2=\frac{\ov v_L {H_L}^0-v_L \ov H_L^0}{\sqrt{v_L^2+\ov v_L^2}}.
\end{equation}

The minimization conditions are given by:
\begin{align}
0&=2 \mu_2^2 v_R + 2 m_4^2 v_R - 2 B_2 \mu_2 \ov v_R +
 \frac12 v_R[ g_R^2 (v_R^2 - \ov v_R^2) -
     g_V^2 (v_L^2 -\ov v_L^2 - v_R^2 +\ov v_R^2)], \notag \\
0&=2 \mu_2^2 \ov v_R + 2 m_6^2 \ov v_R-2 B_2 \mu_2 v_R  +
 \frac12 \ov v_R [g_R^2 (-v_R^2 +\ov v_R^2) +
    g_V^2 (v_L^2 -\ov v_L^2 - v_R^2 +\ov v_R^2)],  \notag \\
0&=2 \mu_1^2 v_L + 2 m_3^2 v_L + 2 B_1 \mu_1 \ov v_L +
 \frac12 v_L [g_L^2 (v_L^2 -\ov v_L^2) +
    g_V^2 (v_L^2 -\ov v_L^2 - v_R^2 +\ov v_R^2)], \notag \\
0&=2 \mu_1^2\ov v_L + 2 m_5^2\ov v_L + 2 B_1 \mu_1 v_L +
 \frac12 \ov v_L [g_L^2 (- v_L^2 +\ov v_L^2) +
    g_V^2 (\ov v_L^2 - v_L^2 + v_R^2 -\ov v_R^2)].
\label{eq:min2b}
\end{align}

Using the potential and minimization equations, we calculate the mass-squared matrix in the basis $({\text{Re}}\rho_1,{\text{Re}}\rho_2,{\text{Re}}{H_R}^0,{\text{Re}}{\ov H_R}^0)$. We get the following matrix:
\begin{equation}
\left[
\begin{array}{cccc}
\frac{(g_L^2+g_V^2)(v_L^2-\ov v_L^2)^2}{2(v_L^2+\ov v_L^2)}&\frac{(g_L^2+g_V^2)(v_L^2-\ov v_L^2)v_L\ov v_L}{(v_L^2+\ov v_L^2)}&-\frac{g_V^2 v_R(v_L^2-\ov v_L^2)}{2\sqrt{(v_L^2+\ov v_L^2)}}&\frac{g_V^2 \ov{v}_R(v_L^2-\ov v_L^2)}{2\sqrt{(v_L^2+\ov v_L^2)}}\\
\frac{(g_L^2+g_V^2)(v_L^2-\ov v_L^2)v_L\ov v_L}{(v_L^2+\ov v_L^2)}&\frac{ 2(g_L^2+g_V^2)(v_L^2 \ov v_L^2+(m_3^2+m_5^2+\mu_1^2)(v_L^2+\ov v_L^2)}{(v1_L^2+\ov v_L^2)}&-\frac{ g_V^2 v_L \ov v_L v_R}{\sqrt{v_L^2+\ov v_L^2}}&\frac{ g_V^2 v_L \ov v_L \ov{v}_R}{\sqrt{v_L^2+\ov v_L^2}} \\
-\frac{g_V^2 v_R(v_L^2-\ov v_L^2)}{2\sqrt{(v_L^2+\ov v_L^2)}}&-\frac{ g_V^2 v_L \ov v_L v_R}{\sqrt{v_L^2+\ov v_L^2}}&\frac{(g_R^2+g_V^2)v_R^3+2B_2 \mu_2\ov{v}_R}{2v_R}&\mbox {\footnotesize $ -B_2 \mu_2-\frac{1}{2}(g_R^2+g_V^2)v_R\ov{v}_R$ }\\
\frac{g_V^2 \ov{v}_R(v_L^2-\ov v_L^2)}{2\sqrt{(v_L^2+\ov v_L^2)}}&\frac{ g_V^2 v_L \ov v_L \ov{v}_R}{\sqrt{v_L^2+\ov v_L^2}}&\mbox {\footnotesize $-B_2 \mu_2-\frac{1}{2}(g_R^2+g_V^2)v_R\ov{v}_R $}&\frac{(g_R^2+g_V^2)\ov{v}_R^3+2B_2 \mu_2v_R}{2\ov{v}_R}
\end{array}
\right].
\label{eq:masmat2b}
\end{equation}

Here we have assumed $v_R,\ov{v}_R \neq 0$ in obtaining the mass matrix. We calculate the contribution of the off-diagonal elements to the lightest eigenvalue using the seesaw formula and this gives us the result
\begin{eqnarray}
M_{h_{tree}}^2 = M_Z^2 \cos^2 2 \beta,
\end{eqnarray}
where we have also assumed that the $SU(2)_R$ gauge coupling ($g_R$) is equal to the $SU(2)_L$ gauge coupling ($g_L$), $\tan\beta = \frac{\ov v_L}{v_L}$ and $v^2 = v_L^2+\ov v_L^2$.

The CP-odd Higgs boson mass-squared matrix is a $4\times4$ matrix which has two Goldstone states same as in Sec.~\ref{3.1}. The resulting matrix after eliminating the Goldstone states is a $2\times2$ matrix whose eigenvalues are given as:
\begin{eqnarray}
M^2_{A_1} = m_4^2+m_6^2+2 \mu_2^2, ~~~~~~M^2_{A_2} = m_3^2+m_5^2+2 \mu_1^2.
\end{eqnarray}

The charged Higgs boson matrix is again a $4\times4$ matrix with two Goldstome states which are the same as in Eq.~(\ref{eq:useechgold}). The eigenvalues of the remaining $2\times2$ charged Higgs boson mass-squared matrix in this case are given by:
\begin{eqnarray}
M^2_{h^+_1} &=& m_4^2+m_6^2+\frac{1}{2} g_R^2(v_R^2+\ov v_R^2)+2 \mu_2^2, \notag \\
M^2_{h^+_2} &=& m_3^2+m_5^2+\frac{1}{2} g_L^2(v_L^2+\ov v_L^2)+2 \mu_1^2.
\end{eqnarray}

Here we use the minimization conditions given in Eq.~(\ref{eq:min2b}) to eliminate $B_1$, $B_2$, $\mu_1$ and $\mu_2$. Since the light Higgs boson mass in this case is the same as in MSSM we use a stop squark mass of 1.5 TeV and maximal mixing between the stop squarks to get a Higgs boson mass of 125 GeV. The pseudo-scalar and charged Higgs boson masses are given in Table~\ref{tab:four}.

\vspace*{0.1in}
\noindent
{\large{\bf{Chargino and Neutralino masses}}}
\vspace*{0.1in}

The chargino mass terms in this case is written as
\begin{equation}
{\mathcal{L}}_{chargino} = -\frac{1}{2}\begin{pmatrix}\widetilde{H}_R^+&\widetilde{\ov H}_L^+&\widetilde{W}_R^+ & \widetilde{W}_L^+\end{pmatrix} \begin{pmatrix}\mu_2&0&g_R v_R&0 \\ 0&- \mu_1&0&g_L \ov v_L \\ g_R \ov v_R&0&M_R&0\\0&g_L v_L&0&M_L \end{pmatrix} \begin{pmatrix}\widetilde{\ov H}_R^-\\ \widetilde{H}_L^- \\ \widetilde{W}_R^- \\ \widetilde{W}_L^- \end{pmatrix},
\label{eq:2bch}
\end{equation}
and the neutralino mass matrix in the basis $\begin{pmatrix}\widetilde{H}_R^0&\widetilde{H}_L^0&\widetilde{\ov H}_R^0&\widetilde{\ov H}_L^0&\widetilde{B}&\widetilde{W}_{R_3}&\widetilde{W}_{L_3}\end{pmatrix}$ is given as
\begin{equation}
M_{n} = \begin{pmatrix} 0&0&-\mu_2&0&\frac{g_V v_R}{\sqrt{2}}&-\frac{g_R v_R}{\sqrt{2}}&0 \\ 0&0&0&\mu_1&-\frac{g_V v_L}{\sqrt{2}}&0&\frac{g_L v_L}{\sqrt{2}} \\ -\mu_2 &0&0&0&-\frac{g_V \ov v_R}{\sqrt{2}}&\frac{g_R \ov v_R}{\sqrt{2}}&0 \\ 0&\mu_1&0&0&\frac{g_V \ov v_L}{\sqrt{2}}&0&-\frac{g_L \ov v_L}{\sqrt{2}} \\ \frac{g_V v_R}{\sqrt{2}}&-\frac{g_V v_L}{\sqrt{2}}&-\frac{g_V \ov v_R}{\sqrt{2}}&\frac{g_V \ov v_L}{\sqrt{2}}&M_1&0&0 \\ -\frac{g_R v_R}{\sqrt{2}}&0&\frac{g_R \ov v_R}{\sqrt{2}}&0&0&M_R&0 \\ 0&\frac{g_L v_L}{\sqrt{2}}&0&-\frac{g_L \ov v_L}{\sqrt{2}}&0&0&M_L\end{pmatrix},
\end{equation}
where $M_R, M_L$ and $M_1$ are given in Eq.~{\ref{eq:gluino}. The numerical values of the masses for the chosen sample point are given in Table~{\ref{tab:four}.

\begin{center}
\begin{table}[h!]
\centering
\begin{tabular}{|C{2.7cm}|C{2.7cm}|C{2.8cm}|C{2.9cm}|C{2.9cm}|} \hline
{\bf{Scalar Higgs boson masses}} & {\bf{Pseudo-scalar Higgs boson masses}} & {\bf{Single charged Higgs boson masses}}& \bf{Chargino masses}&{\bf{Neutralino masses}}  \\ \hline
$M_{H_1}$=5.71 TeV, $M_{H_2}$=2.32 TeV, $M_{H_3}$=360 GeV & $M_{A_1}$=5.07 TeV, $M_{A_2}$=2.32 TeV & $M_{H_1^+}$=5.50 TeV, $M_{H_2^+}$=2.32 TeV & $M_{\wdt \chi_1^+}$=4.57 TeV, $M_{\wdt \chi_2^+}$=1.11 TeV, $M_{\wdt \chi_3^+}$=792 GeV, $M_{\wdt \chi_4^+}$=389 GeV & $M_{\wdt \chi_1^0}$=4.98 TeV, $M_{\wdt \chi_2^0}$=3.39 TeV, $M_{\wdt \chi_{3}^0}$=1.11 TeV, $M_{\wdt \chi_4^0}$=1.10 TeV, $M_{\wdt \chi_4^0}$=948 GeV, $M_{\wdt \chi_6^0}$=793 GeV, $M_{\wdt \chi_7^0}$=554 GeV \\ \hline
    \end{tabular}
    \caption{{Higgs boson, chargino and neutralino masses for Universal seesaw model using a sample point with parameters given as:
    $v_L$=10 GeV, $\ov v_L$=173.71 GeV, $v_R$=3 TeV, $\ov v_R$=3.5 TeV, $m_3^2$=4 $\text{TeV}^2$, $m_4^2$=4 $\text{TeV}^2$, $m_5^2$=-1 $\text{TeV}^2$, $m_6^2$=-1 $\text{TeV}^2$, $M_R$=800 GeV, $M_L$=800 GeV, $M_1$=400 GeV.}}
\label{tab:four}
\end{table}
\end{center}

\section{$E_6$ Inspired Left-right Supersymmetric model} \label{e6}

The Higgs spectrum for this model is discussed in Eq.~(\ref{eq:2ah}). The relevant terms in the superpotential involving the $H_L$, $H_R$ and $\Phi$ fields are given as:
\begin{equation}
W = \lambda{H_L}^T \tau_2 \Phi \tau_2 H_R + \mu {\text{Tr}}\left[\Phi \tau_2 \Phi^T \tau_2\right],
\end{equation}
where the parameter $\lambda$ and $\mu$ must be real for the superpotential to be invariant under parity transformation.

The Higgs potential consisting of the $V_F$, $V_D$ and $V_{Soft}$ terms will be given as:
\begin{eqnarray}
V_F&=&{\text{Tr}}({|\lambda H_R^T \tau_2 \Phi \tau_2|}^2+{|\lambda H_L^T \tau_2 \Phi \tau_2|}^2)+{\text{Tr}}(|\lambda H_L H_R^T+2 \mu \Phi|^2), \\
V_D&=&\frac{g_L^2}{8}\sum \limits_{a=1}^3 |H_L^\dagger \tau_a H_L+{\text{Tr}}(\Phi^\dagger \tau_a \Phi)|^2+\frac{g_R^2}{8}\sum \limits_{a=1}^3 |H_R^\dagger \tau_a H_R+{\text{Tr}}(\Phi^* \tau_a \Phi^T)|^2\nonumber \\
&+& \frac{g_V^2}{8}|H_R^\dagger H_R-H_L^\dagger H_L|^2, \\
V_{Soft}&=& m_1^2 {\text{Tr}}(\Phi^\dagger \Phi)+ \left[{B \mu \text{Tr}}(\Phi^T \tau_2 \Phi \tau_2)+h.c.\right]+m_3^2 H_L^\dagger H_L + m_4^2 H_R^\dagger H_R \nonumber \\
&+& (A_{\lambda} \lambda H_L^T\tau_2 \Phi \tau_2 H_R+h.c.).
\label{eq:one}
\end{eqnarray}

Using this potential we calculate the Higgs boson mass-squared matrix. We choose the following vacuum structure for the Higgs fields:
\begin{equation}
\left<H_L\right> = \begin{pmatrix}
v_L \\ 0  \end{pmatrix},
\left<H_R\right> = \begin{pmatrix}
0 \\v_R  \end{pmatrix},
\left<\Phi\right> ={\begin{pmatrix}
0 & v_2 \\ v_1 & 0 \end{pmatrix}}
\end{equation}
To easily identify the field corresponding to the lightest eigenvalue, we take a linear combination of the ${H_L}^0,\phi_1^0$ and $\phi_2^0$ fields. We make sure that only one of the newly defined fields get a non-zero vacuum expectation value(or VEV). The field redefinition that we used is:
\begin{eqnarray}
\rho_1&=&\frac{v_L {H_L}^0+v_1 \phi_1^0 +v_2 \phi_2^0}{\sqrt{v_L^2+v_1^2+v_2^2}},
\rho_2=\frac{v_L\phi_1^0-v_1 {H_L}^0}{\sqrt{v_1^2+v_L^2}},\notag \\
\rho_3&=&\frac{v_L v_2 {H_L}^0+v_1 v_2 \phi_1^0-(v_1^2+v_L^2)\phi_2^0}{\sqrt{(v_1^2+v_2^2)(v_1^2+v_2^2+v_L^2)}}.~~~~~~~~
\end{eqnarray}
With this choice, one can verify that only the $\rho_1$ field gets a non-zero vacuum expectation value of $\sqrt{v_1^2+v_2^2+v_L^2}$. We calculate the $4\times4$ mass-squared matrix for the neutral CP-even Higgs boson in the basis $({\text{Re}}\rho_1,{\text{Re}} {H_R}^0,{\text{Re}} \rho_2,{\text{Re}} \rho_3)$. It is easy to identify the lightest mass eigenvalue in this new basis. We use the minimization condition for the potential to express the soft SUSY breaking masses and the coefficient $\mu$ in terms of the other parameters in the model. The minimization conditions and mass-squared matrix is given in Appendix.
We assume that $v_R >> v_1,v_2,v_L$ and using this assumption we can get he lightest eigenvalue of the mass-squared matrix. It turns out that we can neglect the corrections from two of the off-diagonal matrix elements as they are of order of $\sim \frac{v_1^4}{v_R^2}$. So we effectively have a $2\times2$ matrix. Diagonalizing this matrix, we get the lightest neutral CP-even Higgs mass given by:
\begin{align}
M_{h_{tree}}^2 &= [g_R^2 (v_1^2 - v_2^2)^2 + g_V^2 v_L^4 + g_L^2 (-v_1^2 + v_2^2 + v_L^2)^2 +
 8 v_1^2 v_L^2 \lambda^2 \notag \\
 & - (g_V^2 v_L^2 + g_R^2 (-v_1^2 + v_2^2 + v_L^2) +   4 v_1^2 \lambda^2)^2/(g_R^2 + g_V^2)]/(2 (v_1^2 + v_2^2 + v_L^2)).
\end{align}
We then choose $v_1=v \sin\beta$, $v_2= v \cos\beta \cos\phi$ and $v_L = v \cos\beta \cos\phi$.  Maximizing the resulting expression with respect to $\lambda$ and $\phi$ and choosing $g_R=g_L$, we get:
 \begin{equation}
M_{h_{tree}}^2 =2 M_W^2 \cos^2 2\beta.
\end{equation}
This result is exactly the same as in {{Section~\ref{heavysinglet}}} and has been discussed in details in that section.

The pseudo-scalar Higgs boson mass-squared matrix is obtained by eliminating the Goldstone states and choosing the following basis:
\begin{equation}
\rho_1 = \frac{v_1 v_L H_R^0+v_1 v_R H_L^0+v_L v_R \phi_1^0}{\sqrt{v_1^2v_L^2+v_1^2v_R^2+v_L^2 v_R^2}},~~~~\rho_2 =\frac{v_2 \phi_1^0+v_1 \phi_2^0}{\sqrt{v_1^2+v_2^2}}.
\end{equation}
The $2\times2$ mass-squared matrix in the basis (Im$\rho_1$,Im$\rho_2$) can be written as
\begin{equation}
\begin{pmatrix}
\frac{\l(A_\l v_1-2\mu v_2)[v_L^2v_R^2+v_1^2(v_L^2+v_R^2)]}{v_1^2v_Lv_R}&\frac{2\l \mu \sqrt{(v_1^2+v_2^2)[v_L^2v_R^2+v_1^2(v_L^2+v_R^2)]}}{v_1^2}\\\frac{2\l \mu \sqrt{(v_1^2+v_2^2)[v_L^2v_R^2+v_1^2(v_L^2+v_R^2)]}}{v_1^2}&\frac{2(v_1^2+v_2^2)(\mu_2^2v_1-\l \mu v_L v_R)}{v_1^2 v_2}
\end{pmatrix}
\end{equation}

The charged Higgs boson mass-squared matrix is a $4\times4$ matrix of which there are two Goldstone states. In the original basis of (${H_L^-}^*,H_R^+,\phi_1^+,{\phi_2^-}^*$) the mass-squared matrix is given as:
\begin{eqnarray}
M_{11}&=&m_3^2 +\frac14 \left[ g_L^2(v_1^2-v_2^2+v_L^2)+g_V^2(v_L^2-v_R^2)\right]+\l^2(v_2^2+v_R^2),\notag \\
M_{12}&=&\l(-A_\l v_2+2 \mu v_1), \notag \\
M_{13}&=&\frac12 g_L^2 v_1 v_L+\l( A_\l v_R-\l v_1 v_L),\notag \\
M_{14}&=&\frac12 g_L^2 v_2 v_L-\l(\l v_2 v_L-2\mu v_R),\notag \\
M_{22}&=&m_3^2 +\frac14 \left[ g_R^2(v_1^2-v_2^2+v_R^2)+g_V^2(-v_L^2+v_R^2)\right]+\l^2(v_2^2+v_L^2),\notag \\
M_{23}&=&\frac12 g_R^2 v_2 v_R+\l(-\l v_2 v_R+2\mu v_L),\notag \\
M_{24}&=&\frac12 g_R^2 v_1 v_R-\l(\l v_1 v_R-A_\l v_L),\notag \\
M_{33}&=&m_1^2 +\frac14 \left[g_L^2(v_1^2+v_2^2+v_L^2)+g_R^2(v_1^2+v_2^2-v_R^2)\right]+\l^2 v_R^2 +4 \mu^2,\notag \\
M_{34}&=&\frac{(g_L^2+g_R^2)v_1 v_2+4 B \mu}{2}, \notag \\
M_{44}&=&m_1^2 +4 \mu^2+\frac14 \left[ g_L^2(v_1^2+v_2^2-v_L^2)+g_R^2(v_1^2+v_2^2 +v_R^2)\right]+\l^2 v_L^2.
\end{eqnarray}

We use the minimization conditions given in Eq.~(\ref{eq:e6min}) to eliminate $B, m_1, m_3$ and $m_4$. To get the correct CP-even lightest Higgs boson mass of 125 GeV, we choose a stop squark mass of 600 GeV and $X_t$ = 1. The numerical values of the masses of the Higgs boson physical states are given in Table~\ref{tab:five}.


\vspace*{0.1in}
\noindent
{\large{\bf{Chargino and Neutralino masses}}}
\vspace*{0.1in}

The higgsinos and the gauginos mix to form the charginos and the neutralinos. The chargino mass term in this case is written as
\begin{equation}
{\mathcal{L}}_{chargino} = -\frac{1}{2}\begin{pmatrix}\widetilde{H}_R^+&\widetilde{\phi}_1^+&\widetilde{W}_R^+ & \widetilde{W}_L^+\end{pmatrix} \begin{pmatrix}-\l v_2&\l v_L&g_R v_R&0 \\ \l v_R&2 \mu&g_R v_1&g_L v_1 \\ 0&g_R v_2&M_R&0\\g_L v_L&g_L v_2&0&M_L \end{pmatrix} \begin{pmatrix}\widetilde{H}_L^-\\ \widetilde{\phi}_2^- \\ \widetilde{W}_R^- \\ \widetilde{W}_L^- \end{pmatrix},
\end{equation}
and the neutralino mass matrix in the basis $\begin{pmatrix}\widetilde{H}_R^0&\widetilde{H}_L^0&\widetilde{\phi}_1^0&\widetilde{\phi}_2^0&\widetilde{B}&\widetilde{W}_{R_3}&\widetilde{W}_{L_3}\end{pmatrix}$ is given as
\begin{equation}
M_{n} = \begin{pmatrix} 0&-\l v_1&-\l v_L&0&\frac{g_V v_R}{\sqrt{2}}&-\frac{g_R v_R}{\sqrt{2}}&0 \\ -\l v_1&0&-\l v_R&0&-\frac{g_V v_L}{\sqrt{2}}&0&\frac{g_L v_L}{\sqrt{2}} \\ -\l v_L &-\l v_R&0&-2 \mu&0&-\frac{g_R v_1}{\sqrt{2}}&-\frac{g_L v_1}{\sqrt{2}} \\ 0&0&-2 \mu&0&0&\frac{g_R v_2}{\sqrt{2}}&\frac{g_L v_2}{\sqrt{2}} \\ \frac{g_V v_R}{\sqrt{2}}&-\frac{g_V v_L}{\sqrt{2}}&0&0&M_1&0&0 \\ -\frac{g_R v_R}{\sqrt{2}}&0&-\frac{g_R v_1}{\sqrt{2}}&\frac{g_R v_2}{\sqrt{2}}&0&M_R&0 \\ 0&\frac{g_L v_L}{\sqrt{2}}&-\frac{g_L v_1}{\sqrt{2}}&\frac{g_L v_2}{\sqrt{2}}&0&0&M_L\end{pmatrix}.
\end{equation}
where $M_R, M_L$ and $M_1$ are given in Eq.~{\ref{eq:gluino}. The numerical values of the masses for the chosen sample point are given in Table~{\ref{tab:five}.

\begin{center}
\begin{table}[h!]
\centering
\begin{tabular}{|C{2.7cm}|C{2.7cm}|C{2.8cm}|C{2.9cm}|C{2.9cm}|} \hline
{\bf{Scalar Higgs boson masses}} & {\bf{Pseudo-scalar Higgs boson masses}} & {\bf{Single charged Higgs boson masses}}& \bf{Chargino masses}&{\bf{Neutralino masses}}  \\ \hline
$M_{H_1}$=3.04 TeV, $M_{H_2}$=1.72 TeV, $M_{H_3}$=890 GeV & $M_{A_1}$=3.05 TeV, $M_{A_2}$=888 GeV & $M_{H_1^+}$=3.04 TeV, $M_{H_2^+}$=898 GeV & $M_{\wdt \chi_1^+}$=2.22 TeV, $M_{\wdt \chi_2^+}$=2.09 TeV, $M_{\wdt \chi_3^+}$=799 GeV, $M_{\wdt \chi_4^+}$=2.53 GeV & $M_{\wdt \chi_1^0}$=2.21 TeV, $M_{\wdt \chi_2^0}$=2.20 TeV, $M_{\wdt \chi_{3}^0}$=2.00 TeV, $M_{\wdt \chi_4^0}$=1.63 TeV, $M_{\wdt \chi_4^0}$=799 GeV, $M_{\wdt \chi_6^0}$=24.3 GeV, $M_{\wdt \chi_7^0}$=4.96 GeV \\ \hline
    \end{tabular}
    \caption{{Higgs boson, chargino and neutralino masses for Universal seesaw model using parameters given as:
    $\l$=0.3, $v_L$=20 GeV, $v_1$=172.5 GeV, $v_2$=11 GeV, $v_R$=3 TeV, $\m$=-1 TeV, $A_{\l}$=1 TeV, $M_R$=-800 GeV, $M_L$=-800 GeV, $M_1$=400 GeV.}}
\label{tab:five}
\end{table}
\end{center}

\section{Doubly-charged Higgs boson mass from loop corrections } \label{2chhiggs}

In the models discussed under section~\ref{case1}, the $SU(2)_R$ symmetry breaking is achieved by triplet Higgs bosons. Each triplet Higgs boson has a doubly-charged particle which should be relatively easy to detect experimentally if they can be produced at the colliders. These doubly-charged particles, if seen, can tell us a lot about the symmetry breaking pattern and their properties can help identify the underlying model.  It turns out that in the minimal models, the doubly charged
Higgs boson remains light with a mass below a TeV regardless of the scale of $SU(2)_R$ breaking.  This arises owing to an enhanced symmetry of the tree-level Higgs potential of the model.  In this section we present a complete calculation of
the one-loop induced mass of this scalar proportional to its Majorana Yukawa coupling.  This  completes the calculation
initiated in Ref. \cite{bm0}.   

We focus on a realistic left-right supersymmetric model where the $SU(2)_R \times U(1)_{B-L}$ symmetry is broken into $U(1)_Y$ by triplet Higgs boson field $\D^c$, and then the $SU(2)_L \times U(1)_Y$ symmetry breaking is achieved via bidoublet field $\Phi$. The chiral matter sector of this model is given in Eq.~(\ref{eq:matter}).
The Higgs boson sector is given in Eq.~(\ref{eq:triphig}). A singlet field $S$ is introduced so that the $SU(2)_R \times U(1)_{B_L}$ symmetry breaking can be achieved in the supersymmetric limit.

The superpotential of the model is given as:
\begin{eqnarray}
W&=& Y_{u} Q^{T} \tau_{2}\Phi_{1}\tau_{2}Q^{c} + Y_{d} Q^{T} \tau_{2}\Phi_{2}\tau_{2}Q^{c} +Y_{\nu} L^{T} \tau_{2}\Phi_{1}\tau_{2}L^{c} +Y_{l} L^{T} \tau_{2}\Phi_{2}\tau_{2}L^{c} \nonumber\\
&+&i(\frac{f}{2}^{*} L^{T}\tau_{2} \Delta L+\frac{f}{2} L^{c^{T}}\tau_{2}\Delta^{c}L^{c} ) \nonumber\\
&+&S[Tr(\lambda^{*} \Delta \overline{\Delta}+\lambda \Delta^{c}\overline{\Delta}^{c}) + \lambda_{ab}^{'} Tr(\Phi_{a}^{T}\tau_{2}\Phi_{b}\tau_{2}) -M_{R}^{2}] + W'
\label{LRSsup}
\end{eqnarray}
where
\begin{equation}
W'=\left[M_{\Delta} Tr(\Delta\overline{\Delta})+M_{\Delta}^* Tr(\Delta^c\overline{\Delta}^c)\right]+\mu_{ab} Tr\left(\Phi^T_a\tau_2 \Phi_b\tau_2\right)+M_{S}S^2+\lambda_{S}S^3.
\end{equation}

\noindent Here $Y_{u,d}$ and $Y_{\nu,l}$ are the Yukawa couplings for quarks and leptons respectively and $f$ is the Majorana neutrino Yukawa coupling matrix. This is the most general superpotential. $R$-parity is automatically preserved in this case. Putting $W'=0$ gives an enhanced $U(1)$ $R$-symmetry in the theory. Under this $R$-symmetry, $Q,Q^C,L,L^C$ fields have a charge of +1, $S$ has charge +2 and all other fields have charge zero with $W$ carrying a charge $+2$. Putting $W'=0$ also helps in understanding the $\mu$-problem. The doubly-charged left-handed and right-handed Higgsinos would be degenerate in mass in this case.

We will study the case where $W^{\prime}=0$. The left-handed triplets do not get any VEV and hence the masses of their doubly-charged particles are heavy. Thus we will concentrate on the right-handed Higgs boson triplet sector from here on. The Higgs potential consists of $F$ term, $D$ term and soft supersymmetry breaking terms which in this case are then given as
\begin{eqnarray}
V_F &=& \left|\lambda {\rm Tr}(\Delta^c \overline{\Delta}^c) +
\lambda'_{ab} {\rm Tr} \left( \Phi_a^T \tau_2 \Phi_b \tau_2
\right)-{\cal M}_R^2\right|^2 + |\lambda|^2 |S|^2 \left|{\rm
Tr}(\Delta^c \Delta^{c \dagger}) + {\rm
Tr}(\overline{\Delta}^c~\overline{\Delta}^{c \dagger}) \right|
\nonumber \\
V_{\rm soft} &=&  M_1^2 {\rm Tr} (\Delta^{c \dagger} \Delta^c) +
M_2^2 {\rm Tr} (\overline{\Delta}^{c \dagger} \overline{\Delta}^c) +M_S^2 |S|^2\nonumber \\
&+& \{A_\lambda \lambda S {\rm Tr}(\Delta^c \Delta^{c \dagger}) -
C_\lambda {\cal M}_R^2 S + h.c.\} \nonumber \\
 V_D &=& {g_R^2 \over
8}\sum_{a}\left|{\rm Tr}(2 \Delta^{c \dagger} \tau_a \Delta^c +
2 \overline{\Delta}^{c \dagger} \tau_a \overline{\Delta}^c + \Phi_a \tau_a^T \Phi_a^\dagger)\right|^2  \nonumber \\
&+& {g'^2 \over 8} \left|{\rm Tr}(2 \Delta^{c \dagger}  \Delta^c +
2 \overline{\Delta}^{c \dagger} \overline{\Delta}^c )\right|^2~.
\end{eqnarray}

If we consider a charged breaking vacuum structure for the $\D^c$ and $\ov \D^c$ fields given as
\begin{equation}
\left<\D^c\right> = \begin{pmatrix}
0 & v_R \\ v_R & 0  \end{pmatrix},~~~~
\left<\ov\D^c\right> = \begin{pmatrix}
0&\ov v_R\\\ov{v}_R & 0  \end{pmatrix},
\end{equation}
it can be shown that the Higgs potential is lower compared to the charge conserving vacuum given in Eq.~(\ref{eq:vev1a}) \cite{km}. The $F$ term and the soft SUSY breaking terms will be the same for both vacuua whereas the $D$ term of the potential will vanish for the charged breaking vacuum while being positive definite for the charge conserving one. This would lead to a charge breaking vacuum to be the stable one which is not desirable. This is also the root cause for the doubly charged scalar of the model
receiving negative squared mass, which is unacceptable.  The solution to these problems lies in the calculation of the loop correction to the Higgs potential which can make the mass of the doubly charged field positive and at the same time reverse
the roles of charge breaking and charge conserving vacuua.  

The tree-level doubly-charged Higgs mass-squared matrix in the basis $(\d^{c^{--^*}},\ov \d^{c^{++}})$ is given as
\begin{eqnarray}
M^2_{\delta^{++}} = \begin{pmatrix} -2 g_R^2(|v_R|^2-|\overline{v}_R|^2)-{\overline{v}_R^*
\over v_R} Y^* & Y \\ Y^* & 2 g_R^2(|v_R|^2 - |\overline{v}_R|^2) - {v_R \over \overline{v}_R^*}Y \end{pmatrix}
\end{eqnarray}
where $Y = \lambda A_\lambda S +\left| \lambda \right| ^2 \left( v_R \ov v_R - \frac{M^2_R}{\lambda} \right)$ and the electroweak vev has been neglected. It can be easily seen that if the gauge couplings are neglected, then this matrix will have a massless mode. Thus in this limit, the loop corrections to this massless mode should remain finite \cite{Weinberg:1973ua}. We proceed to compute the one-loop corrections to the would-be Goldstone boson mass arising from its Majorana Yukawa couplings.

We first identify the eigenstate corresponding to the Goldstone state. It is given as
\begin{equation}
G^{++} = \frac{v_R^* \d^{c^{--^*}}+\ov v_R \ov \d^{c^{++}}}{\sqrt{v_R^2+\ov v_R^2}}.
\end{equation}
The couplings that we would need to consider include the direct coupling of the doubly-charged particles to the electron and selectron fields, doubly-charged Higgs coupling to the neutral Higgs triplet and singlet Higgs bosons and the coupling of these neutral fields to the neutrino and sneutrino fields. These are the fields that appear in one-loop diagrams that
induce a finite mass for $G^{++}$. We also need to calculate the masses of each of these particles.
 
 We assume that the Majorana coupling $f$ of Eq. (\ref{LRSsup}) is significant for one generation of leptons.  This coupling involves
an almost massless electron, a heavy right-handed neutrino, two degenerate selectrons and two sneutrinos. If we denote
\begin{equation}
\widetilde \nu^c = \frac{n_1+i n_2}{\sqrt{2}},~~~~\widetilde \nu^{c^*} = \frac{n_1-i n_2}{\sqrt{2}},
\end{equation}
then the masses of all the particles are then given as
\begin{align}
& M_{e^c} \approx 0,~~~~ M^2_{\widetilde{e}_{1,2}^c}= m_{L^c}^2,~~~~ M_{\nu ^c} = f v_R,\notag \\
&M^2_{n_{1,2}} = m_{L^c}^2 + \left[ f^2 v_R^2 \pm (f \l \ov v_R v_S+f A_f v_R) \right]
\end{align}
where $m^2_{L^c}$ is the soft mass for the sleptons and $A_f$ is the trilinear coupling associated with the Majorana 
Yukawa coupling $f$.

The neutral Higgs sector relevant for our calculation would include the $\d^{c^0}, \ov \d^{c^0}$ and $S$ fields.
Let us write them as
\begin{equation}
\d^{c^0} = \frac{X_1+i Y_1}{\sqrt{2}},~~~~ \ov \d^{c^0} = \frac{X_2+i Y_2}{\sqrt{2}},~~~~ S = \frac{X_3+i Y_3}{\sqrt{2}}.
\end{equation}
If we choose all the couplings and the VEVs to be real, then we will get two $3\times 3$ mass-squared matrices for these fields-- one for the real part and another for the imaginary part. We only need to consider the real fields as the imaginary fields will have no relevant cubic couplings to the Goldstone field $G^{++}$. The relevant interaction terms in the Lagrangian which would be necessary for our calculation are given as
\begin{eqnarray}
-\mathcal{L}_{int} &=& G^{++} G^{--} \left[ (|\tilde{e}^c_1|^2+|\tilde{e}^c_2|^2) \frac{f^2 v_R^2}{v_R^2+\ov v_R^2} + \sqrt{2} \frac{\l ^2 v_R \ov v_R^2}{v_R^2+\ov v_R^2} X_1+\sqrt{2} \frac{\l ^2 v_R^2 \ov v_R}{v_R^2+\ov v_R^2} X_2 \right. \notag \\
&+& \left.\sqrt{2}\left(\l ^2 v_S + \frac{\l A_\l v_R \ov v_R}{v_R^2+\ov v_R^2} \right) X_3 \right]-\left[\frac{f A_f v_R+f \l \ov v_R v_S}{2 \sqrt{v_R^2+\ov v_R^2}}(\tilde{e}^c_1 \tilde{e}^c_1+\tilde{e}^c_2\tilde{e}^c_2)G^{--}\right]\notag \\
&+&\left[ \frac{f A_f}{2 \sqrt{2}}({n_1^2}-{ n_2^2})+\frac{f^2 v_R}{\sqrt{2}}({ n_1^2}+{ n_2^2})\right]X_1 \notag \\&+&\frac{f \l v_S}{2 \sqrt{2}}(n_1^2-n_2^2)X_2 + \frac{f \l \ov v_R}{2 \sqrt{2}}(n_1^2-n_2^2)X_3.
\end{eqnarray}

The mass-squared matrix for the neutral scalar Higgs bosons is given as
\begin{equation}
M_H^2 = \begin{pmatrix}
M_1^2+\l^2(v_S^2+\ov v_R^2)&\l^2 v_R \ov v_R+\l A_\l  v_S - \l^2 M_R^2&2 \l^2 v_S v_R+\l A_\l\ov v_R \\
\l^2 v_R \ov v_R+\l A_\l  v_S - \l^2 M_R^2&M_2^2 +\l^2 (v_S^2+v_R^2)&2 \l^2 v_S \ov v_R+\l A_\l v_R \\
2 \l^2 v_S v_R+\l A_\l\ov v_R & 2 \l^2 v_S \ov v_R+\l A_\l v_R &M_S^2+\l^2(v_R^2+\ov v_R^2)\\
\end{pmatrix}.
\end{equation}
Usually one would need to diagonalize this mass-squared matrix and identify the mass eigenstates. Fortunately that is not the case here. Let us choose a basis given as
\begin{equation}
\hat{X} = V^T X
\end{equation}
where $X = \begin{pmatrix}X_1 & X_2 & X_3 \end{pmatrix}^T$, $V$ is an orthogonal transformation matrix and $\hat{X}$ represents the mass eigenbasis. Then the diagonal mass-squared matrix is given as
\begin{equation}
D^2 = V^T M_H^2 V.
\end{equation}
All the couplings of the neutral Higgs bosons can now be written as
\begin{equation}
-\mathcal{L}_{\hat{X}} = P_i V_{ij}\hat{X}_j G^{++}G^{--} + Q_i V_{ij}\hat{X}_j n_1^2 + R_i V_{ij}\hat{X}_j n_2^2 + T_i V_{ij}\hat{X}_j \nu^c \nu^c
\end{equation}
where $P,Q,R$ and $T$ are vectors given as
\begin{eqnarray}
P&=&\begin{bmatrix} \sqrt{2} \frac{\l ^2 v_R \ov v_R^2}{v_R^2+\ov v_R^2},&\sqrt{2} \frac{\l ^2 v_R^2 \ov v_R}{v_R^2+\ov v_R^2},&\sqrt{2}\left(\l ^2 v_S + \frac{\l A_\l v_R \ov v_R}{v_R^2+\ov v_R^2}\right)\end{bmatrix},\notag \\
Q&=&\begin{bmatrix} \frac{f A_f}{2 \sqrt{2}}+\frac{f^2 v_R}{\sqrt{2}},&\frac{f \l v_S}{2 \sqrt{2}},&\frac{f \l \ov v_R}{2 \sqrt{2}}\end{bmatrix},\notag \\
R&=&\begin{bmatrix} \frac{-f A_f}{2 \sqrt{2}}+\frac{f^2 v_R}{\sqrt{2}},&-\frac{f \l v_S}{2 \sqrt{2}},&-\frac{f \l \ov v_R}{2 \sqrt{2}}\end{bmatrix},\notag \\
T&=&\begin{bmatrix}\frac{f}{\sqrt{2}},&0,&0\end{bmatrix}.
\end{eqnarray}

\begin{figure}[h!]
\begin{center}
\includegraphics[width=1.6in]{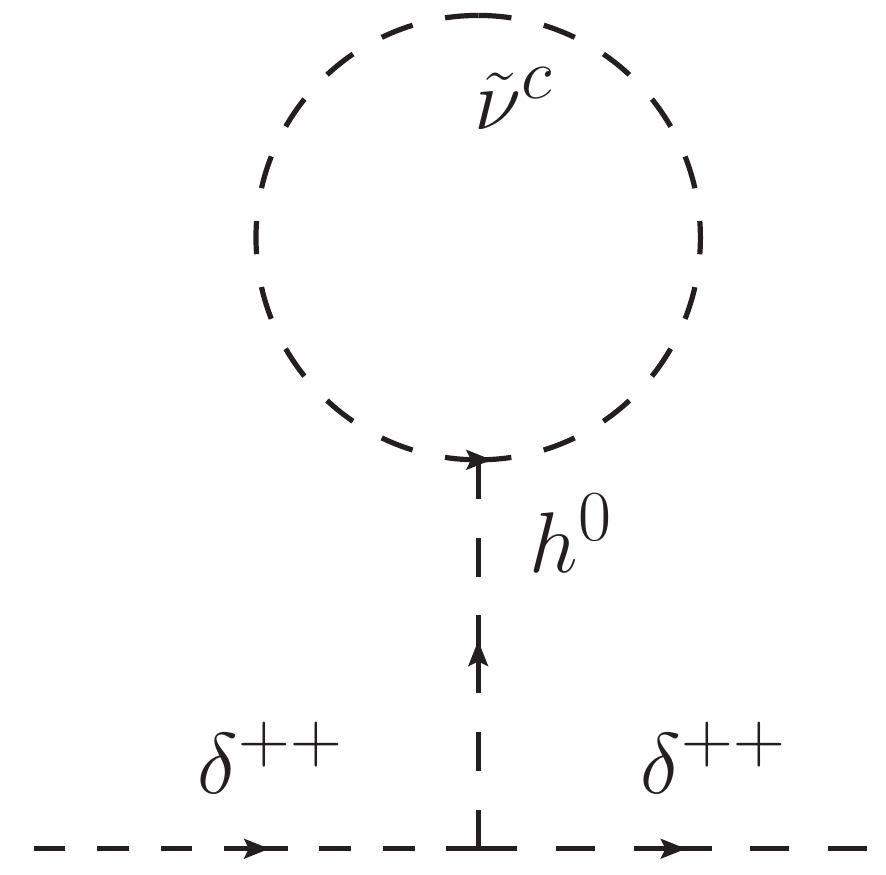}
\includegraphics[width=1.6in]{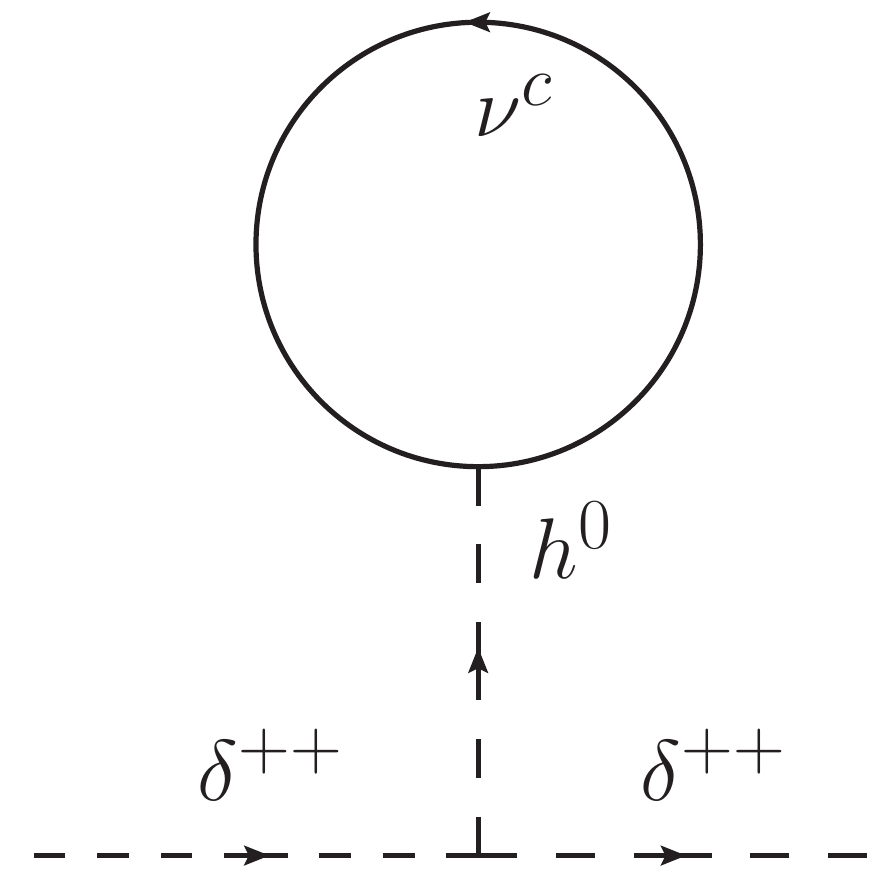}
\end{center}
\caption{One-loop Feynman diagrams inducing finite mass for doubly charged Higgs boson with sneutrino and neutrino exchange.}
\label{fig:dc2}
\end{figure}

We can now calculate the one-loop corrections to the doubly-charged Higgs boson mass. The corrections coming from the right-handed neutrino and sneutrino sector are given by the Feynman diagrams in Fig.~{\ref{fig:dc2}}.
The corresponding amplitudes are given as
\begin{eqnarray}
M_1 &=& -\frac{i}{2}\left[ P^T M_h^{-2}Q \int \frac{d^4k}{(2\pi)^4} \frac{1}{k^2-m_{n_1}^2} + P^T M_h^{-2}R \int \frac{d^4k}{(2\pi)^4} \frac{1}{k^2-m_{n_2}^2} \right], \notag \\
M_2 &=&2 i M_{\nu^c} P^T M_h^{-2} T \int \frac{d^4k}{(2\pi)^4} \text{Tr}\left( \frac{\slashed{k}+M_{\nu^c}}{k^2 -M_{\nu^c}^2}  \right).
\end{eqnarray}
\begin{figure}[h!]
\includegraphics[width=2.1in]{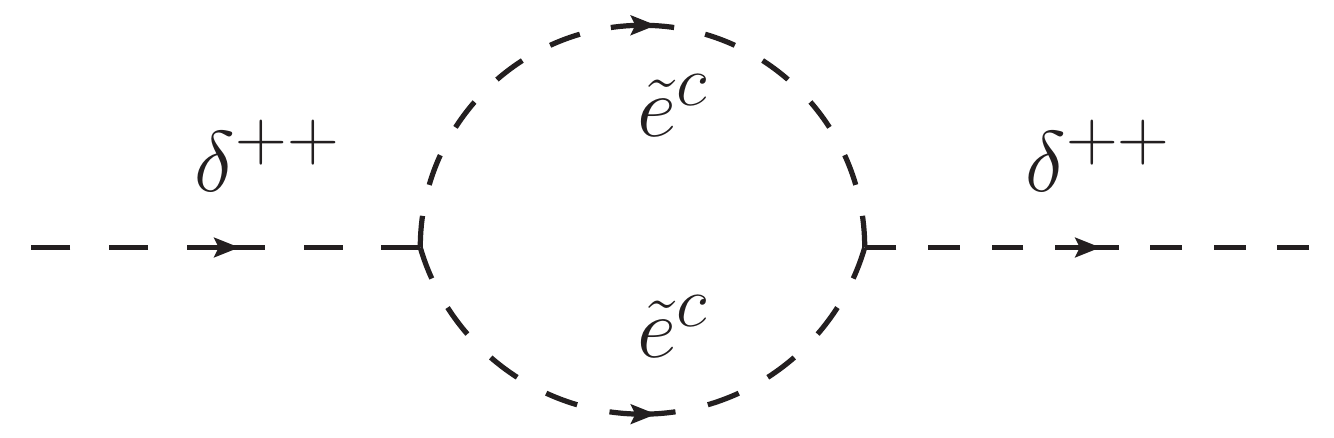}
\includegraphics[width=2.1in]{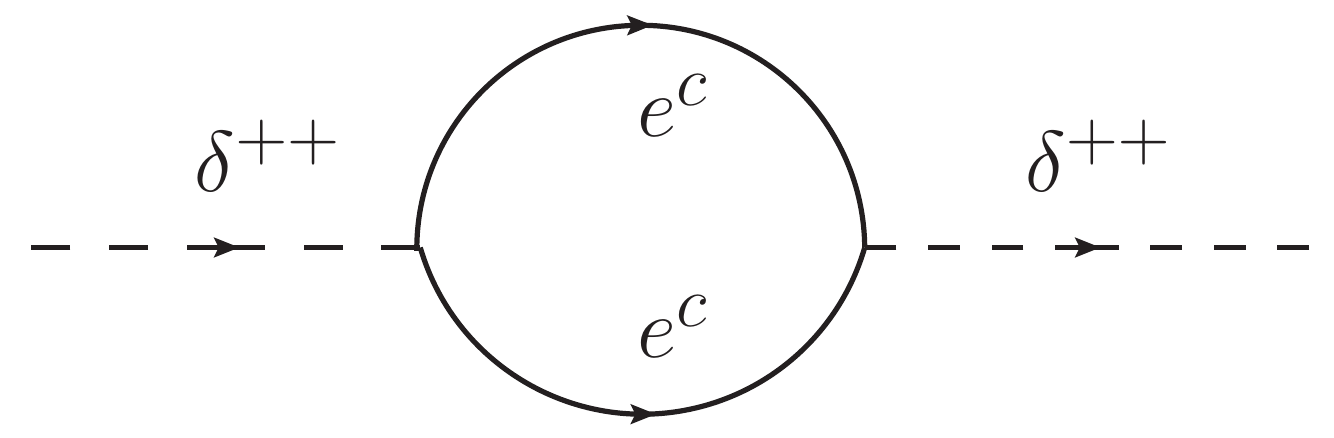}
\includegraphics[width=1.3in]{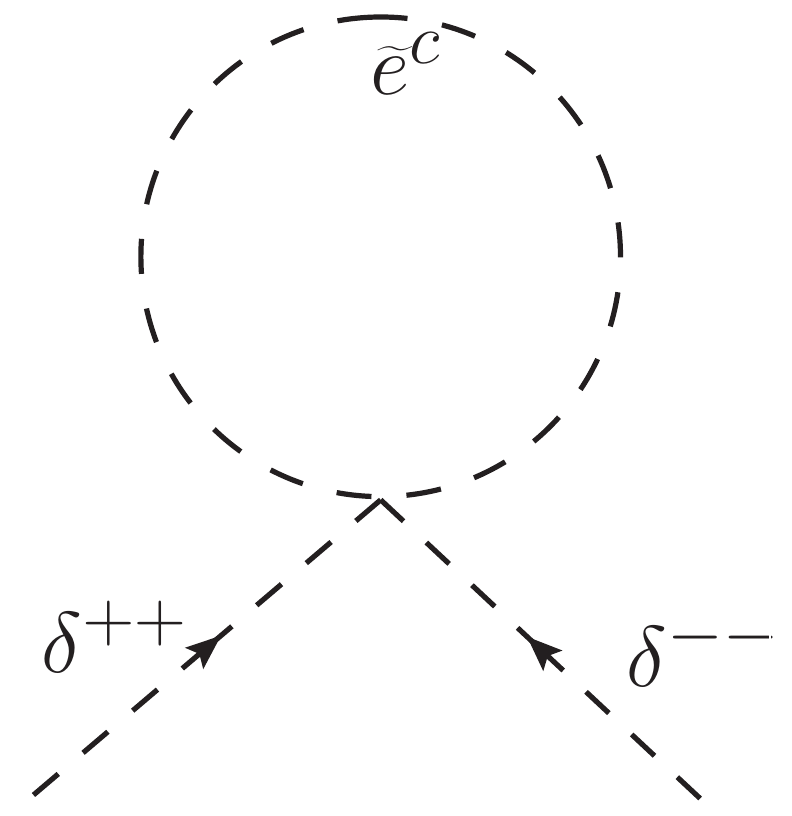}
\caption{Feynman diagrams for electron and selectron one-loop correction}
\label{fig:dc2a}
\end{figure}
The Feynman diagrams for the electron and selectron corrections are given in Fig.~\ref{fig:dc2a} and the corresponding amplitudes are given as
\begin{eqnarray}
M_3 &=& -\frac{i}{2}\frac{(f A_f v_R+f \l \ov v_R v_S)^2}{v_R^2+\ov v_R^2} \int \frac{d^4k}{(2\pi)^4} \frac{1}{k^2-m_{\tilde e^c}^2}, \notag \\
M_4 &=&-\frac{i f^2 v_R^2}{v_R^2+\ov v_R^2} \int \frac{d^4k}{(2\pi)^4} \frac{1}{k^2},\notag \\
M_5 &=& \frac{i f^2 v_R^2}{v_R^2+\ov v_R^2} \int \frac{d^4k}{(2\pi)^4} \frac{1}{k^2-m_{\tilde e^c}^2}
\end{eqnarray}

Summing over all the correction to the doubly-charged Higgs boson mass coming from these diagrams we get
\begin{eqnarray}
\Delta M_{G^{++}}^2&=& \frac{1}{16 \pi^2 \left( v_R^2 + \ov v_R^2 \right)} \left[ f^2 v_R^2 m^2_{\tilde{e}^c} \text{ln} \left( \frac{m^2_{\tilde{e}^c}}{M^2_{\nu^c}}\right) + \frac{f^2}{2} \left( \lambda \ov v_R v_S+A_f v_R \right)^2 \text{ln} \left(\frac{m^2_{\tilde{e}^c}}{M^2_{\nu^c}} +1 \right) \right. \notag \\
&-&\frac{f}{4} \left( A_f v_R+2 f v_R^2+\lambda \ov v_R v_S \right) m^2_{n_1} \text{ln} \left( \frac{m^2_{n_1}}{M^2_{\nu^c}} \right) \notag \\
&-&\left. \frac{f}{4} \left(- A_f v_R+2 f v_R^2-\lambda \ov v_R v_S \right) m^2_{n_2} \text{ln} \left( \frac{m^2_{n_2}}{M^2_{\nu^c}} \right) \right].
\label{eq:corr2}
\end{eqnarray}
A nontrivial check of the calculation is finiteness of the sum, although individual diagrams diverge.  

It is interesting to see what happens to the mass as $v_R, \ov v_R \gg M_{SUSY}$ is taken.  In this limit Eq. (\ref{eq:corr2})
reduces to
\begin{equation}
\Delta M_{G^{++}}^2 \simeq \frac{f^2}{16\pi^2}\frac{v_R^2}{v_R^2+\ov v_R^2}\left[m^2_{\tilde{e}^c}+\frac{1}{2}
\frac{(\lambda \ov v_R v_S + A_f v_R)^2} {v_R^2 + \ov v_R^2}  \right]\left( {\rm ln} \frac{m^2_{\tilde{e}^c}}{M_{\nu^c}^2}-1\right)~.
\end{equation}
Since $m^2_{\tilde{e}^c}$ is of order SUSY breaking scale and $M_{\nu^c}^2$ is of order $v_R^2$, in this limit we
see that the loop correction to the doubly charged mass is negative!  This suggests that the $SU(2)_R$ breaking 
scale cannot be much above the SUSY breaking scale for consistency.  When the two scales are comparable, the loop
correction can make the doubly charged Higgs boson squared mass to be positive for various choice of parameters.
One would expect the mass to be below a TeV, owing to the suppression factor $f^2/(16 \pi^2)$.

\vskip0.2in

\section{Discussion and Conclusion}
\label{sec:conclusion}

In this paper we have carried out a systematic investigation of the Higgs boson spectra in a variety of
supersymmetric left-right models.  We have focussed on the lightest CP even Higgs boson mass and found its
theoretical upper limit at tree level deviates significantly from $M_h < M_Z$ of MSSM.  Several variations
relax this limit to $M_h < \sqrt{2} m_W$, while other variations make it even weaker.  Our analysis focussed on
two basic classes of models, one which uses Higgs triplets to break $SU(2)_R$ gauge symmetry, and the other
which uses doublet for this purpose.  In the latter case additional fermion fields are needed in order to
generate realistic fermion masses.  We studied models with inverse seesaw for neutrino masses, universal seesaw
model for fermion masses, and an $E_6$ inspired left-right model.  The Higgs sectors of these models were analyzed
with or without a gauge singlet Higgs field present. The relaxed limit on $M_h$ suggests that large supersymmetric
contribution from the top--stop sector is not required, and fine tuning may be minimized compared to MSSM.

In the model with $SU(2)_R$ triplet Higgs fields, a doubly charged scalar remains light below a TeV, regardless
of the scale of $SU(2)_R$ breaking. We have computed one-loop corrections to its mass arising from Majorana Yukawa
couplings.  For these corrections to be positive, the $SU(2)_R$ breaking scale should be not much above the
SUSY breaking scale.

SUSYLR models with Higgs triplet fields can only be extrapolated to energy scales of order $10^{12}$ GeV, at which
point some new dynamics should appear.  On the other hand, models with Higgs doublets and bidoublets can be extrapolated all the way to the GUT scale.   


\section*{Acknowledgements}

This research is supported in part by the National Science Foundation under Grant No. NSF PHY11-25915 (KSB)
and in part by the US Department of Energy Grant No. de-sc0010108.
AP was supported in part by the the NRF grant funded by Korea government of the MEST (No.2014R1A1A2057665).

\newpage

\begin{appendix}
\begin{center}
{\bf APPENDIX}
\end{center}
Here we summarize various expressions that were relevant for the Higgs boson mass spectra
in variations of the SUSYLR models.

{\bf{Minimization conditions and scalar Higgs boson mass-squared matrix}}

{\underline{$E_6$ Inspired LRSUSY model}}

The minimization conditions for the potential are given as:
\begin{eqnarray}
 0&=& 2 m_1^2 v_1 +
 \frac{1}{8}(-4 g_L^2 v_1 (-v_1^2 + v_2^2 + v_L^2) -
    4 g_R^2 v_1 (-v_1^2 + v_2^2 + v_R^2)) - 2 \l A_\l v_L v_R +
 2 v_1 (v_L^2 + v_R^2) \l^2 \notag \\
 &-& 4 B v_2 \mu + 8 v_1 \mu^2, \notag \\
 0&=& 2 m_3^2 v_L +
 \frac12 v_L \left[ g_L^2 (-v_1^2 + v_2^2 + v_L^2) + g_V^2 (v_L^2 - v_R^2)\right] +
 2 \l (\l v_1^2 v_L + \l v_L v_R^2  + 2\mu v_2 v_R - A_\l v_1 v_R), \notag \\
 0&=&2 m_4^2 v_R +
 \frac12 v_R \left[g_R^2 (-v_1^2 + v_2^2 + v_R^2) + g_V^2 (-v_L^2 + v_R^2)\right] +
 2 \l( \l v_1^2 v_R + v_L^2 v_R \l + 2 \mu v_2 v_L-A_\l v_1 v_L), \notag \\
 0&=&2 m_1^2 v_2 +
 \frac12 v_2 (g_L^2 (-v_1^2 + v_2^2 + v_L^2) + g_R^2 (-v_1^2 + v_2^2 + v_R^2))+ 4 \mu (\l v_L v_R  + 2 \mu v_2 -B v_1)
  \label{eq:e6min}
\end{eqnarray}

The mass-squared matrix elements $M_{ij}(=M_{ji})$ in this case in the basis $({\text{Re}}\rho_1,{\text{Re}} {H_R}^0,{\text{Re}} \rho_2,{\text{Re}} \rho_3)$  is given by:
\begin{eqnarray}
M_{11}&=&\frac{g_R^2(v_1^2-v_2^2)^2+g_V^2v_L^4+g_L^2(v_1^2-v_2^2-v_L^2)^2+8\lambda^2 v_1^2v_L^2}{2(v_1^2+v_2^2+v_L^2)},\nonumber \\
M_{12}&=&\frac{g_R^2v_R(v_2^2-v_1^2)-g_V^2v_Rv_L^2+4\lambda\left\{-A_{\lambda}v_1v_L+\lambda v_R(v_1^2+v_L^2)+2\mu v_2 v_L\right\}}{2\sqrt{v_1^2+v_2^2+v_L^2}},\nonumber \\
M_{13}&=&\frac{v_1v_L\left[g_R^2(v_2^2-v_1^2)+g_V^2v_L^2+2g_L^2(v_2^2-v_1^2+v_L^2)+4\lambda^2(v_1^2-v_L^2)\right]}{2\sqrt{(v_1^2+v_2^2+v_L^2)(v_1^2+v_L^2)}},\nonumber \\
M_{14}&=&\frac{v_2\left[g_V^2v_L^4+2g_L^2v_1^2(v_1^2-v_2^2-v_L^2)+g_R^2(v_1^2-v_2^2)(2v_1^2+v_L^2)+8\lambda^2v_1^2v_L^2\right]}{2(v_1^2+v_2^2+v_L^2)\sqrt{(v_1^2+v_L^2)}},\nonumber \\
M_{22}&=&\frac{g_R^2v_R^3+g_V^2v_R^3+2\lambda A_{\lambda}v_1v_L-4\lambda \mu v_2 v_L}{2v_R},\nonumber \\
M_{23}&=&\frac{(g_R^2-g_V^2)v_1v_Lv_R+2\lambda\left[A_\lambda(v_L^2-v_1^2)+2\mu v_1v_2\right]}{2 \sqrt{v_1^2+v_L^2}},\nonumber \\
M_{24}&=&\frac{v_2v_R\left[-2g_R^2v_1^2-(g_R^2+g_V^2)v_L^2+4\lambda^2(v_1^2+v_L^2)\right]-4\lambda v_L\left[A_\lambda v_1v_2+\mu(v_1^2-v_2^2+v_L^2)\right]}{2\sqrt{(v_1^2+v_2^2+v_L^2)(v_1^2+v_L^2)}},\nonumber \\
M_{33}&=&\frac{(4g_L^2+g_R^2+g_V^2)v_1^3v_L^3+4B \mu v_2v_L^3+2\lambda A_\lambda v_R(v_1^2+v_L^2)^2-8\lambda^2v_1^3v_L^3-4\lambda \mu v_1^3v_2v_R}{2v_1v_L(v_1^2+v_L^2)},\nonumber \\
M_{34}&=&[-4m_2^2v_L(v_1^2+v_2^2+v_L^2)+v_1v_2v_L (4g_L^2v_1^2-g_V^2v_L^2+g_R^2(2v_1^2+v_L^2))+4\lambda^2 v_2v_L(v_L^2-v_1^2),\nonumber \\
&+& 4\lambda \mu v_R(v_1^2+v_2^2+v_L^2)]/ \left[2(v_1^2+v_L^2)\sqrt{v_1^2+v_2^2+v_L^2)}\right] \nonumber \\
M_{44}&=& \left[-g_R^2 \left( v_1^6 + 10 v_1^4 v_2^2 - v_1^2 v_2^4 - 2 v_1^4 v_L^2 + 10 v_1^2 v_2^2 v_L^2 - v_1^2 v_L^4 +   3 v_2^2 v_L^4+ v_1^4 v_R^2 - v_1^2 v_2^2 v_R^2 +
  2 v_1^2 v_L^2 v_R^2+ v_L^4 v_R^2 \right)\right. \notag \\
&+& g_L^2 \left\{ -v_1^6 + v_1^4 (10 v_2^2 - v_L^2) + v_L^2 (v_2^2 + v_L^2)^2 +
     v_1^2 (-v_2^4 + v_L^4)\right\} +g_V^2 (3 v_2^2 v_L^4 - v_2^2 v_L^2 v_R^2)+4 M_3^2 v_2^2 v_L^2 \notag \\
  &+&  4 M_1^2 \left\{ v_1^4 + v_L^4 + v_1^2 (v_2^2 + 2 v_L^2)\right\} - 8 \l A_\l v_1 v_2^2 v_L v_R+ 4 \l^2( 6 v_1^2 v_2^2 v_L^2  + v_1^2 v_2^2 v_R^2 + v_2^2 v_L^2 v_R^2)
  \notag \\
     &+& \left. 16 \mu  v_2 (v_1^2 + v_L^2) (B v_1 - \l v_L v_R )   +
  16 \mu^2 \left\{ v_1^4 + v_L^4 + v_1^2 (v_2^2 + 2 v_L^2)\right\} \right]
 \Big{/}\left[8 (v_1^2 + v_L^2) (v_1^2 + v_2^2 + v_L^2)\right].
\end{eqnarray}

{\underline{Case with a pair of triplets and two bidoublets}}

The minimization conditions for this case are given as:
\begin{eqnarray}
0&=&4 m_{12}^2 v_{{d_1}} + 4 {B_{12}\mu_{12}} v_{{d_2}} + 4 m_{11}^2 v_{{u_1}} + 8 {B_{11} \mu_{11}} v_{{u_2}} +
  g_{L}^2 v_{{u_1}} (v_{{d_1}}^2 - v_{{d_2}}^2 + v_{{u_1}}^2 - v_{{u_2}}^2) \notag \\
  &+& g_{R}^2 v_{{u_1}} (v_{{d_1}}^2 - v_{{d_2}}^2 + 2 {v_R}^2 - 2 \ov{v}_R^2 + v_{{u_1}}^2 - v_{{u_2}}^2) +
  4 \left[ v_{{u_1}} (\mu_{11}^2 + \mu_{12}^2) + v_{{d_1}} \mu_{12} (\mu_{11} + \mu_{22}) \right],\notag \\
0&=&4 {B_{12}\mu_{12}} v_{{d_1}} + 4 m_{12}^2 v_{{d_2}} + 8 {B_{11}\mu_{11}} v_{{u_1}} + 4 m_{11}^2 v_{{u_2}} +
  g_{L}^2 v_{{u_2}} (-v_{{d_1}}^2 + v_{{d_2}}^2 - v_{{u_1}}^2 + v_{{u_2}}^2) \notag \\
  &+&
  g_{R}^2 v_{{u_2}} (-v_{{d_1}}^2 + v_{{d_2}}^2 - 2 {v_R}^2 + 2 \ov{v}_R^2 - v_{{u_1}}^2 + v_{{u_2}}^2) +
  4 \left[ v_{{u_2}} (\mu_{11}^2 + \mu_{12}^2) + v_{{d_2}} \mu_{12} (\mu_{11} + \mu_{22})\right],\notag \\
0&=&  4 m_{22}^2 v_{{d_1}} + 8 {B_{22} \mu_{22}} v_{{d_2}} + 4 m_{12}^2 v_{{u_1}} + 4 {B_{12} \mu_{12}} v_{{u_2}} +
  g_{L}^2 v_{{d_1}} (v_{{d_1}}^2 - v_{{d_2}}^2 + v_{{u_1}}^2 - v_{{u_2}}^2)\notag \\
  & +&
  g_{R}^2 v_{{d_1}} (v_{{d_1}}^2 - v_{{d_2}}^2 + 2 {v_R}^2 - 2 \ov{v}_R^2 + v_{{u_1}}^2 - v_{{u_2}}^2 +
  4 \left[v_{{u_1}} \mu_{12} (\mu_{11} + \mu_{22}) + v_{{d_1}} (\mu_{12}^2 + \mu_{22}^2)\right],\notag \\
0&=&  8 {B_{22}\mu_{22}} v_{{d_1}} + 4 m_{22}^2 v_{{d_2}} + 4 {B_{12} \mu_{12}} v_{{u_1}} + 4 m_{12}^2 v_{{u_2}} +
  g_{L}^2 v_{{d_2}} (-v_{{d_1}}^2 + v_{{d_2}}^2 - v_{{u_1}}^2 + v_{{u_2}}^2)\notag \\
  & + &
  g_{R}^2 v_{{d_2}} (-v_{{d_1}}^2 + v_{{d_2}}^2 - 2 {v_R}^2 + 2 \ov{v}_R^2 - v_{{u_1}}^2 + v_{{u_2}}^2) +
  4 \left[ v_{{u_2}} \mu_{12} (\mu_{11} + \mu_{22}) + v_{{d_2}} (\mu_{12}^2 + \mu_{22}^2))\right],\notag \\
0&=&  2 B_2 \mu_2 \ov{v}_R +
 {v_R} (2 m_5^2 + 2 \mu_2^2 + 2 g_{V}^2 ({v_R}^2 - {\ov{v}_R}^2) +
    g_{R}^2 (v_{{d_1}}^2 - v_{{d_2}}^2 + 2 {v_R}^2 - 2 \ov{v}_R^2 + v_{{u_1}}^2 - v_{{u_2}}^2),\notag \\
0&=&    2 B_2 \mu_2 {v_R} +
 \ov{v}_R (2 m_6^2 + 2 \mu_2^2 + 2 g_{V}^2 (\ov{v}_R^2-{v_R}^2) +
 g_R^2 (v_{{d_2}}^2 - v_{{d_1}}^2 - 2 {v_R}^2 + 2 \ov{v}_R^2 - v_{{u_1}}^2 + v_{{u_2}}^2).~~~~~~~~~~
\end{eqnarray}

{\underline{Universal seesaw model with a singlet}}

The mass-squared matrix elements are given by:
\begin{eqnarray}
M_{11} &=& \frac{g_L^2 \left( {v_L}^2- {\overline{v}_L}^2\right)^2+ { {g_V} }^2 \left( {v_L}^2- {\overline{v}_L}^2\right)^2+8 {v_L}^2  {\overline{v}_L}^2 \lambda ^2}{2 \left( {v_L}^2+ {\overline{v}_L}^2\right)},  \nonumber \\
 M_{12} &=& \frac{ {v_L}  {\overline{v}_L} \left( {v_L}^2- {\overline{v}_L}^2\right) \left( { g_L}^2+ { {g_V} }^2-2 \lambda ^2\right)}{ {v_L}^2+ {\overline{v}_L}^2},  \nonumber \\
 {M_{13}} &=& \frac{- { {g_V} }^2 \left( {v_L}^2- {\overline{v}_L}^2\right) \left( {v_R}^2- {\overline{v}_R}^2\right)-8  {v_L}  {\overline{v}_L}  {v_R}
 {\overline{v}_R} \lambda ^2}{2 \sqrt{ {v_L}^2+ {\overline{v}_L}^2} \sqrt{ {v_R}^2+ {\overline{v}_R}^2}} , \nonumber \\
 {M_{14}} &=& \frac{ { {g_V} }^2 \left(- {v_L}^2+ {\overline{v}_L}^2\right)  {v_R}  {\overline{v}_R}+2  {v_L}  {\overline{v}_L} \left( {v_R}^2- {\overline{v}_R}^2\right)
\lambda ^2}{\sqrt{ {v_L}^2+ {\overline{v}_L}^2} \sqrt{ {v_R}^2+ {\overline{v}_R}^2}},  \nonumber \\
 {M_{15}} &=& \frac{2 \lambda  \left( { A_{\lambda}  }  {v_L}  {\overline{v}_L}+\left( {v_L}^2+ {\overline{v}_L}^2\right)  {v_S} \lambda \right)}{\sqrt{ {v_L}^2+ {\overline{v}_L}^2}},  \nonumber \\
 {M_{22}} &=& 2  [{ g_L}^2  {v_L}^2  {\overline{v}_L}^2+2  { {g_V} }^2  {v_L}^2  {\overline{v}_L}^2+ {m_3}^2 \left( {v_L}^2+ {\overline{v}_L}^2\right)+ {m_5}^2
\left( {v_L}^2+ {\overline{v}_L}^2\right)\nonumber \\
&+& {v_L}^4 \lambda ^2-2  {v_L}^2  {\overline{v}_L}^2 \lambda ^2+ {\overline{v}_L}^4 \lambda ^2+2  {v_L}^2  {v_S}^2
\lambda ^2+2  {\overline{v}_L}^2  {v_S}^2 \lambda ^2]/\left( {v_L}^2+ {\overline{v}_L}^2\right),  \nonumber \\
 {M_{23}} &=& \frac{ { {g_V} }^2  {v_L}  {\overline{v}_L} \left(- {v_R}^2+ {\overline{v}_R}^2\right)+2 \left( {v_L}^2- {\overline{v}_L}^2\right)  {v_R}
 {\overline{v}_R} \lambda ^2}{\sqrt{ {v_L}^2+ {\overline{v}_L}^2} \sqrt{ {v_R}^2+ {\overline{v}_R}^2}} , \nonumber \\
 {M_{24}} &=& \frac{-2  { {g_V} }^2  {v_L}  {\overline{v}_L}  {v_R}  {\overline{v}_R}-\left( {v_L}^2- {\overline{v}_L}^2\right) \left( {v_R}^2- {\overline{v}_R}^2\right)
\lambda ^2}{\sqrt{ {v_L}^2+ {\overline{v}_L}^2} \sqrt{ {v_R}^2+ {\overline{v}_R}^2}},  \nonumber \\
 {M_{25}} &=& -\frac{ { A_{\lambda}  } \left( {v_L}^2- {\overline{v}_L}^2\right) \lambda }{\sqrt{ {v_L}^2+ {\overline{v}_L}^2}} , \nonumber \\
 {M_{33}} &=& \frac{ {g_R}^2 \left( {v_R}^2- {\overline{v}_R}^2\right)^2+ { {g_V} }^2 \left( {v_R}^2- {\overline{v}_R}^2\right)^2+8  {v_R}^2
 {\overline{v}_R}^2 \lambda ^2}{2 \left( {v_R}^2+ {\overline{v}_R}^2\right)},  \nonumber \\
 {M_{34}} &=& \frac{ {v_R}  {\overline{v}_R} \left( {v_R}^2- {\overline{v}_R}^2\right) \left( {g_R}^2+ { {g_V} }^2-2 \lambda ^2\right)}{ {v_R}^2+ {\overline{v}_R}^2},  \nonumber \\
 {M_{35}} &=& \frac{\lambda  \left(-2  { A_{\lambda}  }  {v_R}  {\overline{v}_R}+2 \left( {v_R}^2+ {\overline{v}_R}^2\right)  {v_S} \lambda \right)}{\sqrt{ {v_R}^2+ {\overline{v}_R}^2}} , \nonumber \\
 {M_{44}} &=& [2  {g_R}^2  {v_R}^2  {\overline{v}_R}^2+2  { {g_V} }^2  {v_R}^2  {\overline{v}_R}^2+ {m_4}^2 \left( {v_R}^2+ {\overline{v}_R}^2\right)+ {m_6}^2
\left( {v_R}^2+ {\overline{v}_R}^2\right)\nonumber \\
&+& {v_R}^4 \lambda ^2-2  {v_R}^2  {\overline{v}_R}^2 \lambda ^2+ {\overline{v}_R}^4 \lambda ^2+2  {v_R}^2  {v_S}^2
\lambda ^2+2  {\overline{v}_R}^2  {v_S}^2 \lambda ^2]/\left( {v_R}^2+ {\overline{v}_R}^2\right),  \nonumber \\
 {M_{45}} &=& \frac{ { A_{\lambda} } \left( {v_R}^2- {\overline{v}_R}^2\right) \lambda }{\sqrt{ {v_R}^2+ {\overline{v}_R}^2}},  \nonumber \\
 {M_{55}} &=&  m_S^2 + (v_L^2 + \ov v_L^2 + v_R^2 +\ov v_R^2).
\end{eqnarray}

\end{appendix}

\end{document}